\documentclass[journal, 10pt, twocolumn, letterpaper, oneside, final]{IEEEtran}

\usepackage{ifpdf}

\usepackage[noadjust]{cite}

\ifCLASSINFOpdf
  \usepackage[pdftex]{graphicx}
\else
  \usepackage[dvips]{graphicx}
\fi

\usepackage{tikz}
\usetikzlibrary{
	positioning,
	arrows.meta,
	fadings,
	shapes.arrows,
	shapes.geometric,
	shadows,
	decorations.pathmorphing,
	decorations.pathreplacing,
	angles,
	quotes,
	fit
}
\tikzfading[name=arrowfading, top color=transparent!0, bottom color=transparent!95]
\tikzset{
  arrowfill/.style={
     top color=blue!20,
     bottom color=blue,
     general shadow={fill=black, shadow yshift=-0.8ex, path fading=arrowfading}
  },
  arrowstyle/.style={
     draw=red,
     arrowfill,
     single arrow,
     minimum height=#1,
     single arrow head extend=.4cm
   }
}

\usepackage[cmex10]{amsmath}
\usepackage{array}

\ifCLASSOPTIONcompsoc
  \usepackage[caption=false,font=normalsize,labelfont=sf,textfont=sf]{subfig}
\else
  \usepackage[caption=false,font=footnotesize]{subfig}
\fi

\usepackage{fixltx2e}

\usepackage{url}

\usepackage[utf8]{inputenc} 
\usepackage[T1]{fontenc}
\usepackage{ifthen}

\usepackage{mathtools}
\usepackage{amssymb}
\usepackage{relsize}
\usepackage{bbm}
\usepackage{xfrac}
\usepackage{amsthm}
\theoremstyle{plain}
\newtheorem{definition}{Definition}
\newtheorem{lemma}{Lemma}

\newtheorem{theorem}{Theorem}
\newtheorem{corollary}{Corollary}
\newtheorem{remark}{Remark}

\newtheorem{proposition}{Proposition}
\usepackage[varg, cmbraces]{newtxmath}

\usepackage{xcolor}
\definecolor{burgundy}{rgb}{0.545098,0,0}
\definecolor{navyblue}{rgb}{0.0, 0.0, 0.5}
\definecolor{leafgreen}{rgb}{0.290196, 0.470588, 0.0}
\definecolor{bluegreen}{rgb}{0, 0.470588, 0.415686}
\definecolor{zuhl}{rgb}{0.1875, 0.26171875, 0.46484375}
\definecolor{orange}{rgb}{1, 0.6470588235, 0}
\definecolor{red}{rgb}{1, 0, 0}

\usepackage{color}
\usepackage{overpic}
\usepackage{pict2e}
\usepackage{booktabs}

\newcommand{\bvec}[1]{\boldsymbol{#1}}

\newcommand{\sgn}{\operatorname{sgn}}

\newcommand{\figref}[1]{Fig.~\ref{#1}}

\newcommand{\lemref}[1]{Lemma~\ref{#1}}
\newcommand{\propref}[1]{Proposition~\ref{#1}}
\newcommand{\thref}[1]{Theorem~\ref{#1}}

\newcommand{\sectref}[1]{Section~\ref{#1}}
\newcommand{\remref}[1]{Remark~\ref{#1}}

\newcommand{\appref}[1]{Appendix~\ref{#1}}

\allowdisplaybreaks[3]

\usepackage{array}
\usepackage[linesnumbered, algoruled, boxed, lined, vlined]{algorithm2e}
\SetKwRepeat{Do}{do}{while}

\usepackage{hyperref}
\hypersetup{
    colorlinks,
    linkcolor = {red!60!black},
    citecolor = {blue!60!black},
    urlcolor = {blue!50!black}
}
\usepackage{microtype}

\begin{document}

\title{Variable-Length Source Dispersions Differ \\ under Maximum and Average Error Criteria}

\IEEEoverridecommandlockouts

\author{%
\IEEEauthorblockN{%
Yuta~Sakai,~\IEEEmembership{Member,~IEEE,} and
Vincent~Y.~F.~Tan,~\IEEEmembership{Senior~Member,~IEEE}}%
\thanks{This work is supported by a Singapore National Research Foundation (NRF) Fellowship (R-263-000-D02-281).
This article was presented in part at the 2020 IEEE International Symposium on Information Theory (ISIT2020), Los Angels, CA, USA \cite{sakai_tan_isit2020}. \emph{(Corresponding author: Yuta Sakai.)}}%
\thanks{Y.~Sakai is with the Department of Electrical and Computer Engineering, National University of Singapore, Singapore, Email: \{\url{eleyuta@nus.edu.sg}, \url{yuta.sakai@ieee.org}, \url{yuta.sakai@m.ieice.org}\}.}%
\thanks{V.~Y.~F.~Tan is with the Department of Electrical and Computer Engineering and the Department of Mathematics, National University of Singapore, Singapore, Email: \url{vtan@nus.edu.sg}.}}%

\maketitle

\begin{abstract}
Variable-length compression without prefix-free constraints and with side-information available at both encoder and decoder is considered.
Instead of requiring the code to be error-free, we allow for it to have a non-vanishing error probability.
We derive one-shot bounds on the optimal average codeword length by proposing two new information quantities; namely, the conditional and unconditional $\varepsilon$-cutoff entropies.
Using these one-shot bounds, we obtain the second-order asymptotics of the problem under two different formalisms---the average and maximum probabilities of error over the realization of the side-information.
While the first-order terms in the asymptotic expansions for both formalisms are identical, we find that the source dispersion under the average error formalism is, in most cases, strictly smaller than its maximum error counterpart.
Applications to a certain class of guessing problems, previously studied by Kuzuoka [\emph{{IEEE} Trans.\ Inf.\ Theory}, vol.~66, no.~3, pp.~1674--1690, 2020], are also discussed.
\end{abstract}


\IEEEpeerreviewmaketitle

\section{Introduction}

\IEEEPARstart{I}{n} this paper, we are concerned with the problem of variable-length compression without prefix-free constraints. In the simplest version of this problem, a source $X$ is to be compressed to finite-length binary strings.
The objective is to ensure that the average codeword length is minimized under the condition that the source code is one-to-one.
One-to-one codes have been studied by several researchers (see \cite{drmota_szpankowski_2016} and references therein).
Specifically, Wyner \cite{wyner_1972} and Alon and Orlitsky \cite{alon_orlitsky_1994} derived the following upper and lower bounds:
\begin{align}
H(X) - \log(H(X) + 1) - \log \mathrm{e}
\le
L^{\ast}(X)
\le
H(X) ,
\end{align}
respectively, where $\log$ stands for the base-2 logarithm and $L^{\ast}(X)$ stands for the minimum average codeword length of the one-to-one codes for the source $X$. A direct consequence of these bounds is that for a stationary memoryless source $X^n=(X_1,\ldots, X_n)$, one has
\begin{align}
L^{\ast}( X^{n} )
=
n \, H(X) + \mathrm{O}( \log n )
\qquad (\mathrm{as} \ n \to \infty) .
\label{eq:asympt_one-to-one}
\end{align}

The above-mentioned studies and results assume that the code is not allowed to commit any error. In practical latency-constrained applications, occasional errors are often tolerable. Hence, it is worthwhile to study the counterparts to the above zero-error results when one allows the code to have a decoding error probability $\varepsilon$ that is non-vanishing. Towards this end, Kostina, Polyanskiy, and Verd\'{u} \cite{kostina_polyanskiy_verdu_2015} showed that the fundamental limit on the average codeword length $L^{\ast}( \varepsilon, X^{n} )$, again without the prefix-free constraint, admits the following  asymptotic expansion:
\begin{align}
\!\!
L^{\ast}( \varepsilon, X^{n} )
=
n \, (1 - \varepsilon) \, H(X) - \sqrt{ n \, V(X) } \, f_{\mathrm{G}}( \varepsilon ) + \mathrm{O}( \log n ) \!\!
\label{eq:KPV}
\end{align}
as $n \to \infty$ for every $0 \le \varepsilon \le 1$.
In this expression, the quantity $V(X)$ denotes the variance of the information density of the source $X$ (often referred to as the varentropy \cite{kontoyiannis_verdu_2014}) and the map $f_{\mathrm{G}} : [0, 1] \to [0, 1/\sqrt{2 \pi}]$ is defined as
\begin{align}
f_{\mathrm{G}}( s )
& :=
\begin{cases}
\varphi( \Phi^{-1}( s ) )
& \mathrm{if} \ 0 < s < 1 ,
\\
0
& \mathrm{if} \ s = 0 \ \mathrm{or} \ s = 1 ,
\end{cases}
\label{def:Gaussian_f} \\
\varphi( t )
& :=
\frac{ 1 }{ \sqrt{ 2 \pi } } \mathrm{e}^{-t^{2}/2} ;
\end{align}
and $\Phi^{-1} : (0, 1) \to  \mathbb{R}$ denotes the inverse function of the Gaussian cumulative distribution function
\begin{align}
\Phi( u ) := \int_{-\infty}^{u} \varphi( t ) \, \mathrm{d} t .
\label{def:Gaussian_CDF}
\end{align}
This is the first instance of the second-order asymptotics \cite{strassen_1962, hayashi_2008, hayashi_2009, polyanskiy_thesis, polyanskiy_poor_verdu_2010} for variable-length compression.
It is worth noting, for our subsequent considerations, that the first-order term in  \eqref{eq:KPV} is $(1-\varepsilon)\, H(X)$ and so the strong converse property, in the sense of Wolfowitz \cite{wolfowitz_1978}, does not hold.
Additionally, since the second-order term $\sqrt{ V(X) } \, f_{\mathrm{G}}( \varepsilon )$ in \eqref{eq:KPV} is nonnegative for all $0<\varepsilon<1$, the fundamental limit $L^{\ast}( \varepsilon, X^{n} )$ for variable-length compression is always \emph{smaller} than the first-order optimal coding rate $n \, (1 - \varepsilon) \, H(X)$.
This is in contrast to, say, almost lossless fixed-length source coding \cite{strassen_1962, hayashi_2008} in which if the tolerable error probability $\varepsilon$ is less than $1/2$, the second-order term is positive, which means that the optimal code rate at a finite blocklength $n$ is \emph{larger} than the first-order term.

\subsection{Main Contributions}

In this paper, we extend this setting and result by considering the presence of side-information $Y^n$ at both encoder and decoder.
In this case, the notion of the error probability can take one of two different forms. One can consider the \emph{maximum} error probability in which we would like%
\footnote{Strictly speaking, Eq.~\eqref{eq:max-error_X_vs_Xhat} means error probability constraints except on a null set.
In other words, the error constraint is imposed on the \emph{essential supremum} of the left-hand side of \eqref{eq:max-error_X_vs_Xhat}.
However, we call it the \emph{maximum} error probability as usual.}
\begin{align}
\mathbb{P} \{ X^{n} \neq \hat{X}^{n} \mid Y^n \}
\le
\varepsilon
\qquad (\mathrm{a.s.}) .
\label{eq:max-error_X_vs_Xhat}
\end{align}
Said differently, we require that the reconstructed source $\hat{X}^{n}$ is equal to the original source sequence $X^{n}$ with probability at least $1-\varepsilon$ almost surely with respect to the side-information $Y^{n}$.
This is obviously a more stringent criterion than the \emph{average} error probability criterion in which one simply requires that
\begin{align}
\mathbb{P}\{ X^{n} \neq \hat{X}^{n} \}
\le
\varepsilon .
\end{align}
Here, the error probability is \emph{averaged} over the realizations of $Y^n$.
Clearly, the rate of compression under the maximum error criterion is at least as large as the average error criterion.
In this paper, we quantify this gap precisely in terms of the second-order asymptotics, i.e., the analogue of the term scaling $\sqrt{n}$ in \eqref{eq:KPV}.
We show that the first-order terms in the asymptotic expansions are identical and equal to $n \, (1 - \varepsilon) \, H(X \mid Y)$, but the source dispersion for the maximum error case is smaller than that of its average error counterpart.
That is, the backoff from $n\, (1-\varepsilon) \, H(X \mid Y)$ is smaller for the former, more stringent, case compared to the latter.
In fact, the maximum (resp.\ average) error source dispersion is the conditional (resp.\ unconditional) information variance of the conditional information density.
By the law of total variance, the conditional information variance is not larger than its unconditional counterpart.
It is easy to show that the difference is non-zero for most sources.
En route to proving our second-order results, we develop new and novel one-shot bounds for both these error probability formalisms.
We introduce two new information measures, namely the \emph{unconditional} and \emph{conditional $\varepsilon$-cutoff entropies}; in the $n$-shot setting, these characterize the fundamental compression limits up to terms scaling as $\mathrm{O}( \log n )$.
Finally, we discuss applications of the second-order asymptotic results to guessing problems with a ``giving-up'' policy; this class of problems was recently introduced by Kuzuoka \cite{kuzuoka_2019}.

\subsection{Related Works}

\subsubsection{Prefix-Free Codes}

The problem of variable-length compression allowing errors was initiated by Han \cite{han_2000} who considered the fundamental limits of prefix-free codes with vanishing error probability.
Han \cite{han_2000} derived a general formula for the normalized average codeword length.
A general formula allowing for non-vanishing error probabilities was derived by Koga and Yamamoto \cite{koga_yamamoto_2005}.
For a stationary memoryless source $X^{n}$ on a finite alphabet $\mathcal{X}^{n}$, Koga and Yamamoto's general formula can be reduced to the first-order term $n \, (1 - \varepsilon) \, H(X)$, which coincides with \eqref{eq:KPV} up to a term scaling as $\mathrm{o}( n )$.
In fact, as mentioned by Kuzuoka and Watanabe \cite[Remark~2]{kuzuoka_watanabe_2015}, the asymptotics of the prefix-free codes and the one-to-one codes are equal up to a constant factor.

\subsubsection{Guessing Problems}

One-to-one codes with side-information are essentially equivalent to strategies for guessing problems \cite{massey_isit1994, arikan_1996} via Campbell's source coding problem \cite{campbell_1965} without prefix-free constraints (cf.\ \cite{kosut_sankar_2017, sason_verdu_2018, sason_2018}).
Kuzuoka \cite{kuzuoka_2019} generalized the guessing problem by allowing positive error probabilities.
The guesser can also give up guessing at each stage; in this case, an error is declared.
Kuzuoka \cite{kuzuoka_2019} derived general formulas of both Campbell's source coding problems and guessing problems with non-vanishing error probability by introducing the conditional smooth R\'{e}nyi entropy and by exploiting its properties.

\subsubsection{Conditional Rate-Distortion Theory and State-Dependent Channels}

A related topic to our present considerations is the conditional rate-distortion problem \cite{gray_1972, gray_1973}.
Gray considered the problem of lossy compression with common side-information at both encoder and decoder.
The duality between source coding and state-dependent channel coding problems with side-information available at both encoder and decoder have been characterized by Cover and Chiang \cite{cover_chiang_2002} and Pradhan, Chou, and Ramchandran \cite{pradhan_chou_ramchandran_2003}.

\subsubsection{Variable-Length Slepian--Wolf Coding}

He, Lastras-Monta\~{n}o, Yang, Jagmohan, and Chen \cite{he_montano_yang_jagmohan_chen_2009} investigated fixed- and variable-length Slepian--Wolf coding problems \cite{slepian_wolf_1973} with error probabilities that vanish but not exponentially fast.
They derived the second-order coding rates and showed that variable-length Slepian--Wolf coding has a better second-order term compared to its fixed-length counterpart.
These are characterized by some forms of the conditional and unconditional information variances, and the superiority of variable-length Slepian--Wolf coding is characterized by these differences.
Variable-length Slepian--Wolf coding problems were also investigated by Kimura and Uyematsu \cite{kimura_uyematsu_2004} and Kuzuoka and Watanabe \cite{kuzuoka_watanabe_2015}.

\subsection{Paper Organization}

This paper is organized as follows:
The problem setting is formulated in \sectref{sect:pre}.
\sectref{eq:measure} presents some definitions and notations of information measures for a correlated source $(X, Y)$.
The $\varepsilon$-cutoff entropies are defined in \sectref{sect:cutoff}.
\sectref{sect:code} introduces the variable-length conditional lossless source coding problems.
The main results of this study are given in \sectref{sect:main}.
Specifically, the second-order asymptotics of variable-length compression under maximum and average error formalisms are stated in Theorems~\ref{th:max} and~\ref{th:avg} of \sectref{sect:second-order}, respectively.
Our one-shot coding theorems are stated in Lemmas~\ref{lem:eps-cutoff_max-err} and~\ref{lem:eps-cutoff_avg-err} of Sections~\ref{sect:proof-max} and~\ref{sect:proof-avg}, respectively; those are used to prove Theorems~\ref{th:max} and~\ref{th:avg}, respectively.
Applications of Theorems~\ref{th:max} and~\ref{th:avg} to guessing problems with a ``giving-up policy'' are discussed in \sectref{sect:guessing}.
\sectref{sect:conclusion} concludes this study.

\section{Preliminaries}
\label{sect:pre}

\subsection{Information Measures for Correlated Sources}
\label{eq:measure}

Assume throughout that the underlying probability space $(\Omega, \mathcal{F}, \mathbb{P})$ is rich enough so that all random variables (r.v.'s) are well-defined on the space.
Consider a countably infinite alphabet%
\footnote{In this study, assume that the $\sigma$-algebra on a countable alphabet is always the power set of the alphabet, as usual.}
$\mathcal{X} = \{ 1, 2, \dots \}$ and an abstract alphabet $\mathcal{Y}$.
Let $X$ be an $\mathcal{X}$-valued r.v.\ and $Y$ a $\mathcal{Y}$-valued r.v.
Then, the pair $(X, Y)$ can be thought of as a correlated source pair.

In the conditional source coding, the second source $Y$ plays the role of the side-information for the first source $X$.
We now introduce several information quantities.
Let $P_{X|Y}( x \mid Y )$ be a version of the conditional probability $\mathbb{P}\{ X = x \mid Y \}$ for each $x \in \mathcal{X}$.%
\footnote{Note that $P_{X|Y}(\cdot \mid Y)$ is a probability measure on $\mathcal{X}$ almost surely (a.s.) because the conditional probability is $\sigma$-additive.}
Denote by
\begin{align}
\iota(X \mid Y)
=
\iota_{X|Y}(X \mid Y)
\coloneqq
\log \frac{ 1 }{ P_{X|Y}(X \mid Y) }
\end{align}
the conditional information density of $X$ given $Y$.
Define three $\sigma(Y)$-measurable information measures of $X$ given $Y$ as follows:
\begin{align}
\mathcal{H}(X \mid Y)
& \coloneqq
\mathbb{E}[ \iota(X \mid Y) \mid Y ] ,
\label{def:mathcal-H} \\
\mathcal{V}(X \mid Y)
& \coloneqq
\mathbb{E}[ (\iota(X \mid Y) - \mathcal{H}(X \mid Y))^{2} \mid Y ] ,
\label{def:mathcal-V} \\
\mathcal{T}(X \mid Y)
& \coloneqq
\mathbb{E}[ |\iota(X \mid Y) - \mathcal{H}(X \mid Y)|^{3} \mid Y ] ,
\label{def:mathcal-T}
\end{align}
where $\mathbb{E}[Z \mid W]$ stands for the conditional expectation of a real-valued r.v.\ $Z$ given a sub-$\sigma$-algebra $\sigma(W)$ generated by a r.v.\ $W$.
When $\mathcal{Y}$ is countable, we analogously define
\begin{align}
\mathcal{H}(X \mid y)
& \coloneqq
\sum_{x \in \mathcal{X}} P_{X|Y}(x \mid y) \log \frac{ 1 }{ P_{X|Y}(x \mid y) } ,
\\
\mathcal{V}(X \mid y)
& \coloneqq
\sum_{x \in \mathcal{X}} P_{X|Y}(x \mid y) \left( \log \frac{ 1 }{ P_{X|Y}(x \mid y) } - \mathcal{H}(X \mid y) \right)^{2} ,
\\
\mathcal{T}(X \mid y)
& \coloneqq
\sum_{x \in \mathcal{X}} P_{X|Y}(x \mid y) \left| \log \frac{ 1 }{ P_{X|Y}(x \mid y) } - \mathcal{H}(X \mid y) \right|^{3} ,
\end{align}
for each $y \in \mathcal{Y}$, provided that $P_{Y}( y ) \coloneqq \mathbb{P}\{ Y = y \} > 0$, where $P_{X|Y}(x \mid y) \coloneqq \mathbb{P}\{ X = x \mid Y = y \}$ stands for the conditional probability given the event $\{ Y = y \}$.
Moreover, we define four information measures of $X$ given $Y$ as follows:
\begin{align}
H(X \mid Y)
& \coloneqq
\mathbb{E}[ \mathcal{H}(X \mid Y) ] ,
\label{def:cond_H} \\
V_{\mathrm{c}}(X \mid Y)
& \coloneqq
\mathbb{E}[ \mathcal{V}(X \mid Y) ] ,
\label{def:cond_inf_variance} \\
V_{\mathrm{u}}(X \mid Y)
& \coloneqq
\mathbb{E}[ (\iota(X \mid Y) - H(X \mid Y))^{2} ] ,
\label{def:uncond_inf_variance} \\
T_{\mathrm{u}}(X \mid Y)
& \coloneqq
\mathbb{E}[ |\iota(X \mid Y) - H(X \mid Y)|^{3} ] ,
\label{def:uncond_T}
\end{align}
where $\mathbb{E}[ Z ]$ stands for the expectation of a real-valued r.v.\ $Z$.

\begin{remark}
The somewhat unconventional notations $\mathcal{H}(X \mid Y)$, $\mathcal{V}(X \mid Y)$, and $\mathcal{T}(X \mid Y)$ defined in \eqref{def:mathcal-H}--\eqref{def:mathcal-T}, respectively, are introduced to indicate that they are \emph{$\sigma(Y)$-measurable r.v.'s,} i.e., they are not deterministic quantities like $H(X \mid Y)$, $V_{\mathrm{c}}(X \mid Y)$, $V_{\mathrm{u}}(X \mid Y)$, and $T_{\mathrm{u}}(X \mid Y)$.
The notation $\mathcal{H}(X \mid Y)$ was also adopted in \cite{berlin_nakiboglu_rimoldi_telatar_2009}.
\end{remark}

The quantity $H(X \mid Y)$ is the well-known conditional Shannon entropy of $X$ given $Y$.
In this study, we respectively call $V_{\mathrm{c}}(X \mid Y)$ and $V_{\mathrm{u}}(X \mid Y)$ the \emph{conditional} and \emph{unconditional information variances}%
\footnote{These terminologies are inspired by Polyanskiy's second-order asymptotic analysis in the channel coding problem \cite[Equations~(3.97)--(3.100)]{polyanskiy_thesis}.}
of $X$ given $Y$.
It follows by the law of total variance that
\begin{align}
V_{\mathrm{u}}(X \mid Y)
=
V_{\mathrm{c}}(X \mid Y) + \mathbb{E}[ (\mathcal{H}(X \mid Y) - H(X \mid Y))^{2} ] .
\label{eq:total_variance}
\end{align}
Thus, the unconditional information variance $V_{\mathrm{u}}(X \mid Y)$ is larger than $V_{\mathrm{c}}(X \mid Y)$ by the term $\mathbb{E}[ (\mathcal{H}(X \mid Y) - H(X \mid Y))^{2} ]$, and these variances coincide if and only if $\mathcal{H}(X \mid Y)$ is almost surely constant.

\subsection{$\varepsilon$-Cutoff Entropies}
\label{sect:cutoff}

Given a real-valued r.v.\ $Z$, define the (unconditional) \emph{$\varepsilon$-cutoff transformation action of $Z$} by
\begin{align}
\langle Z \rangle_{\varepsilon}
\coloneqq
\begin{cases}
Z
& \mathrm{if} \ Z < \eta ,
\\
B \, Z
& \mathrm{if} \ Z = \eta ,
\\
0
& \mathrm{if} \ Z > \eta , 
\end{cases}
\label{def:cutoff}
\end{align}
where $B$ denotes a Bernoulli r.v.\ with parameter $(1 - \beta)$ in which the independence $B \Perp Z$ holds, and two real parameters $\eta \in \mathbb{R}$ and $0 \le \beta < 1$ are chosen so that
\begin{align}
\mathbb{P}\{ Z > \eta \} + \beta \, \mathbb{P}\{ Z = \eta \}
=
\varepsilon .
\label{eq:eta_unconditional}
\end{align}
This is the same definition as \cite[Equation~(13)]{kostina_polyanskiy_verdu_2015}, and the notation $\langle Z \rangle_{\varepsilon}$ is consistent with that used in \cite{kostina_polyanskiy_verdu_2015}.
In addition, given a real-valued r.v.\ $Z$ and an arbitrary r.v.\ $W$, define the \emph{conditional $\varepsilon$-cutoff transformation action of $Z$ given $W$} by
\begin{align}
\langle Z \mid W \rangle_{\varepsilon}
\coloneqq
\begin{cases}
Z
& \mathrm{if} \ Z < \eta_{W} ,
\\
B_{W} \, Z
& \mathrm{if} \ Z = \eta_{W} ,
\\
0
& \mathrm{if} \ Z > \eta_{W} , 
\end{cases}
\label{def:cond_cutoff}
\end{align}
where $B_{W}$ denotes a Bernoulli r.v.\ in which the conditional independence $B_{W} \Perp Z \mid W$ holds and
\begin{align}
\mathbb{P}\{ B_{W} = 0 \mid W \}
=
\beta_{W}
\qquad (\mathrm{a.s.}) ,
\end{align}
and two $\sigma(W)$-measurable real-valued r.v.'s $\eta_{W} \in \mathbb{R}$ and $0 \le \beta_{W} < 1$ are chosen so that
\begin{align}
\mathbb{P}\{ Z > \eta_{W} \mid W \} + \beta_{W} \, \mathbb{P}\{ Z = \eta_{W} \mid W \}
=
\varepsilon
\qquad (\mathrm{a.s.}) .
\label{eq:eta_W}
\end{align}
In this paper, we call $\eta$ and $\eta_{W}$ the \emph{cutoff points} of $\langle Z \rangle_{\varepsilon}$ and $\langle Z \mid W \rangle_{\varepsilon}$, respectively.

Using these cutoff operations, we now define the \emph{unconditional} and \emph{conditional $\varepsilon$-cutoff entropies} as follows:
\begin{align}
\mathfrak{C}_{\mathrm{u}}^{\varepsilon}(X \mid Y)
& \coloneqq
\mathbb{E}[ \langle \iota(X \mid Y) \rangle_{\varepsilon} ] ,
\label{def:unconditional_eps-entropy} \\
\mathfrak{C}_{\mathrm{c}}^{\varepsilon}(X \mid Y)
& \coloneqq
\mathbb{E}[ \langle \iota(X \mid Y) \mid Y \rangle_{\varepsilon} ] ,
\label{def:conditional_eps-entropy}
\end{align}
respectively.
Note that these $\varepsilon$-cutoff entropies are not additive in general.

Finally, the following proposition gives some basic properties of the $\varepsilon$-cutoff transformation actions.

\begin{proposition}
\label{prop:expectation_cutoff}
Let $Z$ be a nonnegative-valued r.v., $W$ a $\mathcal{W}$-valued r.v.\ with an abstract alphabet $\mathcal{W}$, and $0 \le \varepsilon \le 1$ a real number.
Then, it holds that
\begin{align}
\mathbb{E}[ \langle Z \rangle_{\varepsilon} ]
& =
\min_{\epsilon : \mathbb{E}[ \epsilon(Z) ] \le \varepsilon} \mathbb{E}[ (1 - \epsilon(Z)) \, Z ] ,
\label{eq:exp_uncond-cutoff_min} \\
\mathbb{E}[ \langle Z \mid W \rangle_{\varepsilon} ]
& =
\min_{\epsilon : \mathbb{E}[ \epsilon(Z, W) \mid W ] \le \varepsilon \, (\mathrm{a.s.})} \mathbb{E}[ (1 - \epsilon(Z, W)) \, Z ] ,
\label{eq:exp_cond-cutoff_min}
\end{align}
where the minimization in \eqref{eq:exp_uncond-cutoff_min} (resp.\ \eqref{eq:exp_cond-cutoff_min}) is taken over the measurable maps $\epsilon : [0, \infty) \to [0, 1]$ (resp.\ the measurable maps $\epsilon : [0, \infty) \times \mathcal{W} \to [0, 1]$) satisfying $\mathbb{E}[ \epsilon(Z) ] \le \varepsilon$ (resp.\ $\mathbb{E}[ \epsilon(Z, W) \mid W ] \le \varepsilon$ a.s.).
Moreover, the following identities hold:
\begin{align}
\mathbb{E}[ \langle Z \rangle_{\varepsilon} ]
& =
(1 - \varepsilon) \, \mathbb{E}[ Z ]
\notag \\
& \quad {}
- \int_{\eta}^{\infty} \mathbb{P}\{ Z > t \} \, \mathrm{d} t - \varepsilon \, (\eta - \mathbb{E}[ Z ]) ,
\label{eq:uncond-cutoff_int-spectrum} \\
\mathbb{E}[ \langle Z \mid W \rangle_{\varepsilon} \mid W ]
& =
(1 - \varepsilon) \, \mathbb{E}[ Z \mid W ]
\notag \\
& \quad {}
- \int_{\eta_{W}}^{\infty} \mathbb{P}\{ Z > t \mid W \} \, \mathrm{d} t
\notag \\
& \qquad {}
- \varepsilon \, (\eta_{W} - \mathbb{E}[Z \mid W])
\qquad (\mathrm{a.s.}) ,
\label{eq:cond-cutoff_int-spectrum}
\end{align}
where $\eta$ and $\eta_{W}$ are given in \eqref{eq:eta_unconditional} and \eqref{eq:eta_W} respectively.
In addition, the following inequality holds:
\begin{align}
\mathbb{E}[ \langle Z \rangle_{\varepsilon} ]
\le
\mathbb{E}[ \langle Z \mid W \rangle_{\varepsilon} ] .
\label{eq:inequality_cutoff_cond}
\end{align}
Furthermore, given two nonnegative-valued r.v.'s $Z_{1}$ and $Z_{2}$, it holds that
\begin{align}
&
\mathbb{P}\{ Z_{1} \le t \} \le \mathbb{P}\{ Z_{2} \le t \} \ (\forall t > 0)
\notag \\
& \ \Longrightarrow \quad
\mathbb{E}[ \langle Z_{1} \rangle_{\varepsilon} ] \ge \mathbb{E}[ \langle Z_{2} \rangle_{\varepsilon} ] ,
\label{eq:stochastic-dominance_uncond-cutoff} \\
&
\mathbb{P}\{ Z_{1} \le t \mid W \} \le \mathbb{P}\{ Z_{2} \le t \mid W \} \ (\mathrm{a.s.} \ \forall t > 0)
\notag \\
& \ \Longrightarrow \quad
\mathbb{E}[ \langle Z_{1} \mid W \rangle_{\varepsilon} \mid W] \ge \mathbb{E}[ \langle Z_{2} \mid W \rangle_{\varepsilon} \mid W] \ (\mathrm{a.s.}) .
\label{eq:stochastic-dominance_cond-cutoff}
\end{align}
\end{proposition}

\begin{IEEEproof}[Proof of \propref{prop:expectation_cutoff}]
See \appref{app:expectation_cutoff}.
\end{IEEEproof}

Note that \eqref{eq:uncond-cutoff_int-spectrum} and \eqref{eq:cond-cutoff_int-spectrum} in \propref{prop:expectation_cutoff} are useful in the subsequent second-order asymptotic analysis of the $\varepsilon$-cutoff entropies $\mathfrak{C}_{\mathrm{u}}^{\varepsilon}(X \mid Y)$ and $\mathfrak{C}_{\mathrm{c}}^{\varepsilon}(X \mid Y)$, respectively.
The identities \eqref{eq:exp_uncond-cutoff_min} and \eqref{eq:exp_cond-cutoff_min} will be used in the proofs of one-shot bounds stated in Lemmas~\ref{lem:eps-cutoff_avg-err} and \ref{lem:eps-cutoff_max-err}, respectively, of Sections~\ref{sect:proof-max} and \ref{sect:proof-avg}, respectively.
It follows from \eqref{eq:inequality_cutoff_cond} of \propref{prop:expectation_cutoff} that
\begin{align}
\mathfrak{C}_{\mathrm{u}}^{\varepsilon}(X \mid Y)
\le
\mathfrak{C}_{\mathrm{c}}^{\varepsilon}(X \mid Y) .
\end{align}

\subsection{Variable-Length Compression Under Two Error Criteria}
\label{sect:code}

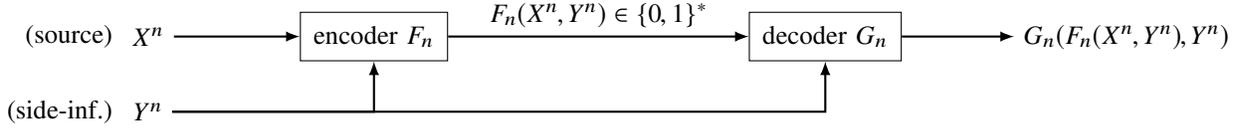
\begin{figure*}[!t]
\centering
\begin{tikzpicture}
\node (x) at (0, 0) {$X^{n}$};
\node (y) at (0, -1) {$Y^{n}$};
\node (f) at (3, 0) [draw, rectangle, inner sep = 5pt] {encoder $F_{n}$};
\node (g) at (9, 0) [draw, rectangle, inner sep = 5pt] {decoder $G_{n}$};
\node (hat-x) at (13, 0) {$G_{n}(F_{n}(X^{n}, Y^{n}), Y^{n})$};
\draw [-latex, thick] (x) node [left = 1em] {(source)} -- (f);
\draw [-latex, thick] (f) -- node [above] {$F_{n}(X^{n}, Y^{n}) \in \{ 0, 1 \}^{\ast}$} (g);
\draw [-latex, thick] (y) node [left = 1em] {(side-inf.)} -| (f);
\draw [-latex, thick] (y) -| (g);
\draw [-latex, thick] (g) -- (hat-x);
\end{tikzpicture}
\caption{Variable-length conditional source coding}
\label{fig:FV-code}
\end{figure*}

Given an integer $n \ge 1$, denote by $(X_{1}, Y_{1}), (X_{2}, Y_{2}), \dots, (X_{n}, Y_{n})$ $n$ i.i.d. copies of the source pair $(X, Y)$.
Then, we may think of $(X^{n}, Y^{n})$ as a sequence of outputs from the stationary memoryless correlated source $(X, Y)$, where $X^{n} = (X_{1}, \dots, X_{n})$ and $Y^{n} = (Y_{1}, \dots, Y_{n})$.
In this subsection, we formalize the variable-length conditional (almost) lossless source coding problems.
Let $\{ 0, 1 \}^{\ast}$ be the set of finite-length binary strings containing the empty string $\varnothing$.
For each $n \ge 1$, consider two random maps $F_{n} : \mathcal{X}^{n} \times \mathcal{Y}^{n} \to \{ 0, 1 \}^{\ast}$ and $G_{n} : \{ 0, 1 \}^{\ast} \times \mathcal{Y}^{n} \to \mathcal{X}^{n}$ in which both $F_{n}(X^{n}, Y^{n})$ and $G_{n}(F_{n}(X^{n}, Y^{n}), Y^{n})$ are $\mathcal{F}$-measurable.
Then, we call the pair $(F_{n}, G_{n})$ a \emph{variable-length stochastic code} for the source $X^{n}$ with side-information $Y^{n}$ available at both encoder $F_{n}$ and decoder $G_{n}$.
We illustrate this compression scheme in \figref{fig:FV-code}.

\begin{remark}
\label{eq:another_stochastic-code}
Another way to consider a variable-length stochastic code is to design a $\{ 0, 1 \}^{\ast}$-valued r.v.\ $B_{n}$ and an $\mathcal{X}^{n}$-valued r.v.\ $\hat{X}^{n}$ in which those probability laws are determined by versions of the conditional probabilities $\mathbb{P}\{ B_{n} = \bvec{b} \mid X^{n}, Y^{n} \}$ for $\bvec{b} \in \{ 0, 1 \}^{\ast}$ and $\mathbb{P}\{ \hat{X}^{n} = \bvec{x} \mid B_{n}, Y^{n} \}$ for $\bvec{x} \in \mathcal{X}^{n}$, respectively.
Kostina \emph{et al.}\ \cite{kostina_polyanskiy_verdu_2015} studied variable-length stochastic codes without side-information $Y^{n}$ in this manner.
\end{remark}

Let $\ell : \{ 0, 1 \}^{\ast} \to \mathbb{N} \cup \{ 0 \}$ be the length function of a finite-length binary string; e.g., $\ell( \varnothing ) = 0$, $\ell( 0 ) = \ell( 1 ) = 1$, $\ell( 00 ) = \ell( 01 ) = \ell( 10 ) = \ell( 11 ) = 2$, $\ell( 000 ) = \ell( 001 ) = \ell( 010 ) = \ell( 011 ) = \ell( 100 ) = \ell( 101 ) = \ell( 110 ) = \ell( 111 ) = 3$, and so on.
Given a variable-length stochastic code $(F_{n}, G_{n})$, we are interested in the average codeword length $\mathbb{E}[ \ell( F_{n}(X^{n}, Y^{n}) ) ]$ to measure the efficiency of the data compressor for the source $X^{n}$ with side-information $Y^{n}$.

\begin{definition}[Maximum error criterion]%
\label{def:sup}
Let $n \ge 1$ be an integer, and $L \ge 0$ and $0 \le \varepsilon \le 1$ real numbers.
Given a source $X$ with side-information $Y$, an \emph{$(n, L, \varepsilon)_{\max}$-code} is a variable-length stochastic code $(F_{n}, G_{n})$ satisfying
\begin{align}
\mathbb{E}[ \ell( F_{n}( X^{n}, Y^{n} ) ) ]
& \le
L ,
\\
\mathbb{P}\{ X^{n} \neq G_{n}( F_{n}( X^{n}, Y^{n} ), Y^{n} ) \mid Y^{n} \}
& \le
\varepsilon 
\qquad (\mathrm{a.s.}) .
\end{align}
\end{definition}

\begin{definition}[Average error criterion]
\label{def:avg}
Let $n \ge 1$ be an integer, and $L \ge 0$ and $0 \le \varepsilon \le 1$ real numbers.
Given a source $X$ with side-information $Y$, an \emph{$(n, L, \varepsilon)_{\mathrm{avg}}$-code} is a variable-length stochastic code $(F_{n}, G_{n})$ satisfying
\begin{align}
\mathbb{E}[ \ell( F_{n}( X^{n}, Y^{n} ) ) ]
& \le
L ,
\\
\mathbb{P}\{ X^{n} \neq G_{n}( F_{n}( X^{n}, Y^{n} ), Y^{n} ) \}
& \le
\varepsilon .
\end{align}
\end{definition}

Given a probability of error $0 \le \varepsilon \le 1$, this study deals with the fundamental limits of the average codeword length under these error criteria.
Specifically, we will investigate the two operational quantities $L_{\max}^{\ast}(n, \varepsilon, X, Y)$ and $L_{\mathrm{avg}}^{\ast}(n, \varepsilon, X, Y)$ defined in \eqref{def:L_sup} and \eqref{def:L_avg}, respectively, and presented at the top of the next page.

\begin{figure*}[!t]
\hrule
\begin{align}
L_{\max}^{\ast}(n, \varepsilon, X, Y)
& \coloneqq
\inf \{ L > 0 \mid \text{there exists an $(n, L, \varepsilon)_{\max}$-code for the source $X$ with side-information $Y$} \}
\label{def:L_sup} \\
L_{\mathrm{avg}}^{\ast}(n, \varepsilon, X, Y)
& \coloneqq
\inf \{ L > 0 \mid \text{there exists an $(n, L, \varepsilon)_{\mathrm{avg}}$-code for the source $X$ with side-information $Y$} \}
\label{def:L_avg}
\end{align}
\hrule
\end{figure*}

\section{Second-Order Asymptotics and \\ One-Shot Bounds}
\label{sect:main}

\subsection{Statements of Second-Order Asymptotic Results}
\label{sect:second-order}

Recall the definition of the function $f_{\mathrm{G}} : [0, 1] \to [0, 1/\sqrt{2 \pi}]$ in \eqref{def:Gaussian_f}.

\begin{theorem}[Under maximum error criterion]
\label{th:max}
Suppose that the following two hypotheses hold:%
\footnote{We say that a real-valued r.v.\ $Z$ is \emph{bounded away from zero} (resp.\ \emph{infinity}) \emph{almost surely} if there exists a positive constant $c$ satisfying $\mathbb{P}\{ |Z| > c \} = 1$ (resp.\ $\mathbb{P}\{ |Z| < c \} = 1$).}
\begin{itemize}
\item[(a)]
$\mathcal{V}(X \mid Y)$ is bounded away from zero almost surely; and
\item[(b)]
$\mathcal{T}(X \mid Y)$ is bounded away from infinity almost surely.
\end{itemize}
Then, it holds that
\begin{align}
L_{\max}^{\ast}(n, \varepsilon, X, Y)
& =
n \, (1 - \varepsilon) \, H(X \mid Y)
\notag \\
& \qquad {}
- \sqrt{ n \, V_{\mathrm{c}}(X \mid Y) } \, f_{\mathrm{G}}( \varepsilon ) + \mathrm{O}( \log n )
\label{eq:max_asympt}
\end{align}
as $n \to \infty$ for every $0 \le \varepsilon \le 1$.
\end{theorem}

\begin{IEEEproof}[Proof of \thref{th:max}]
See \sectref{sect:proof-max}.
\end{IEEEproof}

\begin{theorem}[Under average error criterion]
\label{th:avg}
Suppose that $T_{\mathrm{u}}(X \mid Y)$ is finite.
Then, it holds that
\begin{align}
L_{\mathrm{avg}}^{\ast}(n, \varepsilon, X, Y)
& =
n \, (1 - \varepsilon) \, H(X \mid Y)
\notag \\
& \qquad {}
- \sqrt{ n \, V_{\mathrm{u}}(X \mid Y) } \, f_{\mathrm{G}}( \varepsilon ) + \mathrm{O}( \log n )
\label{eq:avg_asympt}
\end{align}
as $n \to \infty$ for every $0 \le \varepsilon \le 1$.
\end{theorem}

\begin{IEEEproof}[Proof of \thref{th:avg}]
See \sectref{sect:proof-avg}.
\end{IEEEproof}

Since an $(n, L, \varepsilon)_{\max}$-code is an $(n, L, \varepsilon)_{\mathrm{avg}}$-code, it is clear that
\begin{align}
L_{\mathrm{avg}}^{\ast}(n, \varepsilon, X, Y)
\le
L_{\max}^{\ast}(n, \varepsilon, X, Y) .
\label{eq:trivial-ineq_avg_sup}
\end{align}
Theorems~\ref{th:max} and~\ref{th:avg} state that the first-order optimal coding rates are the same under both the maximum and average error criteria; they are equal to $n \, (1 - \varepsilon) \, H(X \mid Y)$.
On the other hand, we see from \eqref{eq:total_variance} that unless $\mathcal{H}(X \mid Y)$ is almost surely constant, the optimal second-order coding rates differ under maximum and average error criteria.
Since $f_{\mathrm{G}}( \varepsilon ) \ge 0$ with equality if and only if either $\varepsilon = 0$ or $\varepsilon = 1$, note in Theorems~\ref{th:max} and~\ref{th:avg} that a larger dispersion implies a shorter average codeword length on the $\sqrt{n}$ scale for every fixed $0 < \varepsilon < 1$.
In particular, it follows from \eqref{eq:total_variance} that the variable-length source dispersion $V_{\mathrm{u}}(X \mid Y)$ under the average error criterion is larger than that $V_{\mathrm{c}}(X \mid Y)$ under the maximum error criterion by the term $\mathbb{E}[ (\mathcal{H}(X \mid Y) - H(X \mid Y))^{2} ]$.

\begin{remark}
In channel coding, the counterparts of both conditional and unconditional information variances coincide for every capacity-achieving input distribution (cf.\ \cite[Lemma~46]{polyanskiy_thesis} or \cite[Lemma~62]{polyanskiy_poor_verdu_2010}).
Since the first-order term determines the choice of input distribution (cf.\ \cite[Lemma~48]{polyanskiy_thesis}), the $\varepsilon$-channel dispersion is determined by a capacity-achieving input distribution.
Therefore, there is no difference between the conditional and unconditional information variances in channel coding in the absence of input cost constraints.
On the other hand, the conditional and unconditional information variances are different for the problem at hand as there is no optimization over input distributions.
Thus, the variable-length source dispersion under the maximum and average error formalisms are different.
\end{remark}

\begin{remark}
In Theorems~\ref{th:max} and~\ref{th:avg}, the code is allowed to be stochastic.
Namely, an encoder (resp.\ a decoder) outputs a compressed binary string $B$ (resp.\ the reconstructed source $\hat{X}$) stochastically according to some probability law given a source $X$ (resp.\ a compressed binary string $B$) and the side-information $Y$ (see \remref{eq:another_stochastic-code}).
Since the average codeword length of a stochastic code is almost equal to that of a deterministic code up to a constant additive term of $(\log \mathrm{e})/\mathrm{e}$ (cf. \cite[Section~II-A]{kostina_polyanskiy_verdu_2015}), our asymptotic analysis is the same as that if we assumed the code is deterministic.
See also \cite[Remark~6]{kuzuoka_2019}.
\end{remark}

In \appref{app:ex}, we show that even if the side-information alphabet $\mathcal{Y}$ is countably infinite, there is a correlated source $(X, Y)$ that $\mathcal{V}(X \mid Y)$ and $\mathcal{T}(X \mid Y)$ are respectively not bounded away from zero and infinity a.s., but $T_{\mathrm{u}}(X \mid Y)$ is finite.
Therefore, Hypotheses~(a) and~(b) in \thref{th:max} are stronger than the hypothesis in \thref{th:avg} in general.
On the other hand, these hypotheses are no longer needed if the alphabets of the source and the side-information are finite.

\begin{proposition}
\label{prop:finite-X}
If $X$ is supported on some finite subalphabet $\mathcal{A} \subset \mathcal{X}$, then Hypothesis \emph{(b)} in \thref{th:max} holds.
\end{proposition}

\begin{proposition}
\label{prop:finite-Y}
If $\mathcal{Y}$ is finite, then \eqref{eq:max_asympt} in \thref{th:max} holds without Hypothesis \emph{(a)}.
\end{proposition}

\begin{IEEEproof}[Proof of Propositions~\ref{prop:finite-X} and~\ref{prop:finite-Y}]
See \appref{app:relaxation}.
\end{IEEEproof}

\subsection{Proof of \thref{th:max}}
\label{sect:proof-max}

Firstly, we establish one-shot bounds on the fundamental limit $L_{\max}^{\ast}(\varepsilon, X, Y) \coloneqq L_{\max}^{\ast}(1, \varepsilon, X, Y)$; namely, we consider the case in which $n = 1$.
The following lemma gives us a formula for $L_{\max}^{\ast}(\varepsilon, X, Y)$.

\begin{lemma}
\label{lem:optimal-code_max}
Given $0 \le \varepsilon \le 1$, it holds that
\begin{align}
L_{\max}^{\ast}(\varepsilon, X, Y)
=
\mathbb{E}[ \langle \lfloor \log \varsigma_{Y}^{-1}( X ) \rfloor \mid Y \rangle_{\varepsilon} ] ,
\label{eq:optimal-code_max}
\end{align}
where $\varsigma_{Y}$ stands for a random permutation on $\mathcal{X} \coloneqq \{ 1, 2, \dots \}$ satisfying
\begin{align}
P_{X|Y}(\varsigma_{Y}( 1 ) \mid Y)
\ge
P_{X|Y}(\varsigma_{Y}( 2 ) \mid Y)
\ge
\cdots
\quad (\mathrm{a.s.}) ,
\label{def:sigmaY}
\end{align}
which rearranges the probability masses in $P_{X|Y}(\cdot \mid Y)$ in non-increasing order.
\end{lemma}

\begin{IEEEproof}[Proof of \lemref{lem:optimal-code_max}]
See \appref{app:optimal-code_max}.
\end{IEEEproof}

\lemref{lem:optimal-code_max} tells us that in an optimal variable-length stochastic code $(F^{\ast}, G^{\ast})$ achieving \eqref{eq:optimal-code_max}, given the side-information $Y$, the set of source symbols that result in an error are those $x$ with the smallest $P_{X|Y}(x \mid Y)$.
Moreover, we see that the cutoff point $\eta_{Y}$ of $\langle \lfloor \log \varsigma_{Y}^{-1}( X ) \rfloor \mid Y \rangle_{\varepsilon}$ depends on the side-information $Y$, and the conditional overflow probability of this optimal code can be evaluated as
\begin{align}
\mathbb{P}\{ \ell( F^{\ast}(X, Y) ) \le \eta_{Y} \mid Y \}
\ge
1 - \varepsilon
\qquad (\mathrm{a.s.}) .
\end{align}

Using \lemref{lem:optimal-code_max}, we provide the following one-shot bounds on $L_{\max}^{\ast}(\varepsilon, X, Y)$ in terms of the conditional $\varepsilon$-cutoff entropy $\mathfrak{C}_{\mathrm{c}}^{\varepsilon}$ defined in \eqref{def:conditional_eps-entropy}.

\begin{lemma}
\label{lem:eps-cutoff_max-err}
For every $0 \le \varepsilon \le 1$, it holds that
\begin{align}
&
\mathfrak{C}_{\mathrm{c}}^{\varepsilon}(X \mid Y) - \log( H(X \mid Y) + 1 ) - \log \mathrm{e}
\notag \\
& \qquad \qquad \qquad \qquad \le
L_{\max}^{\ast}(\varepsilon, X, Y)
\le
\mathfrak{C}_{\mathrm{c}}^{\varepsilon}(X \mid Y) .
\label{eq:eps-cutoff_max-err_one-shot}
\end{align}
\end{lemma}

\begin{IEEEproof}[Proof of \lemref{lem:eps-cutoff_max-err}]
See \appref{app:eps-cutoff_max-err}.
\end{IEEEproof}

Note that \lemref{lem:eps-cutoff_max-err} holds without Hypotheses (a) and (b) in \thref{th:max}.
Since $H(X \mid Y) < \infty$ if Hypothesis (b) in \thref{th:max} holds, and since $H(X^{n} \mid Y^{n}) = n \, H(X \mid Y)$ for each $n \ge 1$, \lemref{lem:eps-cutoff_max-err} implies that
\begin{align}
L_{\max}^{\ast}(n, \varepsilon, X, Y)
=
\mathfrak{C}_{\mathrm{c}}^{\varepsilon}(X^{n} \mid Y^{n}) + \mathrm{O}( \log n )
\label{eq:hypothesis_max}
\end{align}
as $n \to \infty$.
Thus, it suffices to provide an appropriate asymptotic estimate on $\mathfrak{C}_{\mathrm{c}}^{\varepsilon}(X^{n} \mid Y^{n})$.

\begin{lemma}
\label{lem:asympt_cond_eps-entropy}
Given a fixed $0 \le \varepsilon \le 1$, it holds that
\begin{align}
\mathfrak{C}_{\mathrm{c}}^{\varepsilon}(X^{n} \mid Y^{n})
& =
n \, (1-\varepsilon) \, H(X \mid Y)
\notag \\
& \qquad {}
- \mathbb{E}\Big[ \sqrt{ \mathcal{V}(X^{n} \mid Y^{n}) } \Big] \, f_{\mathrm{G}}( \varepsilon ) + \mathrm{O}(1)
\label{eq:asympt_cond_eps-entropy}
\end{align}
as $n \to \infty$, provided that Hypotheses \emph{(a)} and \emph{(b)} in \thref{th:max} hold.
\end{lemma}

\begin{IEEEproof}[Proof of \lemref{lem:asympt_cond_eps-entropy}]
See \appref{app:asympt_cond_eps-entropy}.
\end{IEEEproof}

Unfortunately, obtaining an exact single-letter expression for the $n$-letter ``dispersion'' term $\mathbb{E}[ \sqrt{ \mathcal{V}(X^{n} \mid Y^{n}) } ]$ that appears in \eqref{eq:asympt_cond_eps-entropy} is difficult unless $\mathcal{V}(X \mid Y)$ is almost surely constant.
In fact, it can be verified by Jensen's inequality that
\begin{align}
\mathbb{E}\Big[ \sqrt{ \mathcal{V}(X^{n} \mid Y^{n}) } \Big]
\le
\sqrt{ n \, V_{\mathrm{c}}(X \mid Y) } .
\label{eq:Jensen_V}
\end{align}
with equality if and only if $\mathcal{V}(X \mid Y)$ is almost surely constant, because $\mathcal{V}(X_{1} \mid Y_{1}), \mathcal{V}(X_{2} \mid Y_{2}), \dots, \mathcal{V}(X_{n} \mid Y_{n})$ will then be $n$ i.i.d.\ copies of $\mathcal{V}(X \mid Y)$.
Hence, Lemmas~\ref{lem:eps-cutoff_max-err} and~\ref{lem:asympt_cond_eps-entropy} can be readily reduced to Kostina \emph{et al.}'s result \cite[Theorem~4]{kostina_polyanskiy_verdu_2015} in \eqref{eq:KPV}, provided  that $X$ and $Y$ are independent.

The following lemma provides an asymptotic estimate of $\mathbb{E}[ \sqrt{ \mathcal{V}(X^{n} \mid Y^{n}) } ]$.

\begin{lemma}
\label{lem:max_V}
If $\mathcal{V}(X \mid Y)$ is bounded away from infinity almost surely, then
\begin{align}
\mathbb{E}\Big[ \sqrt{ \mathcal{V}(X^{n} \mid Y^{n}) } \Big]
=
\sqrt{ n \, V_{\mathrm{c}}(X \mid Y) } + \mathrm{O}( \sqrt{ \log n } )
\label{eq:Vmax_asympt}
\end{align}
as $n \to \infty$.
\end{lemma}

\begin{IEEEproof}[Proof of \lemref{lem:max_V}]
See \appref{app:max_V}.
\end{IEEEproof}

Hypothesis (b) in \thref{th:max} implies that $\mathcal{V}(X \mid Y)$ is bounded away from infinity almost surely; therefore, Lemmas~\ref{lem:eps-cutoff_max-err}--\ref{lem:max_V} yield \thref{th:max}, as desired.
\hfill\IEEEQEDhere

\subsection{Proof of \thref{th:avg}}
\label{sect:proof-avg}

Firstly, we establish one-shot bounds on the fundamental limits $L_{\mathrm{avg}}^{\ast}(\varepsilon, X, Y) \coloneqq L_{\mathrm{avg}}^{\ast}(1, \varepsilon, X, Y)$, namely, consider the case in which $n = 1$.
The following lemma gives us a formula for $L_{\mathrm{avg}}^{\ast}(\varepsilon, X, Y)$.

\begin{lemma}
\label{lem:optimal-code_avg}
Given $0 \le \varepsilon \le 1$, it holds that
\begin{align}
L_{\mathrm{avg}}^{\ast}(\varepsilon, X, Y)
=
\mathbb{E}[ \langle \lfloor \log \varsigma_{Y}^{-1}( X ) \rfloor \rangle_{\varepsilon} ] ,
\label{eq:optimal-code_avg}
\end{align}
where $\varsigma_{Y}$ is defined in \eqref{def:sigmaY}.
\end{lemma}

\begin{IEEEproof}[Proof of \lemref{lem:optimal-code_avg}]
See \appref{app:optimal-code_avg}.
\end{IEEEproof}

Unlike \lemref{lem:optimal-code_max}, \lemref{lem:optimal-code_avg} tells us that in an optimal variable-length stochastic code $(F^{\ast}, G^{\ast})$ achieving \eqref{eq:optimal-code_avg}, the set of source symbols that result in an error are those with the smallest $\mathbb{P}\{ X = \varsigma_{Y}(x) \}$.
Moreover, we see that the cutoff point $\eta$ of $\langle \lfloor \log \varsigma_{Y}^{-1}( X ) \rfloor \rangle_{\varepsilon}$ is independent of the side-information $Y$, and the overflow probability of this optimal code can be evaluated as
\begin{align}
\mathbb{P}\{ \ell( F^{\ast}(X, Y) ) \le \eta \}
\ge
1 - \varepsilon .
\end{align}

Using \lemref{lem:optimal-code_avg}, we provide the following one-shot bounds on $L_{\mathrm{avg}}^{\ast}(\varepsilon, X, Y)$ in terms of the unconditional $\varepsilon$-cutoff entropy $\mathfrak{C}_{\mathrm{u}}^{\varepsilon}$ defined in \eqref{def:unconditional_eps-entropy}.

\begin{lemma}
\label{lem:eps-cutoff_avg-err}
For every $0 \le \varepsilon \le 1$, it holds that
\begin{align}
&
\mathfrak{C}_{\mathrm{u}}^{\varepsilon}(X \mid Y) - \log( H(X \mid Y) + 1 ) - \log \mathrm{e}
\notag \\
& \qquad \qquad \qquad \qquad \le
L_{\mathrm{avg}}^{\ast}(\varepsilon, X, Y)
\le
\mathfrak{C}_{\mathrm{u}}^{\varepsilon}(X \mid Y) .
\label{eq:eps-cutoff_max-avg}
\end{align}
\end{lemma}

\begin{IEEEproof}[Proof of \lemref{lem:eps-cutoff_avg-err}]
Employing \propref{prop:expectation_cutoff} and \lemref{lem:optimal-code_avg}, we can prove \lemref{lem:eps-cutoff_avg-err} by the same manner as the proof of \lemref{lem:eps-cutoff_max-err}; see \appref{app:eps-cutoff_max-err} for details.
This completes the proof of \lemref{lem:eps-cutoff_avg-err}.
\end{IEEEproof}

Note that \lemref{lem:eps-cutoff_avg-err} holds without Hypotheses (a) and (b) in \thref{th:avg}.
Since $H(X \mid Y) < \infty$ if $T_{\mathrm{u}}(X \mid Y) < \infty$, and since $H(X^{n} \mid Y^{n}) = n \, H(X \mid Y)$ for each $n \ge 1$, \lemref{lem:eps-cutoff_avg-err} tells us that
\begin{align}
L_{\mathrm{avg}}^{\ast}(n, \varepsilon, X, Y)
=
\mathfrak{C}_{\mathrm{u}}^{\varepsilon}(X^{n} \mid Y^{n}) + \mathrm{O}( \log n )
\label{eq:hypothesis_avg}
\end{align}
as $n \to \infty$.
Thus, it suffices to provide an appropriate asymptotic estimate on $\mathfrak{C}_{\mathrm{u}}^{\varepsilon}(X^{n} \mid Y^{n})$.

\begin{lemma}
\label{lem:asympt_uncond_eps-entropy}
Given a fixed $0 \le \varepsilon \le 1$, it holds that
\begin{align}
\mathfrak{C}_{\mathrm{u}}^{\varepsilon}(X^{n} \mid Y^{n})
& =
n \, (1 - \varepsilon) \, H(X \mid Y)
\notag \\
& \qquad {}
- \sqrt{ n \, V_{\mathrm{u}}(X \mid Y) } \, f_{\mathrm{G}}( \varepsilon ) + \mathrm{O}( 1 )
\end{align}
as $n \to \infty$, provided that $T_{\mathrm{u}}(X \mid Y) < \infty$.
\end{lemma}

\begin{IEEEproof}[Proof of \lemref{lem:asympt_uncond_eps-entropy}]
Since $\iota(X^{n} \mid Y^{n}) = \iota(X_{1} \mid Y_{1}) + \iota(X_{2} \mid Y_{2}) + \cdots + \iota(X_{n} \mid Y_{n})$, and since $\iota(X_{1} \mid Y_{1})$, $\iota(X_{2} \mid Y_{2})$, \dots, $\iota(X_{n} \mid Y_{n})$ are $n$ i.i.d.\ real-valued r.v.'s, a na\"{i}ve application of \cite[Lemma~1]{kostina_polyanskiy_verdu_2015} readily proves \lemref{lem:asympt_uncond_eps-entropy}.
For the readers' convenience, we now only give a sketch of the proof as follows:
If $V_{\mathrm{c}}(X \mid Y) = 0$, then we readily see that
\begin{align}
\mathfrak{C}_{\mathrm{u}}^{\varepsilon}(X^{n} \mid Y^{n})
=
n \, (1 - \varepsilon) \, H(X \mid Y) .
\end{align}
Thus, it suffices to consider the case where $V_{\mathrm{c}}(X \mid Y) > 0$.
It follows from \eqref{eq:uncond-cutoff_int-spectrum} of \propref{prop:expectation_cutoff} that
\begin{align}
\mathfrak{C}_{\mathrm{u}}^{\varepsilon}(X^{n} \mid Y^{n})
& =
n \, (1 - \varepsilon) \, H(X \mid Y)
\notag \\
& \quad {}
- \int_{\eta_{n}}^{\infty} \mathbb{P}\{ \iota(X^{n} \mid Y^{n}) > t \} \, \mathrm{d} t
\notag \\
& \qquad {}
- \varepsilon \, (\eta_{n} - n \, H(X \mid Y)) ,
\label{eq:frak-U_formula}
\end{align}
where $\eta_{n} > 0$ is given so that
\begin{align}
\mathbb{P}\{ \iota(X^{n} \mid Y^{n}) > \eta_{n} \} + \beta_{n} \, \mathbb{P}\{ \iota(X^{n} \mid Y^{n}) = \eta_{n} \}
=
\varepsilon
\end{align}
with an appropriate $0 \le \beta_{n} < 1$.
Then, the uniform Berry--Esseen bound (cf.\ \eqref{eq:unif_Berry-Esseen} in \appref{app:asympt_cond_eps-entropy}) shows that
\begin{align}
\eta_{n}
=
n \, H(X \mid Y) + \sqrt{ n \, V_{\mathrm{u}}(X \mid Y) } \, \Phi^{-1}( 1-\varepsilon ) + \mathrm{O}( 1 )
\label{eq:eta_n}
\end{align}
as $n \to \infty$, provided that $T_{\mathrm{u}}(X \mid Y) < \infty$.
On the other hand, it can be verified by the non-uniform Berry--Esseen bound (cf.\ \lemref{lem:non-unif_Berry-Esseen} in \appref{app:asympt_cond_eps-entropy}) that
\begin{align}
&
\int_{\eta}^{\infty} \mathbb{P}\{ \iota(X^{n} \mid Y^{n}) > t \} \, \mathrm{d} t
\notag \\
& \qquad =
\sqrt{ n \, V_{\mathrm{u}}(X \mid Y) } \, \Big( f_{\mathrm{G}}( \varepsilon ) - \varepsilon \, \Phi^{-1}(1 - \varepsilon) \Big) + \mathrm{O}( 1 )
\label{eq:int-spectrum_unconditional}
\end{align}
as $n \to \infty$, provided that $T_{\mathrm{u}}(X \mid Y) < \infty$.
Therefore, \lemref{lem:asympt_uncond_eps-entropy} can be proven by combining \eqref{eq:frak-U_formula}, \eqref{eq:eta_n}, and \eqref{eq:int-spectrum_unconditional}.
\end{IEEEproof}

The proof of \thref{th:avg} is finally completed by combining Lemmas~\ref{lem:eps-cutoff_avg-err} and~\ref{lem:asympt_uncond_eps-entropy}.
\hfill\IEEEQEDhere

\section{Guessing Problem}
\label{sect:guessing}

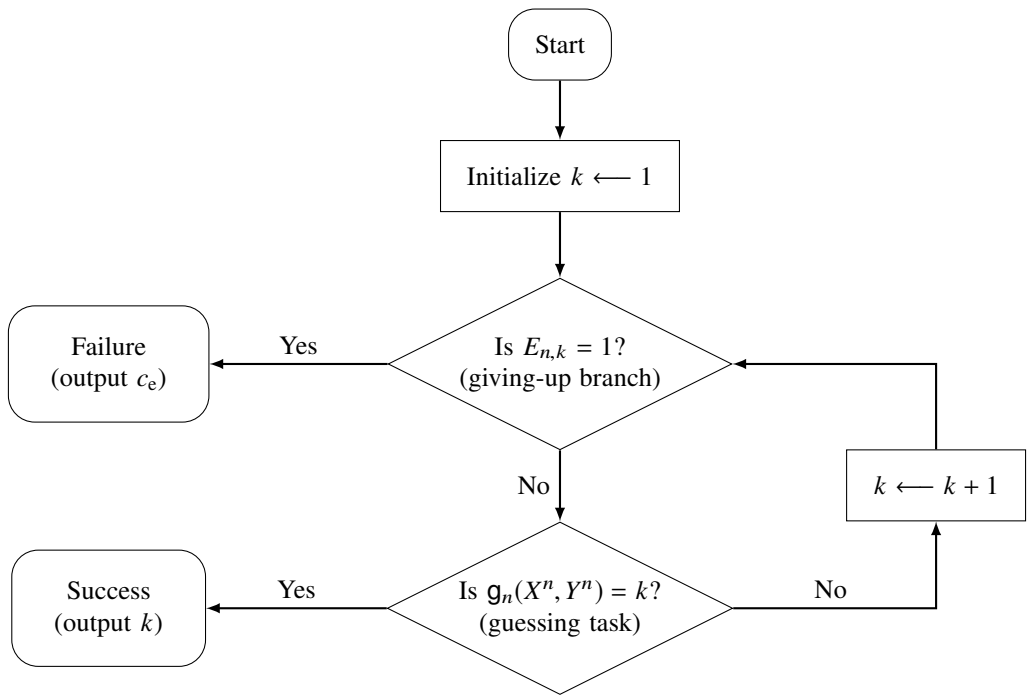
\begin{figure*}[!t]
\centering
\begin{tikzpicture}
\node (start) at (0, 0) [draw, rectangle, rounded corners = 10pt, inner sep = 10pt] {Start};
\node (init) at (0, -1.75) [draw, rectangle, inner sep = 10pt] {Initialize $k \longleftarrow 1$};
\node (error) at (0, -4.25) [draw, diamond, aspect = 2, inner sep = -1pt] {%
\begin{tabular}{c}
Is $E_{n, k} = 1$? \\
(giving-up branch)
\end{tabular}};
\node (error-output) at (-6, -4.25) [draw, rectangle, rounded corners = 10pt, inner sep = 10pt] {%
\begin{tabular}{c}
Failure \\
(output $c_{\mathrm{e}}$)
\end{tabular}};
\node (guessing) at (0, -7.5) [draw, diamond, aspect = 2, inner sep = -1pt] {%
\begin{tabular}{c}
Is $\mathsf{g}_{n}(X^{n}, Y^{n}) = k$? \\
(guessing task)
\end{tabular}};
\node (guessing-output) at (-6, -7.5) [draw, rectangle, rounded corners = 10pt, inner sep = 10pt] {%
\begin{tabular}{c}
Success \\
(output $k$)
\end{tabular}};
\node (add-k) at (5, -5.865) [draw, rectangle, inner sep = 10pt] {$k \longleftarrow k + 1$};
\draw [-latex, thick] (start) -- (init);
\draw [-latex, thick] (init) -- (error);
\draw [-latex, thick] (error) -- node [above] {Yes} (error-output);
\draw [-latex, thick] (error) -- node [left] {No} (guessing);
\draw [-latex, thick] (guessing) -- node [above] {Yes} (guessing-output);
\draw [-latex, thick] (guessing) -| node [above left = 0em and 3em] {No} (add-k);
\draw [-latex, thick] (add-k) |- (error);
\end{tikzpicture}
\caption{Flowchart of a guessing strategy with a giving-up policy $(\mathsf{g}_{n}, \pi_{n})$.}
\label{fig:guessing}
\end{figure*}

Following \cite[Section~III]{kuzuoka_2019}, we now introduce the guessing problem with a ``giving-up'' policy.
Let $n \ge 1$ be an integer.
Consider a stationary memoryless correlated source $(X^{n}, Y^{n})$ as in \sectref{sect:code}.
In the guessing problem, given the side-information $Y^{n}$, a guesser, equipped with a \emph{guessing strategy} $\{ \bvec{x}_{k} \}_{k = 1}^{\infty}$ depending on $Y^{n}$, asks questions of the form ``Is $X^{n} = \bvec{x}_{k}$?'' at each time $k \in \mathbb{N}$.
More precisely, a guessing strategy is induced by a \emph{guessing function} $\mathsf{g}_{n} : \mathcal{X}^{n} \times \mathcal{Y}^{n} \to \mathbb{N}$, which is a deterministic map in which $\mathsf{g}_{n}(\cdot, \bvec{y}) : \mathcal{X}^{n} \to \mathbb{N}$ is bijective for each $\bvec{y} \in \mathcal{Y}^{n}$, as follows:
The guesser asks ``Is $\mathsf{g}_{n}(X^{n}, Y^{n}) = 1$?'' at time $1$; if the answer is ``no,'' the guesser asks ``Is $\mathsf{g}_{n}(X^{n}, Y^{n}) = 2$?'' at time $2$; if the answer is again ``no,'' the guesser asks ``Is $\mathsf{g}_{n}(X^{n}, Y^{n}) = 3$?'' at time $3$, and so on.
By introducing a certain error probability for guessing, the guesser can, instead of committing to a ``yes/no'' answer at each time $k\in\mathbb{N}$, give up at the same time $k$ with a certain probability.
For each $(k, \bvec{y}) \in \mathcal{X} \times \mathcal{Y}^{n}$, let $0 \le \pi_{n}(k \mid \bvec{y}) \le 1$ be a real number.
The collection $\{ \pi_{n}(k \mid \bvec{y}) \mid (k, \bvec{y}) \in \mathbb{N} \times \mathcal{Y}^{n} \}$  plays the role of a \emph{giving-up policy.}
Just before starting on the $k$-th guess, the guesser can give up his guessing task with probability $\pi_{n}(k \mid Y^{n})$.
We call the pair $(\mathsf{g}_{n}, \pi_{n}(\cdot \mid \cdot))$ a \emph{guessing strategy with a giving-up policy.}
Formally, for each $k \ge 1$, the guesser declares an error just before starting on the $k$-th guess if $E_{n, 1} = E_{n, 2} = \dots = E_{n, k-1} = 0$ and $E_{n, k} = 1$, where $\{ E_{n, k} \}_{k = 1}^{\infty}$ denotes a sequence of conditionally (and mutually) independent Bernoulli r.v.'s given $Y^{n}$ in which
\begin{align}
\mathbb{P}\{ E_{n, k} = 1 \mid Y^{n} \}
=
\pi_{n}(k \mid Y^{n})
\qquad (\mathrm{a.s.})
\end{align}
for every $k \ge 1$.
Then, the \emph{giving-up guessing function} $\mathsf{G}_{n} : \mathcal{X}^{n} \times \mathcal{Y}^{n} \to \mathbb{N} \cup \{ c_{\mathrm{e}} \}$ is a random map given as 
\begin{align}
\mathsf{G}_{n}(X^{n}, Y^{n})
& \coloneqq
\begin{cases}
\mathsf{g}_{n}(X^{n}, Y^{n})
& \mathrm{if} \ E_{n, l} = 0
\\
& \mathrm{for} \ \mathrm{all} \ 1 \le l \le \mathsf{g}_{n}(X^{n}, Y^{n}) ,
\\
c_{\mathrm{e}}
& \mathrm{otherwise} ,
\end{cases}
\end{align}
where $c_{\mathrm{e}} > 0$ denotes the cost of making an error.%
\footnote{For simplicity of our analysis, we assume that $c_{\mathrm{e}}$ is not an integer.
This assumption simplifies the guessing error event $\{ \mathsf{G}_{n}(X^{n}, Y^{n}) \neq \mathsf{g}_{n}(X^{n}, Y^{n}) \}$, and does not affect the results in \cite{kuzuoka_2019} and Corollaries~\ref{cor:max_guessing} and~\ref{cor:avg_guessing} under a valid definition of the error event.}
Figure~\ref{fig:guessing} illustrates a flowchart of executing a particular giving-up guessing function $\mathsf{G}_{n} : \mathcal{X}^{n} \times \mathcal{Y}^{n} \to \mathbb{N} \cup \{ c_{\mathrm{e}} \}$.

While Kuzuoka investigated the fundamental limits of the $\rho$-th moment $\mathbb{E}[ \mathsf{G}_{n}(X^{n}, Y^{n})^{\rho} ]$ with a fixed real $\rho > 0$ to evaluate the guessing cost (see \cite[Equation~(33)]{kuzuoka_2019}), we are now interested in the fundamental limits of $\mathbb{E}[ \log \mathsf{G}_{n}(X^{n}, Y^{n}) ]$.
In fact, if $\mathbb{E}[ \mathsf{G}_{n}(X^{n}, Y^{n})^{\rho} ]$ is finite for some $\rho > 0$, then it follows by l'H\^{o}pital's rule and the dominated convergence theorem that
\begin{align}
\lim_{\rho \to 0^{+}} \frac{ 1 }{ \rho } \log \mathbb{E}[ \mathsf{G}_{n}(X^{n}, Y^{n})^{\rho} ]
=
\mathbb{E}[ \log \mathsf{G}_{n}(X^{n}, Y^{n}) ] ,
\label{eq:l'Hopital}
\end{align}
so our study in this section can be thought of as a limiting case of that in \cite{kuzuoka_2019}.
Noting that errors are declared if and only if $\mathsf{G}_{n}(X^{n}, Y^{n}) \neq \mathsf{g}_{n}(X^{n}, Y^{n})$, we define two error formalisms as follows:

\begin{definition}[Maximum error criterion]
Given a source $X$ with side-information $Y$, an \emph{$(n, N, \varepsilon)_{\max}$-guessing strategy} is a guessing strategy $(\mathsf{g}_{n}, \pi_{n}(\cdot \mid \cdot))$ satisfying
\begin{align}
\mathbb{E}[ \log \mathsf{G}_{n}(X^{n}, Y^{n}) ]
& \le
N ,
\\
\mathbb{P}\{ \mathsf{G}_{n}(X^{n}, Y^{n}) \neq \mathsf{g}_{n}(X^{n}, Y^{n}) \mid Y^{n} \}
& \le
\varepsilon
\qquad (\mathrm{a.s.}) .
\end{align}
\end{definition}

\begin{definition}[Average error criterion]
Given a source $X$ with side-information $Y$, an \emph{$(n, N, \varepsilon)_{\mathrm{avg}}$-guessing strategy} is a guessing strategy $(\mathsf{g}_{n}, \pi_{n}(\cdot \mid \cdot))$ satisfying
\begin{align}
\mathbb{E}[ \log \mathsf{G}_{n}(X^{n}, Y^{n}) ]
& \le
N ,
\\
\mathbb{P}\{ \mathsf{G}_{n}(X^{n}, Y^{n}) \neq \mathsf{g}_{n}(X^{n}, Y^{n}) \}
& \le
\varepsilon .
\end{align}
\end{definition}

Given a probability of error $0 \le \varepsilon \le 1$, we investigate the two operational quantities $N_{\max}^{\ast}(n, \varepsilon, X, Y)$ and $N_{\mathrm{avg}}^{\ast}(n, \varepsilon, X, Y)$ defined in \eqref{def:N_sup} and \eqref{def:N_avg}, respectively, and presented at the top of the next page.

\begin{figure*}[!t]
\hrule
\begin{align}
\hspace{-0.5em}
N_{\max}^{\ast}(n, \varepsilon, X, Y)
& \coloneqq
\inf \{ N > 0 \mid \text{there exists an $(n, N, \varepsilon)_{\max}$-guessing strategy for the source $X$ with side-information $Y$} \}
\label{def:N_sup} \\
\hspace{-0.5em}
N_{\mathrm{avg}}^{\ast}(n, \varepsilon, X, Y)
& \coloneqq
\inf \{ N > 0 \mid \text{there exists an $(n, N, \varepsilon)_{\mathrm{avg}}$-guessing strategy for the source $X$ with side-information $Y$} \}
\label{def:N_avg}
\end{align}
\hrule
\end{figure*}

\begin{corollary}[Under maximum error criterion]
\label{cor:max_guessing}
Suppose Hypotheses~\emph{(a)} and~\emph{(b)} in \thref{th:max}.
Then, it holds that
\begin{align}
N_{\max}^{\ast}(n, \varepsilon, X, Y)
& =
n \, (1 - \varepsilon) \, H(X \mid Y)
\notag \\
& \qquad {}
- \sqrt{ n \, V_{\mathrm{c}}(X \mid Y) } \, f_{\mathrm{G}}( \varepsilon ) + \mathrm{O}( \log n )
\label{eq:max_guessing_asympt}
\end{align}
as $n \to \infty$ for every $0 \le \varepsilon \le 1$.
\end{corollary}

\begin{corollary}[Under average error criterion]
\label{cor:avg_guessing}
Suppose that $T_{\mathrm{u}}(X \mid Y)$ is finite.
Then, it holds that
\begin{align}
N_{\mathrm{avg}}^{\ast}(n, \varepsilon, X, Y)
& =
n \, (1 - \varepsilon) \, H(X \mid Y)
\notag \\
& \qquad {}
- \sqrt{ n \, V_{\mathrm{u}}(X \mid Y) } \, f_{\mathrm{G}}( \varepsilon ) + \mathrm{O}( \log n )
\label{eq:avg_guessing_asympt}
\end{align}
as $n \to \infty$ for every $0 \le \varepsilon \le 1$.
\end{corollary}

\begin{IEEEproof}[Proof of Corollaries~\ref{cor:max_guessing} and~\ref{cor:avg_guessing}]
Relying on Theorems~\ref{th:max} and~\ref{th:avg}, it suffices to prove the following lemma:

\begin{lemma}
\label{lem:coding_vs_guessing}
For every $n \ge 1$, every $0 \le \varepsilon \le 1$, and every correlated source $(X, Y)$, it holds that
\begin{align}
| N_{\max}^{\ast}(n, \varepsilon, X, Y) - L_{\max}^{\ast}(n, \varepsilon, X, Y) |
& \le
1 + |\log c_{\mathrm{e}}| ,
\label{eq:Nmax_vs_Lmax} \\
| N_{\mathrm{avg}}^{\ast}(n, \varepsilon, X, Y) - L_{\mathrm{avg}}^{\ast}(n, \varepsilon, X, Y) |
& \le
1 + |\log c_{\mathrm{e}}| .
\label{eq:Navg_vs_Lavg}
\end{align}
\end{lemma}

\lemref{lem:coding_vs_guessing} is proven in \appref{app:guessing}, completing the proof of Corollaries~\ref{cor:max_guessing} and~\ref{cor:avg_guessing}.
\end{IEEEproof}

Note that \lemref{lem:coding_vs_guessing} is analogous to \cite[Lemma~7]{sason_verdu_2018} which is stated in terms of the moment generating functions of $\ell( F_{n}(X^{n}, Y^{n}) )$ and $\mathsf{G}_{n}(X^{n}, Y^{n})$ in the error-free setting (i.e., $\varepsilon = 0$) and in the absence of side-information $Y$.
From \lemref{lem:coding_vs_guessing}, it is worth pointing out that the asymptotic results of Corollaries~\ref{cor:max_guessing} and~\ref{cor:avg_guessing} continue to hold even if the error cost $c_{\mathrm{e}}$ grows polynomially in $n$.

\section{Concluding Remarks}
\label{sect:conclusion}

We considered two variable-length conditional lossless source coding problems in this paper.
We derived one-shot coding theorems and the second-order asymptotic results under two error formalisms: the maximum and the average probabilities of error.
The one-shot bounds of Lemmas~\ref{lem:eps-cutoff_max-err} and \ref{lem:eps-cutoff_avg-err} are stated in terms of the $\varepsilon$-cutoff entropies $\mathfrak{C}_{\mathrm{c}}^{\varepsilon}(X \mid Y)$ and $\mathfrak{C}_{\mathrm{u}}^{\varepsilon}(X \mid Y)$, respectively.
These one-shot bounds are generalizations of Kostina \emph{et al.}'s one-shot coding theorem \cite[Theorem~2]{kostina_polyanskiy_verdu_2015} to the case in which side-information $Y$ is available at both encoder and decoder.
On the other hand, it should be noted that Kostina \emph{et al.}'s argument in the converse proof written in \cite[Section~II-A]{kostina_polyanskiy_verdu_2015} contains a minor error.
Indeed, one can construct a counterexample that Kostina \emph{et al.}'s converse proof fails to hold; see \appref{app:counterexample} for a counterexample.
To circumvent this issue, we explicitly provided our converse proofs of Lemmas~\ref{lem:optimal-code_max} and~\ref{lem:optimal-code_avg} in Appendices~\ref{app:optimal-code_max} and \ref{app:optimal-code_avg}, respectively.
The variable-length source dispersions under the maximum and average error criteria were derived by proving asymptotic estimates on the $\varepsilon$-cutoff entropies $\mathfrak{C}_{\mathrm{c}}^{\varepsilon}(X^{n} \mid Y^{n})$ and $\mathfrak{C}_{\mathrm{u}}^{\varepsilon}(X^{n} \mid Y^{n})$ in Lemmas~\ref{lem:asympt_cond_eps-entropy} and \ref{lem:asympt_uncond_eps-entropy}, respectively.
These two lemmas are again generalizations of Kostina \emph{et al.}'s asymptotic result for the case without side-information in \cite[Lemma~1]{kostina_polyanskiy_verdu_2015} to the case in which side-information $Y^{n}$ is available at both encoder and decoder.
On the other hand, there is a slight subtlety in our proof, in which we have to establish the \emph{uniform boundedness property} of the remainder term, represented by the quantity $+\mathrm{O}( 1 )$, with respect to the side-information $Y^{n}$; see \lemref{lem:remainder_terms} stated in \appref{app:asympt_cond_eps-entropy} for a technical statement of this uniform boundedness property.
In \sectref{sect:guessing}, we showed that our results can be applicable to Kuzuoka's guessing problem \cite[Section~III]{kuzuoka_2019}.

Recently, the present authors \cite{sakai_tan_2020, sakai_tan_smooth-ISIT} derived asymptotic expansions of smooth R\'{e}nyi entropies and their conditional versions and using these expansions, established fundamental limits of various information-theoretic problems including Campbell's source coding problem \cite{campbell_1965}, the guessing problem studied by Massey \cite{massey_isit1994} and Ar{\i}kan \cite{arikan_1996} (see also recent work by Kuzuoka \cite{kuzuoka_2019}), and the task encoding problem \cite{bunte_lapidoth_2014}, all allowing errors.
We then showed that the \emph{first-order terms} of the asymptotic expansions of these fundamental limits differ under average and maximum error criteria, where the the average and maximum are taken with respect to the available side information $Y$, as in Definitions~\ref{def:sup} and~\ref{def:avg}.
The differences between the asymptotic expansions derived in \cite{sakai_tan_2020, sakai_tan_smooth-ISIT} and the main results in this paper arise from the following differences in the problem settings:
While Theorems~\ref{th:max} and~\ref{th:avg} evaluate the ordinary expectation $\mathbb{E}[ \ell( F_{n}(X^{n}, Y^{n}) ) ]$, Campbell's source coding problem considered in \cite{sakai_tan_2020, sakai_tan_smooth-ISIT} evaluates the cumulant generating function of codeword lengths, i.e., $\log \mathbb{E}[ 2^{\rho \ell( F_{n}(X^{n}, Y^{n}) )} ]$ for fixed $\rho > 0$.
Similarly, while Corollaries~\ref{cor:max_guessing} and~\ref{cor:avg_guessing} evaluate the right-hand side of \eqref{eq:l'Hopital}, the guessing problem considered in \cite{sakai_tan_2020, sakai_tan_smooth-ISIT} evaluates the left-hand side of \eqref{eq:l'Hopital} without taking the limit as $\rho \to 0^{+}$.
Moreover, Campbell's source coding problem imposes a prefix-free constraint.
Finally, it is worth mentioning that the difference between the first-order terms derived in \cite{sakai_tan_2020, sakai_tan_smooth-ISIT} under the average and maximum error criteria can be characterized by the law of total variance, as in \eqref{eq:total_variance}, Theorems~\ref{th:max} and~\ref{th:avg}, and Corollaries~\ref{cor:max_guessing} and~\ref{cor:avg_guessing}.

In this study, we investigated the infimum of $L > 0$ in which $\mathbb{E}[ \ell( F_{n}(X^{n}, Y^{n}) ) ] \le L$ under the average error criterion $\mathbb{P}\{ X \neq G_{n}(F_{n}(X^{n}, Y^{n}), Y^{n}) \} \le \varepsilon$ and the maximum error criterion $\mathbb{P}\{ X \neq G_{n}(F_{n}(X^{n}, Y^{n}), Y^{n}) \mid Y^{n} \} \le \varepsilon$ (a.s.); see Definitions~\ref{def:sup} and~\ref{def:avg}, respectively.
A natural avenue for future research is to examine the infimum of $L > 0$ in which $\mathbb{E}[ \ell( F_{n}(X^{n}, Y^{n}) ) \mid Y^{n} ] \le L$ (a.s.) under these error criteria.
In addition, comparing the present setting and results to the variable-length Slepian--Wolf coding problem \cite{he_montano_yang_jagmohan_chen_2009, slepian_wolf_1973, kimura_uyematsu_2004, kuzuoka_watanabe_2015} would be of interest.

\appendices

\section{Proof of \propref{prop:expectation_cutoff}}
\label{app:expectation_cutoff}

\subsection{Proofs of \eqref{eq:exp_uncond-cutoff_min} and \eqref{eq:exp_cond-cutoff_min}}

The identity \eqref{eq:exp_uncond-cutoff_min} is stated in \cite[Equation~(38)]{kostina_polyanskiy_verdu_2015}, and can be thought of as a special case of \eqref{eq:exp_cond-cutoff_min} in which $\sigma(W)$ is the trivial $\sigma$-algebra $\{ \emptyset, \Omega \}$.
Thus, it suffices to prove \eqref{eq:exp_cond-cutoff_min}.

Let $\epsilon^{\ast} : [0, \infty) \times \mathcal{W} \to [0, 1]$ be the measurable map given as
\begin{align}
\epsilon^{\ast}( z, W )
=
\begin{cases}
0
& \mathrm{if} \ z < \eta_{W} ,
\\
\beta_{W}
& \mathrm{if} \ z = \eta_{W} ,
\\
1
& \mathrm{if} \ z > \eta_{W}
\end{cases}
\label{def:randomE-ast}
\end{align}
for each $z \ge 0$, where the $\sigma(W)$-measurable r.v.'s $\eta_{W} \ge 0$ and $0 \le \beta_{W} < 1$ are given in \eqref{eq:eta_W}.
It is clear from \eqref{eq:eta_W} and \eqref{def:randomE-ast} that
\begin{align}
\mathbb{E}[ \epsilon^{\ast}(Z, W) \mid W ]
=
\varepsilon
\qquad (\mathrm{a.s.}) .
\label{eq:E-ast_cond}
\end{align}
After some algebra, we get
\begin{align}
\mathbb{P}\{ \langle Z \mid W \rangle_{\varepsilon} > t \mid W \}
& =
\mathbb{P}\{ (1 - \epsilon^{\ast}(Z, W)) \, Z > t \mid W \}
\notag \\
& =
\begin{cases}
\mathbb{P}\{ Z > t \mid W \}
& \mathrm{if} \ t < 0 ,
\\
\mathbb{P}\{ Z > t \mid W \} - \varepsilon
& \mathrm{if} \ 0 \le t < \eta_{W} ,
\\
0
& \mathrm{if} \ t \ge \eta_{W}
\end{cases}
\label{eq:complement-cdf_eps-cutoff_cond}
\end{align}
a.s.
Therefore, the two r.v.'s $\langle Z \mid W \rangle_{\varepsilon}$ and $(1 - \epsilon^{\ast}(Z, W)) \, Z$ are equal in distribution, which implies that
\begin{align}
\mathbb{E}[ \langle Z \mid W \rangle_{\varepsilon} ]
=
\mathbb{E}[ (1 - \epsilon^{\ast}(Z, W)) \, Z ] .
\label{eq:E-cutoff_E-ast}
\end{align}
Consider an arbitrary measurable map $\epsilon : [0, \infty) \times \mathcal{W} \to [0, 1]$ satisfying
\begin{align}
\mathbb{E}[ \epsilon(Z, W) \mid W ]
\le
\varepsilon
\qquad (\mathrm{a.s.}) .
\label{eq:E_cond}
\end{align}
Denoting by $\bvec{1}_{A}$ the indicator function of $A \subset \Omega$, a direct calculation shows that \eqref{eq:E_vs_E-ast} written in the top of the next page holds,
\begin{figure*}[!t]
\begin{align}
&
\mathbb{E}[ (\epsilon(Z, W) - \epsilon^{\ast}(Z, W)) \, Z \mid W ]
\notag \\
& \qquad =
\mathbb{E}[ (\epsilon(Z, W) - \epsilon^{\ast}(Z, W)) \, Z \, (\bvec{1}_{\{ Z < \eta_{W} \}} + \bvec{1}_{\{ Z = \eta_{W} \}} + \bvec{1}_{\{ Z > \eta_{W} \}}) \mid W ]
\notag \\
& \qquad \overset{\mathclap{\text{(a)}}}{=}
\mathbb{E}[ \epsilon(Z, W) \, Z \, \bvec{1}_{\{ Z < \eta_{W} \}} \mid W ] + \mathbb{E}[ (\epsilon(Z, W) - \beta_{W}) \, Z \, \bvec{1}_{\{ Z = \eta_{W} \}} \mid W ] + \mathbb{E}[ (\epsilon(Z, W) - 1) \, Z \, \bvec{1}_{\{ Z > \eta_{W} \}} \mid W ]
\notag \\
& \qquad \overset{\mathclap{\text{(b)}}}{\le}
\eta_{W} \, \Big( \mathbb{E}[ \epsilon(Z, W) \, \bvec{1}_{\{ Z < \eta_{W} \}} \mid W ] + \mathbb{E}[ (\epsilon(Z, W) - \beta_{W}) \, \bvec{1}_{\{ Z = \eta_{W} \}} \mid W ] + \mathbb{E}[ (\epsilon(Z, W) - 1) \, \bvec{1}_{\{ Z > \eta_{W} \}} \mid W ] \Big)
\notag \\
& \qquad \overset{\mathclap{\text{(c)}}}{=}
\eta_{W} \, \mathbb{E}[ \epsilon(Z, W) - \epsilon^{\ast}(Z, W) \mid W ]
\notag \\
& \qquad \overset{\mathclap{\text{(d)}}}{\le}
0
\qquad (\mathrm{a.s.}) ,
\label{eq:E_vs_E-ast}
\end{align}
\hrule
\end{figure*}
where
\begin{itemize}
\item
(a) and (c) follow from the definition of $\epsilon^{\ast}$ stated in \eqref{def:randomE-ast},
\item
(b) follows from the fact that $\eta_{W}$ is $\sigma(W)$-measurable, and
\item
(d) follows from \eqref{eq:E-ast_cond} and \eqref{eq:E_cond} and the fact that $\eta_{W} \ge 0$.
\end{itemize}
Combining \eqref{eq:E-ast_cond} and \eqref{eq:E-cutoff_E-ast}--\eqref{eq:E_vs_E-ast}, we obtain \eqref{eq:exp_cond-cutoff_min} of \propref{prop:expectation_cutoff}, as desired.
\hfill\IEEEQEDhere

\subsection{Proofs of \eqref{eq:uncond-cutoff_int-spectrum} and \eqref{eq:cond-cutoff_int-spectrum}}

The identity \eqref{eq:uncond-cutoff_int-spectrum} can be shown in the same manner as \cite[Equations~(155)--(156)]{kostina_polyanskiy_verdu_2015}, and can be thought of as a special case of \eqref{eq:cond-cutoff_int-spectrum} in which $\sigma(W)$ is the trivial $\sigma$-algebra.
Thus, it suffices to prove \eqref{eq:cond-cutoff_int-spectrum}.

It can be verified that the following conditional version of \cite[Equation~(157)]{kostina_polyanskiy_verdu_2015} holds:%
\footnote{While \cite[Equation~(157)]{kostina_polyanskiy_verdu_2015} can be verified by applying Tonelli's theorem only once, one can show \eqref{eq:cond-expectation_int} by applying Tonelli's theorem twice.}
\begin{align}
\mathbb{E}[ Z \, \bvec{1}_{\{ Z > z \}} \mid W ]
& =
\int_{z}^{\infty} \mathbb{P}\{ Z > t \mid W \} \, \mathrm{d}t
\notag \\
& \qquad {}
+ z \, \mathbb{P}\{ Z > z \mid W \}
\quad (\mathrm{a.s.})
\label{eq:cond-expectation_int}
\end{align}
for every real number $z \ge 0$.

We have
\begin{align}
&
\mathbb{E}[ \langle Z \mid W \rangle_{\varepsilon} \mid W ]
\notag \\
& \overset{\mathclap{\text{(a)}}}{=}
\mathbb{E}[ Z \, \bvec{1}_{\{ Z < \eta_{W} \}} \mid W ] + \eta_{W} \, (1 - \beta_{W}) \, \mathbb{P}\{ Z = \eta_{W} \mid W \}
\notag \\
& =
\mathbb{E}[ Z \mid W ] - \mathbb{E}[ Z \, \bvec{1}_{\{ Z > \eta_{W} \}} \mid W ]- \eta_{W} \, \beta_{W} \, \mathbb{P}\{ Z = \eta_{W} \mid W \}
\notag \\
& \overset{\mathclap{\text{(b)}}}{=}
\mathbb{E}[ Z \mid W ] - \int_{\eta_{W}}^{\infty} \mathbb{P}\{ Z > t \mid W \} \, \mathrm{d} t
\notag \\
& \qquad {}
- \eta_{W} \, \Big( \mathbb{P}\{ Z > \eta_{W} \mid W \} + \beta_{W} \, \mathbb{P}\{ Z = \eta_{W} \mid W \} \Big)
\notag \\
& \overset{\mathclap{\text{(c)}}}{=}
\mathbb{E}[ Z \mid W ] - \int_{\eta_{W}}^{\infty} \mathbb{P}\{ Z > t \mid W \} \, \mathrm{d} t - \varepsilon \, \eta_{W}
\quad (\mathrm{a.s.}) ,
\end{align}
where
\begin{itemize}
\item
(a) follows from the definition of $\langle \cdot \mid \cdot \rangle_{\varepsilon}$ stated in \eqref{def:cond_cutoff},
\item
(b) follows from \eqref{eq:cond-expectation_int} by noting that $\eta_{W}$ is $\sigma(W)$-measurable, and
\item
(c) follows from \eqref{eq:eta_W}.
\end{itemize}
This completes the proof of \eqref{eq:cond-cutoff_int-spectrum} of \propref{prop:expectation_cutoff}.
\hfill\IEEEQEDhere

\subsection{Proof of \eqref{eq:inequality_cutoff_cond}}

Since the functional $\epsilon \mapsto \mathbb{E}[ (1- \epsilon(Z)) \, Z ]$ of a mapping $\epsilon : [0, \infty) \to [0, 1]$ is linear, we readily see that
\begin{align}
&
\min_{\epsilon : \mathbb{E}[ \epsilon(Z) ] = \varepsilon} \mathbb{E}[ (1 - \epsilon(Z)) \, Z ]
\notag \\
& \qquad \qquad \qquad =
\min_{\epsilon : \mathbb{E}[ \epsilon(Z, W) ] = \varepsilon} \mathbb{E}[ (1 - E(Z)) \, Z ] ,
\end{align}
where the minimization in the left-hand side (resp.\ the right-hand side) is taken over the maps $\epsilon : [0, \infty) \to [0, 1]$ (resp.\ the maps $\epsilon : [0, \infty) \times \mathcal{W} \to [0, 1]$) satisfying $\mathbb{E}[ \epsilon(Z) ] = \varepsilon$ (resp.\ $\mathbb{E}[ \epsilon(Z, W) ] = \varepsilon$).
Therefore, the proof is completed by the identities \eqref{eq:exp_uncond-cutoff_min} and \eqref{eq:exp_cond-cutoff_min} of \propref{prop:expectation_cutoff}, and the fact that $\mathbb{E}[ \epsilon(Z, W) ] = \varepsilon$ if $\mathbb{E}[ \epsilon(Z, W) \mid W ] = \varepsilon$ a.s.
\hfill\IEEEQEDhere

\subsection{Proofs of \eqref{eq:stochastic-dominance_uncond-cutoff} and \eqref{eq:stochastic-dominance_cond-cutoff}}

The identity \eqref{eq:stochastic-dominance_uncond-cutoff} is a special case of \eqref{eq:stochastic-dominance_cond-cutoff} in which $\sigma(W)$ is the trivial $\sigma$-algebra.
Hence, it suffices to prove \eqref{eq:stochastic-dominance_cond-cutoff}.
Choose two $\sigma(W)$-measurable real-valued r.v.'s $\eta_{W}^{(1)}$ and $\eta_{W}^{(2)}$ so that
\begin{align}
\mathbb{P}\{ Z_{1} > \eta_{W}^{(1)} \mid W \} + \beta_{W}^{(1)} \, \mathbb{P}\{ Z_{1} = \eta_{W}^{(1)} \mid W \}
& =
\varepsilon
\quad (\mathrm{a.s.}) ,
\label{eq:eta1_beta1} \\
\mathbb{P}\{ Z_{2} > \eta_{W}^{(2)} \mid W \} + \beta_{W}^{(2)} \, \mathbb{P}\{ Z_{2} = \eta_{W}^{(2)} \mid W \}
& =
\varepsilon
\quad (\mathrm{a.s.})
\end{align}
for some $\sigma(W)$-measurable real-valued r.v.'s $\beta_{W}^{(1)}$ and $\beta_{W}^{(2)}$.
Then, it follows by the premise of \eqref{eq:stochastic-dominance_cond-cutoff} that
\begin{align}
\eta_{W}^{(1)}
\ge
\eta_{W}^{(2)}
\qquad (\mathrm{a.s.}) .
\label{eq:eta1_vs_eta2}
\end{align}
Now, a direct calculation shows
\begin{align}
&
\mathbb{E}[ \langle Z_{1} \mid W \rangle_{\varepsilon} \mid W ]
\notag \\
& \overset{\mathclap{\text{(a)}}}{=}
\int_{0}^{\eta_{W}^{(1)}} \mathbb{P}\{ Z_{1} > t \mid W \} \, \mathrm{d} t - \varepsilon \, \eta_{W}^{(1)}
\notag \\
& \overset{\mathclap{\text{(b)}}}{=}
\int_{0}^{\eta_{W}^{(2)}} \mathbb{P}\{ Z_{1} > t \mid W \} \, \mathrm{d} t + \int_{\eta_{W}^{(1)}}^{\eta_{W}^{(2)}} \mathbb{P}\{ Z_{1} > t \mid W \} \, \mathrm{d} t - \varepsilon \, \eta_{W}^{(1)}
\notag \\
& \overset{\mathclap{\text{(c)}}}{\le}
\int_{0}^{\eta_{W}^{(2)}} \mathbb{P}\{ Z_{1} > t \mid W \} \, \mathrm{d} t + \int_{\eta_{W}^{(1)}}^{\eta_{W}^{(2)}} \varepsilon \, \mathrm{d} t - \varepsilon \, \eta_{W}^{(1)}
\notag \\
& =
\int_{0}^{\eta_{W}^{(2)}} \mathbb{P}\{ Z_{1} > t \mid W \} \, \mathrm{d} t - \varepsilon \, \eta_{W}^{(2)}
\notag \\
& \overset{\mathclap{\text{(d)}}}{\le}
\int_{0}^{\eta_{W}^{(2)}} \mathbb{P}\{ Z_{2} > t \mid W \} \, \mathrm{d} t - \varepsilon \, \eta_{W}^{(2)}
\notag \\
& \overset{\mathclap{\text{(e)}}}{=}
\mathbb{E}[ \langle Z_{2} \mid W \rangle_{\varepsilon} \mid W ]
\end{align}
a.s., where
\begin{itemize}
\item
(a) follows from \eqref{eq:cond-cutoff_int-spectrum},
\item
(b) follows from \eqref{eq:eta1_vs_eta2},
\item
(c) follows from the fact that $\mathbb{P}\{ Z_{1} > t \mid W \} \le \varepsilon$ if $t \ge \eta_{W}^{(1)}$ a.s.\ (see \eqref{eq:eta1_beta1}),
\item
(d) follows by the premise of \eqref{eq:stochastic-dominance_cond-cutoff}, and
\item
(e) follows again from \eqref{eq:cond-cutoff_int-spectrum}.
\end{itemize}
Therefore, the conclusion of \eqref{eq:stochastic-dominance_cond-cutoff} holds, completing the proof.
\hfill\IEEEQEDhere

\section{An Example of Source $(X, Y)$ for the Hypotheses in Theorems~\ref{th:max} and~\ref{th:avg}}
\label{app:ex}

Let $\mathcal{X} = \mathcal{Y} = \{ 1, 2, \dots \}$.
Suppose that
\begin{align}
P_{Y}( y )
& =
\frac{ 6 }{ \pi^{2} \, y^{2} } ,
\\
P_{X|Y}(x \mid y)
& =
\left( 1 - \frac{ 1 }{ y } \right)^{x-1} \frac{ 1 }{ y }
\end{align}
for each $(x, y) \in \mathcal{X} \times \mathcal{Y}$.
Namely, the conditional distribution $P_{X|Y}(\cdot \mid y)$ is the geometric distribution with parameter $1/y$ for each $y \in \mathcal{Y}$.
After some algebra, we have%
\footnote{We adopt the usual convention that $0 \log 0 = 0$ and $0 \log^{s} (1/0) = 0$ for $s \ge 1$.}
\begin{align}
\mathcal{V}(X \mid y)
& =
y \, (y - 1) \left( \log \frac{ y }{ y - 1 } \right)^{2}
\end{align}
for each $y \in \mathcal{Y}$, implying that Hypotheses (a) and (b) in \thref{th:max} fail to holds.
Therefore, \thref{th:max} cannot ensure that \eqref{eq:max_asympt} holds for this source $(X, Y)$.

On the other hand, it can be verified that
\begin{align}
H(X \mid Y)
& \: \le
2 + \frac{ 6 }{ \pi^{2} } \sum_{y = 2}^{\infty} \frac{ \log(y - 1) }{ y^{2} } ,
\\
\tilde{\mathcal{T}}(X \mid y)
& \coloneqq
\sum_{x \in \mathcal{X}} P_{X|Y}(x \mid y) \left( \log \frac{ 1 }{ P_{X|Y}(x \mid y) } \right)^{3}
\notag \\
& \: \le
\log^{3} (y - 1) + 6 \log^{2} (y - 1)
\notag \\
& \qquad {}
+ 12 \log (y - 1) + 12
\label{eq:geometric-T}
\end{align}
for each $y \ge 2$, where $\tilde{\mathcal{T}}(X \mid 1) = 0$.
Thus, we observe that
\begin{align}
T_{\mathrm{u}}(X \mid Y)
& \le
\mathbb{E}[ \tilde{\mathcal{T}}(X \mid Y) ] + H(X \mid Y)^{3}
\notag \\
& \le
\frac{ 6 }{ \pi^{2} } \sum_{y = 2}^{\infty} \frac{ \log^{3}( y - 1 ) + 6 \log^{2}(y - 1) + 12 \log (y - 1) }{ y^{2} }
\notag \\
& \qquad {}
+ 12 + \left( 2 + \frac{ 6 }{ \pi^{2} } \sum_{y = 2}^{\infty} \frac{ \log(y - 1) }{ y^{2} } \right)^{3} ,
\end{align}
implying that $T_{\mathrm{u}}(X \mid Y)$ is finite.
Therefore, it follows from \thref{th:avg} that \eqref{eq:avg_asympt} holds for this source $(X, Y)$.

\section{Proof of \lemref{lem:optimal-code_max}}
\label{app:optimal-code_max}

Throughout \appref{app:optimal-code_max}, we consider one-shot ($n = 1$) variable-length stochastic codes $(F, G)$ as defined in \sectref{sect:code}.
We say that a decoder $G$ is \emph{deterministic} if $G(F(X, Y), X)$ is $\sigma(F(X, Y), X)$-measurable.
To specify the determinism, we use the lower case $g$ to denote a deterministic decoder.

Consider an $(L, \varepsilon)_{\max}$-code $(F, G)$ satisfying
\begin{align}
\mathbb{E}[ \ell( F(X, Y) ) ]
& \le
L ,
\label{eq:L_F0G0} \\
\mathbb{P}\{ X \neq G(F(X, Y), Y) \mid Y \}
& \le
\varepsilon
\qquad (\mathrm{a.s.}) .
\label{eq:max-error_F0G0}
\end{align}
It can be verified that there exists a deterministic decoder $g_{0}$ satisfying%
\footnote{Note that in general, the determinism of decoders is not a necessary condition to be optimal.}
\begin{align}
\!\!
\mathbb{P}\{ X \neq g_{0}(F(X, Y), Y) \mid Y \}
& \le
\mathbb{P}\{ X \neq G(F(X, Y), Y) \mid Y \}
\!\!
\label{eq:max-error_F0g0}
\end{align}
a.s.
In addition, for each $(x, y) \in \mathcal{X} \times \mathcal{Y}$, construct another stochastic encoder $F_{0}$ as
\begin{align}
F_{0}(x, y)
& \coloneqq
\begin{cases}
\varnothing
& \mathrm{if} \ x \neq g_{0}(F(x, y), y) ,
\\
F(x, y)
& \mathrm{otherwise} .
\end{cases}
\label{eq:construction_F0}
\end{align}
As shown later, the new code $(F_{0}, g_{0})$ has a better performance than that of the initial code $(F, G)$.
It is clear that
\begin{align}
F_{0}(X, Y) \neq \varnothing
\quad \Longrightarrow \quad
X = g_{0}(F_{0}(X, Y), Y) .
\label{eq:cond_nonempty-string}
\end{align}

Now, generate a random collection $\mathcal{B}(Y)$ of subsets of $\{ 0, 1 \}^{\ast}$ as
\begin{align}
\mathcal{B}(Y)
\coloneqq
\{ \mathcal{B}(x \mid Y) \mid x \in \mathcal{X} \} \setminus \{ \emptyset \} ,
\end{align}
where the random subset $\mathcal{B}(x \mid Y)$ of $\{ 0, 1 \}^{\ast}$ is defined by
\begin{align}
&
\mathcal{B}(x \mid Y)
\notag \\
& \quad \coloneqq
\{ \bvec{b} \in \{ 0, 1 \}^{\ast} \setminus \{ \varnothing \} \mid \mathbb{P}\{ F_{0}(x, Y) = \bvec{b} \mid Y \} > 0 \}
\end{align}
if $x \neq g_{0}( \varnothing, Y )$; and $\mathcal{B}(x \mid Y) \coloneqq \{ \varnothing \}$ otherwise.
We shall prove the disjointness of the sets $\mathcal{B}(x \mid Y)$, $x \in \mathcal{X}$, in $\mathcal{B}(Y)$ as follows:
Choose $\bvec{b}^{\prime} \in \{ 0, 1 \}^{\ast} \setminus \{ \varnothing \}$ and $x_{1}, x_{2} \in \mathcal{X}$ so that $\mathbb{P}( \mathcal{E}_{1} \cap \mathcal{E}_{2} ) > 0$, where the two events $\mathcal{E}_{1}$ and $\mathcal{E}_{2}$ are given by
\begin{align}
\mathcal{E}_{1}
& \coloneqq
\{ \mathbb{P}\{ F_{0}(x_{1}, Y) = \bvec{b}^{\prime} \mid Y \} > 0 \} ,
\\
\mathcal{E}_{2}
& \coloneqq
\{ \mathbb{P}\{ F_{0}(x_{2}, Y) = \bvec{b}^{\prime} \mid Y \} > 0 \} ,
\end{align}
respectively.
It is clear that $\mathcal{E}_{i} \in \sigma(Y)$ for each $i = 1, 2$.
Moreover, it follows from \eqref{eq:cond_nonempty-string} that $g_{0}(\bvec{b}^{\prime}, Y) = x_{i}$ on the event $\mathcal{E}_{i}$ for each $i = 1, 2$.
Thus, since $g_{0}(\bvec{b}^{\prime}, Y)$ is $\sigma(Y)$-measurable, we observe that $x_{1} = x_{2}$ whenever $\mathbb{P}( \mathcal{E}_{1} \cap \mathcal{E}_{2} ) > 0$.
Therefore, the random collection $\mathcal{B}(Y)$ is disjoint a.s., i.e.,
\begin{align}
\mathbb{P}\{ \mathcal{B}(x_{1}, Y) \cap \mathcal{B}(x_{2}, Y) = \emptyset \ \mathrm{for} \ \mathrm{all} \ x_{1} \neq x_{2} \}
=
1 .
\label{eq:disjoint_B}
\end{align}
By the disjointness of \eqref{eq:disjoint_B}, one can find an index set $\mathcal{I}(Y) = \{ 1, 2, \dots, |\mathcal{B}(Y)| \}$ of the collection $\mathcal{B}(Y)$ so that
\begin{align}
\mathcal{B}(Y)
=
\{ \mathcal{B}_{i}(Y) \mid i \in \mathcal{I}(Y) \}
\label{def:random_collection_B_indexed}
\end{align}
and
\begin{align}
\mathbb{P}\left\{ \!\!
\begin{array}{l}
\mathrm{for} \ \mathrm{all} \ i < j, \ \mathrm{there} \ \mathrm{exists} \  \bvec{b} \in \mathcal{B}_{i}( Y ) \ \mathrm{s.t.} \\
\bvec{b} \prec \tilde{\bvec{b}} \ \mathrm{for} \ \mathrm{all} \ \tilde{\bvec{b}} \in \mathcal{B}_{j}(Y)
\end{array}
\!\! \right\}
=
1 ,
\label{eq:sorted_collection_B}
\end{align}
where the binary relation $\prec$ on $\{ 0, 1 \}^{\ast}$ represents the lexicographical order in $\{ 0, 1 \}^{\ast}$.
Let $\{ \bvec{b}_{i} \}_{i = 1}^{\infty}$ be the lexicographical ordering of the strings in $\{ 0, 1 \}^{\ast}$ so that $\bvec{b}_{i} \prec \bvec{b}_{j}$ whenever $i < j$; e.g., $\bvec{b}_{1} = \varnothing$, $\bvec{b}_{2} = 0$, $\bvec{b}_{3} = 1$, $\bvec{b}_{4} = 00$, $\bvec{b}_{5} = 01$, $\bvec{b}_{6} = 10$, $\bvec{b}_{7} = 11$, $\bvec{b}_{8} = 000$, and so on.
It is trivial that
\begin{align}
\mathbb{P}\{ \mathcal{B}_{1}(Y) = \{ \varnothing \} \}
=
1 ;
\label{eq:B1_is_only_varnothing}
\end{align}
consequently, it follows from \eqref{eq:disjoint_B} that
\begin{align}
\mathbb{P}\{ \varnothing \notin \mathcal{B}_{i}(Y) \ \mathrm{for} \ \mathrm{all} \ i \in \mathcal{I}(Y) \setminus \{ 1 \} \}
=
1 .
\label{eq:Bi_doesnt_contain_varnothing}
\end{align}
Now, define the event
\begin{align}
&
\mathcal{A}_{k}
\notag \\
& \coloneqq
\left\{ \!\!
\begin{array}{l}
\ell( \bvec{b}_{i} ) \le \ell( \bvec{b} ) \ \mathrm{for} \ \mathrm{all} \ 1 \le i \le \min\{ k, |\mathcal{B}(Y)| \} \\
\qquad \qquad \qquad \ \; \mathrm{and} \ \bvec{b} \in \mathcal{B}_{i}(Y)
\end{array}
\!\! \right\}
\end{align}
for each integer $k \ge 1$.
It can be verified from \eqref{eq:sorted_collection_B} and \eqref{eq:B1_is_only_varnothing} by induction that
\begin{align}
\mathbb{P}( \mathcal{A}_{k} )
=
1
\end{align}
for every $k \ge 1$.
Hence, it follows from the monotonicity $\mathcal{A}_{1} \supset \mathcal{A}_{2} \supset \mathcal{A}_{3} \supset \cdots$ that
\begin{align}
\mathbb{P}\{ \ell( \bvec{b}_{i} ) \le \ell( \bvec{b} ) \ \mathrm{for} \ \mathrm{all} \ i \in \mathcal{I}(Y) \ \mathrm{and} \ \bvec{b} \in \mathcal{B}_{i}(Y) \}
=
1 .
\label{eq:bounds_collection_B}
\end{align}

Based on the previous paragraph, define the random map $\Phi_{Y} : \{ 0, 1 \}^{\ast} \to \{ 0, 1 \}^{\ast}$ so that
\begin{align}
\Phi_{Y}(\bvec{b})
\coloneqq
\begin{dcases}
\varnothing
& \mathrm{if} \ \bvec{b} \notin \mathcal{B}_{i}(Y) \ \mathrm{for} \ \mathrm{all} \ i \in \mathcal{I}(Y) ,
\\
\bvec{b}_{i}
& \mathrm{if} \ \bvec{b} \in \mathcal{B}_{i}(Y) \ \mathrm{for} \ \mathrm{some} \ i \in \mathcal{I}(Y) .
\end{dcases}
\label{def:phi}
\end{align}
Moreover, the disjointness of \eqref{eq:disjoint_B} ensures the existence of a random map $\Psi_{Y} : \mathcal{X} \to \mathcal{X} \cup \{ 0 \}$ satisfying
\begin{align}
\Psi_{Y}(i)
=
\begin{cases}
x
& \mathrm{if} \ \mathcal{B}_{i}(Y) = \mathcal{B}(x \mid Y) \ \mathrm{for} \ \mathrm{some} \ x \in \mathcal{X} ,
\\
0
& \mathrm{otherwise} .
\end{cases}
\label{def:psi}
\end{align}
Note that $\Phi_{Y}( \bvec{b} )$ and $\Psi_{Y}( i )$ are $\sigma(Y)$-measurable for each $\bvec{b} \in \{ 0, 1 \}^{\ast}$ and $i \in \mathcal{X}$, respectively.
Then, construct another variable-length stochastic code $(F_{1}, g_{1})$ so that
\begin{align}
F_{1}(x, Y)
& \coloneqq
\Phi_{Y}( F_{0}(x, Y) ) ,
\label{eq:construction_F1} \\
g_{1}(\bvec{b}, Y)
& \coloneqq
\Psi_{Y}(i)
\quad \mathrm{if} \ \bvec{b} = \bvec{b}_{i} \ \mathrm{for} \ \mathrm{some} \ i \ge 1 .
\label{eq:construction_g1}
\end{align}

Now, we shall evaluate the average codeword length of the encoder $F_{1}$.
A direct calculation shows
\begin{align}
&
\mathbb{E}[ \ell( F_{1}(X, Y) ) ]
\notag \\
& \qquad \overset{\mathclap{\text{(a)}}}{=}
\mathbb{E}\left[ \sum_{i = 2}^{|\mathcal{B}(Y)|} \ell( \bvec{b}_{i} ) \, \bvec{1}_{\{ F_{0}(X, Y) \in \mathcal{B}_{i}(Y) \}} \right]
\notag \\
& \qquad \overset{\mathclap{\text{(b)}}}{\le}
\mathbb{E}\left[ \sum_{i = 2}^{|\mathcal{B}(Y)|} \ell( F_{0}(X, Y) ) \, \bvec{1}_{\{ F_{0}(X, Y) \in \mathcal{B}_{i}(Y) \}} \right]
\notag \\
& \qquad \overset{\mathclap{\text{(c)}}}{\le}
\mathbb{E}\Big[ \ell( F_{0}(X, Y) ) \, \bvec{1}_{\{ X \neq g_{0}(\varnothing, Y) \}} \Big]
\notag \\
& \qquad \overset{\mathclap{\text{(d)}}}{=}
\mathbb{E}\Big[ \ell( F(X, Y) ) \, \bvec{1}_{\{ X = g_{0}(F(X, Y), Y) \} \cap \{ X \neq g_{0}(\varnothing, Y) \}} \Big]
\notag \\
& \qquad \le
\mathbb{E}[ \ell( F(X, Y) ) ]
\notag \\
& \qquad \overset{\mathclap{\text{(e)}}}{\le}
L ,
\label{eq:L_F1_max}
\end{align}
where
\begin{itemize}
\item
(a) follows from the definitions of $\mathcal{B}(Y)$ and $\phi$ stated in \eqref{def:random_collection_B_indexed} and \eqref{def:phi}, respectively, and the fact that $\ell(\bvec{b}_{1} = \varnothing) = 0$,
\item
(b) follows from \eqref{eq:bounds_collection_B},
\item
(c) follows from the disjointness of $\mathcal{B}(Y)$ stated in \eqref{eq:disjoint_B} and the fact that
\begin{align}
\hspace{-1em}
\mathbb{P}\{ F_{0}(X, Y) \in \mathcal{B}_{i}(Y) \, \mathrm{for} \, \mathrm{some} \, 1 \le i \le |\mathcal{B}(Y)| \}
=
1 ;
\end{align}
that is,
\begin{align}
&
\sum_{i = 2}^{|\mathcal{B}(Y)|} \bvec{1}_{\{ F_{0}(X, Y) \in \mathcal{B}_{i}(Y) \}}
\notag \\
& \qquad =
\bvec{1}_{\{ F_{0}(X, Y) \in \mathcal{B}_{i}(Y) \, \mathrm{for} \, \mathrm{some} \, 2 \le i \le |\mathcal{B}(Y)| \}}
\notag \\
& \qquad =
\bvec{1}_{\{ F_{0}(X, Y) \neq \varnothing \}}
\notag \\
& \qquad =
\bvec{1}_{\{ X \neq g_{0}(\varnothing, Y) \} \cap \{ X = g_{0}(F_{0}(X, Y), Y) \}}
\notag \\
& \qquad \le
\bvec{1}_{\{ X \neq g_{0}(\varnothing, Y) \}}
\end{align}
a.s.,
\item
(d) follows by the definition of $F_{0}$ stated in \eqref{eq:construction_F0}, and
\item
(e) follows from \eqref{eq:L_F0G0}.
\end{itemize}
Namely, the average codeword length of the encoder $F_{1}$ is shorter than or equal to that of the initial encoder $F$.

Next, we shall evaluate the error probability of the code $(F_{1}, g_{1})$.
We observe that
\begin{align}
&
\mathbb{P}\{ X \neq g_{1}(F_{1}(X, Y), Y) \mid Y \}
\notag \\
& \quad =
\sum_{i = 1}^{\infty} \mathbb{P}\{ F_{1}(X, Y) = \bvec{b}_{i} \ \mathrm{and} \ X \neq g_{1}(\bvec{b}_{i}, Y) \mid Y \}
\notag \\
& \quad \overset{\mathclap{\text{(a)}}}{=}
\sum_{i = 1}^{|\mathcal{B}(Y)|} \mathbb{P}\{ \Phi_{Y}(F_{0}(X, Y)) = \bvec{b}_{i} \ \mathrm{and} \ X \neq \Psi_{Y}(i) \mid Y \}
\notag \\
& \quad \overset{\mathclap{\text{(b)}}}{=}
\sum_{i = 1}^{|\mathcal{B}(Y)|} \mathbb{P}\{ F_{0}(X, Y) \in \mathcal{B}_{i}(Y) \ \mathrm{and} \ X \neq \Psi_{Y}(i) \mid Y \}
\notag \\
& \quad \overset{\mathclap{\text{(c)}}}{=}
\sum_{i = 1}^{|\mathcal{B}(Y)|} \mathbb{P}\{ F_{0}(X, Y) \in \mathcal{B}_{i}(Y) \ \mathrm{and} \ \mathcal{B}(X, Y) \neq \mathcal{B}_{i}(Y) \mid Y \}
\notag \\
& \quad \overset{\mathclap{\text{(d)}}}{=}
\mathbb{P}\{ F_{0}(X, Y) \in \mathcal{B}_{1}(Y) \ \mathrm{and} \ \mathcal{B}(X, Y) \neq \mathcal{B}_{1}(Y) \mid Y \}
\notag \\
& \quad \overset{\mathclap{\text{(e)}}}{=}
\mathbb{P}\{ F_{0}(X, Y) = \varnothing \ \mathrm{and} \ X \neq g_{0}(\varnothing, Y) \mid Y \}
\notag \\
& \quad \overset{\mathclap{\text{(f)}}}{=}
\mathbb{P}\{ X \neq g_{0}(F_{0}(X, Y), Y) \ \mathrm{and} \ X \neq g_{0}(\varnothing, Y) \mid Y \}
\notag \\
& \quad \overset{\mathclap{\text{(g)}}}{=}
\mathbb{P}\{ X \neq g_{0}(F(X, Y), Y) \ \mathrm{and} \ X \neq g_{0}(\varnothing, Y) \mid Y \}
\notag \\
& \quad \le
\mathbb{P}\{ X \neq g_{0}(F(X, Y), Y) \mid Y \}
\notag \\
& \quad \overset{\mathclap{\text{(h)}}}{\le}
\varepsilon
\qquad (\mathrm{a.s.}) ,
\label{eq:Pe_F1g1}
\end{align}
where
\begin{itemize}
\item
(a) follows by the definition of $(F_{1}, g_{1})$ stated in \eqref{eq:construction_F1} and \eqref{eq:construction_g1},
\item
(b) follows by the definition of the random map $\Phi_{Y}(\cdot)$ stated in \eqref{def:phi},
\item
(c) follows by the definition of the random map $\Psi_{Y}(\cdot)$ stated in \eqref{def:psi},
\item
(d) follows from the fact that $F_{0}(X, Y) \in \mathcal{B}_{i}(Y)$ only if $\mathcal{B}(X, Y) = \mathcal{B}_{i}(Y)$ for each $2 \le i \le |\mathcal{B}(Y)|$ a.s.,
\item
(e) follows from \eqref{eq:B1_is_only_varnothing},
\item
(f) follows from the fact that $F_{0}(X, Y) = \varnothing$ if and only if $X = g( \varnothing, Y )$ or $X \neq g_{0}(F_{0}(X, Y), Y)$,
\item
(g) follows by the definition of $F_{0}$ stated in \eqref{eq:construction_F0}, and
\item
(h) follows from \eqref{eq:max-error_F0G0} and \eqref{eq:max-error_F0g0}.
\end{itemize}
In other words, the maximum probability of error for the code $(F_{1}, g_{1})$ is smaller than or equal to that for the initial code $(F, G)$.

Here, we shall prove the converse result of \lemref{lem:optimal-code_max}, i.e., we shall show that the average codeword length of $(F, G)$ satisfying \eqref{eq:max-error_F0G0} is always bounded from below by the right-hand side of \eqref{eq:optimal-code_max}.
We see that
\begin{align}
1 - \varepsilon
& \overset{\mathclap{\text{(a)}}}{\le}
\mathbb{P}\{ X = g_{1}(F_{1}(X, Y), Y) \mid Y \}
\notag \\
& =
\mathbb{P}\{ X = g_{1}(\varnothing, Y) \ \mathrm{and} \ F_{1}(X, Y) = \varnothing \mid Y \}
\notag \\
& \quad {}
+ \mathbb{P}\{ X = g_{1}(F_{1}(X, Y), Y) \ \mathrm{and} \ F_{1}(X, Y) \neq \varnothing \mid Y \}
\notag \\
& \overset{\mathclap{\text{(b)}}}{=}
\mathbb{P}\{ X = \Psi_{Y}(1) \mid Y \}
\notag \\
& \quad {}
+ \mathbb{P}\{ X = g_{1}(F_{1}(X, Y), Y) \ \mathrm{and} \ F_{1}(X, Y) \neq \varnothing \mid Y \}
\notag \\
& =
\mathbb{P}\{ X = \Psi_{Y}(1) \mid Y \}
\notag \\
& \quad {}
+ \mathbb{P}\left\{ \!\!
\begin{array}{l}
X = g_{1}(\bvec{b}_{i}, Y) \ \mathrm{and} \\
F_{1}(X, Y) = \bvec{b}_{i} \ \mathrm{for} \ \mathrm{some} \ i \ge 2
\end{array}
\middle| \ Y
\right\}
\notag \\
& \overset{\mathclap{\text{(c)}}}{=}
\mathbb{P}\{ X = \Psi_{Y}(1) \mid Y \}
\notag \\
& \quad {}
+ \mathbb{P}\{ X = \Psi_{Y}(i) \ \mathrm{for} \ \mathrm{some} \ 2 \le i \le |\mathcal{B}(Y)| \mid Y \}
\notag \\
& \overset{\mathclap{\text{(d)}}}{=}
\sum_{i = 1}^{|\mathcal{B}(Y)|} P_{X|Y}(\Psi_{Y}(i) \mid Y)
\qquad (\mathrm{a.s.}) ,
\label{eq:image_max}
\end{align}
where
\begin{itemize}
\item
(a) follows from \eqref{eq:Pe_F1g1},
\item
(b) and (c) follows from the definition of $(F_{1}, g_{1})$ stated in \eqref{eq:construction_F1} and \eqref{eq:construction_g1}, and
\item
(d) follows from the fact that $\Psi_{Y}(i) \neq \Psi_{Y}(j)$ if $1 \le i < j \le |\mathcal{B}(Y)|$ a.s.
\end{itemize}
Now, define the $\sigma(Y)$-measurable r.v.'s $\xi_{Y}$ and $\zeta_{Y}$ so that%
\footnote{Note that $\xi_{Y} = 0$ if $P_{X|Y}(\psi(1, Y) \mid Y) \ge 1-\varepsilon$; and $\xi_{Y} = \infty$ if $\sum_{i = 1}^{\infty} P_{X|Y}(\psi(i, Y) \mid Y) = 1-\varepsilon$ and $P_{X|Y}(\psi(i, Y) \mid Y) > 0$ for all $i \ge 1$.}
\begin{align}
\xi_{Y}
& \coloneqq
\sup\left\{ k \ge 0 \ \middle| \ \sum_{i = 1}^{k} P_{X|Y}(\Psi_{Y}(i) \mid Y) \le 1-\varepsilon \right\} ,
\\
\zeta_{Y}
& \coloneqq
1 - \varepsilon - \sum_{i = 1}^{\xi_{Y}} P_{X|Y}(\Psi_{Y}(i) \mid Y) ,
\end{align}
respectively.
In addition, define the $\sigma(Y)$-measurable r.v.'s $\kappa_{Y}$ and $\gamma_{Y}$ so that
\begin{align}
\kappa_{Y}
& \coloneqq
\sup\left\{ k \ge 0 \ \middle| \ \sum_{x = 1}^{k} P_{X|Y}(\varsigma_{Y}(x) \mid Y) \le 1 - \varepsilon \right\} ,
\\
\gamma_{Y}
& \coloneqq
1 - \varepsilon - \sum_{x = 1}^{\gamma_{Y}} P_{X|Y}(\varsigma_{Y}(x) \mid Y) ,
\end{align}
respectively, where $\varsigma_{Y}$ is given in \eqref{def:sigmaY}.
Furthermore, define
\begin{align}
p_{1}(k \mid Y)
& \coloneqq
\begin{cases}
P_{X|Y}(\Psi_{Y}(k) \mid Y)
& \mathrm{if} \ 1 \le k \le \xi_{Y} ,
\\
\zeta_{Y}
& \mathrm{if} \ k = \xi_{Y} + 1 ,
\\
0
& \mathrm{if} \ \xi_{Y} + 2 \le k < \infty .
\end{cases}
\\
p_{2}(x \mid Y)
& \coloneqq
\begin{cases}
P_{X|Y}(\varsigma_{Y}(x) \mid Y)
& \mathrm{if} \ 1 \le x \le \kappa_{Y} ,
\\
\gamma_{Y}
& \mathrm{if} \ x = \kappa_{Y} + 1 ,
\\
0
& \mathrm{if} \ \kappa_{Y} + 2 \le x < \infty .
\end{cases}
\label{def:p2}
\end{align}
Then, a direct calculation shows
\begin{align}
\mathbb{E}[ \ell( F_{1}(X, Y) ) \mid Y ]
& \ge
\sum_{k = 1}^{\infty} \lfloor \log k \rfloor \, p_{1}(k \mid Y)
\notag \\
& =
\sum_{j = 0}^{\infty} j \, \sum_{k = 2^{j}}^{2^{j+1}-1} p_{1}(k \mid Y)
\notag \\
& =
\sum_{j = 1}^{\infty} \sum_{k = 2^{j}}^{\infty} p_{1}(k \mid Y) ,
\label{eq:LB_L_F1g1_max_sum}
\\
\mathbb{E}[ \langle \lfloor \log \varsigma_{Y}^{-1}(X) \rfloor \mid Y \rangle_{\varepsilon} \mid Y ]
& =
\sum_{x = 1}^{\infty} \lfloor \log x \rfloor \, p_{2}(x \mid Y)
\notag \\
& =
\sum_{j = 0}^{\infty} j \, \sum_{x = 2^{j}}^{2^{j+1}-1} p_{2}(x \mid Y)
\notag \\
& =
\sum_{j = 1}^{\infty} \sum_{x = 2^{j}}^{\infty} p_{2}(x \mid Y)
\label{eq:cond-eps_optimal_sum}
\end{align}
a.s., respectively, where the second equalities in \eqref{eq:LB_L_F1g1_max_sum} and \eqref{eq:cond-eps_optimal_sum} follow from the fact that $\lfloor \log k \rfloor = j$ if and only if $2^{j} \le k < 2^{j+1}$ for every $k \ge 1$.
On the other hand, since $\varsigma_{Y}$ rearranges the probability masses in $P_{X|Y}(\cdot \mid Y)$ in non-increasing order (see \eqref{def:sigmaY}), it can be verified that $p_{1}(\cdot \mid Y)$ is majorized by $p_{2}(\cdot \mid Y)$ a.s., i.e., it follows that
\begin{align}
\sum_{k = 1}^{l} p_{1}(k \mid Y)
\le
\sum_{x = 1}^{l} p_{2}(x \mid Y)
\qquad (\mathrm{a.s.})
\label{eq:majorization_max}
\end{align}
for every $l \ge 1$, and
\begin{align}
\sum_{k = 1}^{\infty} p_{1}(k \mid Y)
=
\sum_{x = 1}^{\infty} p_{2}(x \mid Y)
=
1 - \varepsilon
\qquad (\mathrm{a.s.}) .
\label{eq:majorization_equality_max}
\end{align}
Therefore, it follows from \eqref{eq:LB_L_F1g1_max_sum}--\eqref{eq:majorization_equality_max} that
\begin{align}
\mathbb{E}[ \ell( F_{1}(X, Y) ) ]
\ge
\mathbb{E}[ \langle \lfloor \log \varsigma_{Y}^{-1}(X) \rfloor \mid Y \rangle_{\varepsilon} ] .
\label{eq:LB_L_F1g1_cutoff-eps}
\end{align}
Combining \eqref{eq:L_F1_max} and \eqref{eq:LB_L_F1g1_cutoff-eps}, we observe that the existence of an $(L, \varepsilon)_{\max}$-code $(F, G)$ implies that
\begin{align}
L
\ge
\mathbb{E}[ \langle \lfloor \log \varsigma_{Y}^{-1}(X) \rfloor \mid Y \rangle_{\varepsilon} ] ,
\label{eq:LB_L_max}
\end{align}
which corresponds to the converse bound of \lemref{lem:optimal-code_max}.

Finally, we shall show the existence of an $(L, \varepsilon)_{\max}$-code meeting the equality in \eqref{eq:LB_L_max}.
In fact, constructing a variable-length stochastic code $(F_{\sup}^{\ast}, g^{\ast})$ so that
\begin{align}
F_{\sup}^{\ast}(x, Y)
& \coloneqq
\begin{cases}
\bvec{b}_{\varsigma_{Y}^{-1}(x)}
& \mathrm{if} \ 1 \le \varsigma_{Y}^{-1}(x) \le \kappa_{Y} ,
\\
B_{\sup}
& \mathrm{if} \ \varsigma_{Y}^{-1}(x) = \kappa_{Y} + 1 ,
\\
\varnothing
& \mathrm{if} \ \kappa_{Y} < \varsigma_{Y}^{-1}(x) < \infty ,
\end{cases}
\label{eq:optimal-F_max}
\\
g^{\ast}(\bvec{b}, Y)
& \coloneqq
x
\quad \mathrm{if} \ \bvec{b} = \bvec{b}_{\varsigma_{Y}^{-1}(x)} \ \mathrm{for} \ \mathrm{some} \ x \in \mathcal{X} ,
\label{eq:optimal-g_max}
\end{align}
where $B_{\sup}$ denotes a $\{ 0, 1 \}^{\ast}$-valued r.v.\ satisfying the condition that it is conditionally independent of $X$ given $Y$ (i.e., that $B_{\sup} \Perp X \mid Y$) and
\begin{align}
\mathbb{P}\{ B_{\sup} = \varnothing \mid Y \}
=
1 - \mathbb{P}\{ B_{\sup} = \bvec{b}_{\kappa_{Y}+1} \mid Y \}
=
1 - \gamma_{Y}
\end{align}
a.s., we readily see that
\begin{align}
\mathbb{E}[ \ell( F_{\sup}^{\ast}(X, Y) ) ]
& =
\mathbb{E}[ \langle \lfloor \log \varsigma_{Y}^{-1}(X) \rfloor \mid Y \rangle_{\varepsilon} ]
\end{align}
and
\begin{align}
\mathbb{P}\{ X \neq g^{\ast}(F_{\sup}^{\ast}(X, Y), Y) \mid Y \}
& =
\varepsilon
\quad (\mathrm{a.s.}) .
\end{align}
This completes the proof of \lemref{lem:optimal-code_max}.
\hfill\IEEEQEDhere

\section{Proof of \lemref{lem:eps-cutoff_max-err}}
\label{app:eps-cutoff_max-err}

Define the event
\begin{align}
\mathcal{C}_{k}
\coloneqq
\left\{ \lfloor \log x \rfloor \le \log \frac{ 1 }{ P_{X|Y}(\varsigma_{Y}(x) \mid Y) } \ \mathrm{for} \ \mathrm{all} \ 1 \le x \le k \right\}
\end{align}
for each integer $k \ge 1$.
Since $\varsigma_{Y}$ rearranges the probability masses in $P_{X|Y}(\cdot \mid Y)$ in non-increasing order (see \eqref{def:sigmaY}), similar to \cite[Theorem~2]{kontoyiannis_verdu_2014}, it can be verified by induction that
\begin{align}
\mathbb{P}( \mathcal{C}_{k} )
=
1
\end{align}
for every $k \ge 1$.
Hence, the monotonicity $\mathcal{C}_{1} \supset \mathcal{C}_{2} \supset \mathcal{C}_{3} \supset \cdots$ implies that
\begin{align}
\mathbb{P}\left\{ \lfloor \log \sigma_{Y}^{-1}(x) \rfloor \le \log \frac{ 1 }{ P_{X|Y}(x \mid Y) } \ \mathrm{for} \ \mathrm{all} \ x \in \mathcal{X} \right\}
=
1 .
\label{eq:log-k_iota-k}
\end{align}
yielding that
\begin{align}
\mathbb{P}\left\{ \log \frac{ 1 }{ P_{X|Y}(X \mid Y) } \le t \ \middle| \ Y \right\}
\le
\mathbb{P}\Big\{ \lfloor \log \sigma_{Y}^{-1}(X) \rfloor \le t \ \Big| \ Y \Big\}
\label{eq:stochastic-dominance_iota-logk}
\end{align}
a.s.\ for all $t > 0$.
Therefore, it follows from \eqref{eq:stochastic-dominance_cond-cutoff} of \propref{prop:expectation_cutoff} that
\begin{align}
\mathbb{E}[ \langle \lfloor \log \varsigma_{Y}^{-1}( X ) \rfloor \mid Y \rangle_{\varepsilon} ]
\le
\mathfrak{C}_{\mathrm{c}}^{\varepsilon}(X \mid Y) .
\end{align}
Thus, it follows from \lemref{lem:optimal-code_max} that the left-hand inequality of \eqref{eq:eps-cutoff_max-err_one-shot} holds.

On the other hand, we observe that \eqref{eq:eps-cutoff_max-err} written in the top of the next page holds,
\begin{figure*}[!t]
\begin{align}
\mathbb{E}[ \langle \lfloor \log \varsigma_{Y}^{-1}( X ) \rfloor \mid Y \rangle_{\varepsilon} ]
& \overset{\mathclap{\text{(a)}}}{=}
\mathbb{E}[ \lfloor \log \varsigma_{Y}^{-1}( X ) \rfloor ] - \max_{\epsilon : \mathbb{E}[ \epsilon(\lfloor \log \varsigma_{Y}^{-1}( X ) \rfloor, Y) \mid Y ] = \varepsilon \, (\mathrm{a.s.})} \mathbb{E}[ \epsilon(\lfloor \log \varsigma_{Y}^{-1}( X ) \rfloor, Y) \, \lfloor \log \varsigma_{Y}^{-1}( X ) \rfloor ]
\notag \\
& \overset{\mathclap{\text{(b)}}}{\ge}
\mathbb{E}[ \lfloor \log \varsigma_{Y}^{-1}( X ) \rfloor ] - \max_{\epsilon : \mathbb{E}[ \epsilon(\iota(X \mid Y), Y) \mid Y ] = \varepsilon \, (\mathrm{a.s.})} \mathbb{E}[ \epsilon(\iota(X \mid Y), Y) \, \iota(X \mid Y) ]
\notag \\
& \overset{\mathclap{\text{(c)}}}{\ge}
\mathbb{E}[ \iota(X \mid Y) ] - \max_{\epsilon : \mathbb{E}[ \epsilon(\iota(X \mid Y), Y) \mid Y ] = \varepsilon \, (\mathrm{a.s.})} \mathbb{E}[ \epsilon(\iota(X \mid Y), Y) \, \iota(X \mid Y) ] - \log( H(X \mid Y) + 1 ) - \log \mathrm{e}
\notag \\
& \overset{\mathclap{\text{(d)}}}{=}
\mathfrak{C}_{\mathrm{c}}^{\varepsilon}(X \mid Y) - \log( H(X \mid Y) + 1 ) - \log \mathrm{e} ,
\label{eq:eps-cutoff_max-err}
\end{align}
\hrule
\end{figure*}
where
\begin{itemize}
\item
(a) follows from \eqref{eq:exp_cond-cutoff_min} of \propref{prop:expectation_cutoff},
\item
(b) follows from \eqref{eq:exp_cond-cutoff_min} and \eqref{eq:stochastic-dominance_cond-cutoff} of \propref{prop:expectation_cutoff} and \eqref{eq:stochastic-dominance_iota-logk},
\item
(c) follows from the fact that the same argument as \cite{alon_orlitsky_1994} proves
\begin{align}
&
\mathbb{E}[ \lfloor \log \varsigma_{Y}^{-1}( X ) \rfloor \mid Y ]
\notag \\
& \quad \ge
\mathcal{H}(X \mid Y) - \log( \mathcal{H}(X \mid Y) + 1 ) - \log \mathrm{e}
\end{align}
a.s., which leads together with Jensen's inequality that
\begin{align}
&
\mathbb{E}[ \lfloor \log \varsigma_{Y}^{-1}( X ) \rfloor ]
\notag \\
& \quad \ge
H(X \mid Y) - \log(H(X \mid Y) + 1) - \log \mathrm{e} ;
\label{eq:LB_Alon_Orlitsky}
\end{align}
and
\item
(d) follows as in (a).
\end{itemize}
Therefore, it follows from \lemref{lem:optimal-code_max} that the right-hand side of \eqref{eq:eps-cutoff_max-err_one-shot} holds.
This completes the proof of \lemref{lem:eps-cutoff_max-err}.
\hfill\IEEEQEDhere

\section{Proof of \lemref{lem:asympt_cond_eps-entropy}}
\label{app:asympt_cond_eps-entropy}

It is clear from the definition of $\mathfrak{C}_{\mathrm{c}}^{\varepsilon}$ stated in \eqref{def:conditional_eps-entropy} that
\begin{align}
\varepsilon = 0
\quad & \Longrightarrow \quad
\mathfrak{C}_{\mathrm{c}}^{\varepsilon}(X^{n} \mid Y^{n}) = n \, H(X \mid Y) ,
\\
\varepsilon = 1
\quad & \Longrightarrow \quad
\mathfrak{C}_{\mathrm{c}}^{\varepsilon}(X^{n} \mid Y^{n}) = 0 .
\end{align}
Hence, it suffices to consider the case where $0 < \varepsilon < 1$.
It follows from \eqref{eq:cond-cutoff_int-spectrum} of \propref{prop:expectation_cutoff} that
\begin{align}
&
\mathbb{E}[ \langle \iota(X^{n} \mid Y^{n}) \mid Y^{n} \rangle_{\varepsilon} \mid Y^{n} ]
\notag \\
& \quad =
(1 - \varepsilon) \, \mathcal{H}(X^{n} \mid Y^{n})
\notag \\
& \qquad {}
- \int_{\eta_{Y^{n}}}^{\infty} \mathbb{P}\{ \iota(X^{n} \mid Y^{n}) > t \mid Y^{n} \} \, \mathrm{d} t
\notag \\
& \quad \qquad {}
- \varepsilon \, \Big( \eta_{Y^{n}} - \mathcal{H}(X^{n} \mid Y^{n}) \Big)
\qquad (\mathrm{a.s.})
\label{eq:first-identity_max}
\end{align}
for every $n \ge 1$, where $\sigma(Y^{n})$-measurable r.v.'s $\eta_{Y^{n}} \ge 0$ and $0 \le \beta_{Y^{n}} < 1$ are given so that
\begin{align}
\mathbb{P}\{ \iota(X^{n} \mid Y^{n}) > \eta_{Y^{n}} \mid Y^{n} \} + \beta_{Y^{n}} \, \mathbb{P}\{ \iota(X^{n} \mid Y^{n}) = \eta_{Y^{n}} \mid Y^{n} \}
=
\varepsilon
\label{eq:epsilon_given-yn}
\end{align}
a.s.
Similar to \cite[Equations~(159)--(165)]{kostina_polyanskiy_verdu_2015}, we see that \eqref{eq:second-identity_max} written in the top of the next page holds,
\begin{figure*}[!t]
\begin{align}
\int_{\eta_{Y^{n}}}^{\infty} \mathbb{P}\{ \iota(X^{n} \mid Y^{n}) > t \mid Y^{n} \} \, \mathrm{d} t
& \overset{\mathclap{\text{(a)}}}{=}
\int_{b_{Y^{n}}}^{\infty} \mathbb{P}\Big\{ \iota(X^{n} \mid Y^{n}) > \mathcal{H}(X^{n} \mid Y^{n}) + \sqrt{ \mathcal{V}(X^{n} \mid Y^{n}) } \, \Phi^{-1}( 1 - \varepsilon ) + t \ \Big| \ Y^{n} \Big\} \, \mathrm{d} t
\notag \\
& \overset{\mathclap{\text{(b)}}}{=}
\int_{0}^{\infty} \mathbb{P}\Big\{ \iota(X^{n} \mid Y^{n}) > \mathcal{H}(X^{n} \mid Y^{n}) + \sqrt{ \mathcal{V}(X^{n} \mid Y^{n}) } \, \Phi^{-1}( 1 - \varepsilon ) + t \ \Big| \ Y^{n} \Big\} \, \mathrm{d} t - B_{Y^{n}}
\notag \\
& \overset{\mathclap{\text{(c)}}}{=}
\sqrt{ \mathcal{V}(X^{n} \mid Y^{n}) } \int_{0}^{\infty} \mathbb{P}\left\{ \frac{ \iota(X^{n} \mid Y^{n}) - \mathcal{H}(X^{n} \mid Y^{n}) }{ \sqrt{ \mathcal{V}(X^{n} \mid Y^{n}) } } > \Phi^{-1}( 1 - \varepsilon ) + r \ \middle| \ Y^{n} \right\} \, \mathrm{d} r - B_{Y^{n}}
\notag \\
& =
\sqrt{ \mathcal{V}(X^{n} \mid Y^{n}) } \int_{\Phi^{-1}(1 - \varepsilon)}^{\infty} \mathbb{P}\left\{ \frac{ \iota(X^{n} \mid Y^{n}) - \mathcal{H}(X^{n} \mid Y^{n}) }{ \sqrt{ \mathcal{V}(X^{n} \mid Y^{n}) } }  > r \ \middle| \ Y^{n} \right\} \, \mathrm{d} r - B_{Y^{n}}
\notag \\
& \overset{\mathclap{\text{(d)}}}{=}
\sqrt{ \mathcal{V}(X^{n} \mid Y^{n}) } \int_{\Phi^{-1}(1 - \varepsilon)}^{\infty} (1 - \Phi( r )) \, \mathrm{d} r - B_{Y^{n}} + D_{Y^{n}}
\notag \\
& \overset{\mathclap{\text{(e)}}}{=}
\sqrt{ \mathcal{V}(X^{n} \mid Y^{n}) } \left( \int_{\Phi^{-1}(1 - \varepsilon)}^{\infty} r \, \varphi( r ) \, \mathrm{d} r - \varepsilon \, \Phi^{-1}(1 - \varepsilon) \right) - B_{Y^{n}} + D_{Y^{n}}
\notag \\
& \overset{\mathclap{\text{(f)}}}{=}
\sqrt{ \mathcal{V}(X^{n} \mid Y^{n}) } \, \Big( f_{\mathrm{G}}(1 - \varepsilon) - \varepsilon \, \Phi^{-1}(1 - \varepsilon) \Big)- B_{Y^{n}} + D_{Y^{n}}
\qquad (\mathrm{a.s.})
\label{eq:second-identity_max}
\end{align}
\hrule
\end{figure*}
for every $n \ge 1$, where
\begin{itemize}
\item
(a) follows by the definition
\begin{align}
b_{Y^{n}}
& \coloneqq
\eta_{Y^{n}} - \mathcal{H}(X^{n} \mid Y^{n})
\notag \\
& \qquad {}
- \sqrt{ \mathcal{V}(X^{n} \mid Y^{n}) } \, \Phi^{-1}( 1 - \varepsilon ) ,
\label{def:b_yn}
\end{align}
\item
(b) follows by the definition of $B_{Y^{n}}$ (see \eqref{def:capital-B_yn} written in the top of the next page)
\begin{figure*}
\begin{align}
B_{Y^{n}}
\coloneqq
\sgn( b_{Y^{n}} ) \int_{\min\{ 0, b_{Y^{n}} \}}^{\max\{ 0, b_{Y^{n}} \}} \mathbb{P}\Big\{ \iota(X^{n} \mid Y^{n}) > \mathcal{H}(X^{n} \mid Y^{n}) + \sqrt{ \mathcal{V}(X^{n} \mid Y^{n}) } \, \Phi^{-1}( 1 - \varepsilon ) + t \ \Big| \ Y^{n} \Big\} \, \mathrm{d} t
\label{def:capital-B_yn}
\end{align}
\hrule
\end{figure*}
with the sign function $\sgn : \mathbb{R} \to \{ -1, 0, 1 \}$ defined by
\begin{align}
\sgn( u )
\coloneqq
\begin{cases}
-1
& \mathrm{if} \ u < 0 ,
\\
0
& \mathrm{if} \ u = 0 ,
\\
1
& \mathrm{if} \ u > 0 ,
\end{cases}
\end{align}
\item
(c) follows by the substitution rule for integrals with
\begin{align}
t
=
r \sqrt{ \mathcal{V}(X^{n} \mid Y^{n}) } ,
\end{align}
\item
(d) follows by the definition of $D_{Y^{n}}$ (see \eqref{def:capital-G_yn} written in the top of the next page),
\begin{figure*}[!t]
\begin{align}
D_{Y^{n}}
\coloneqq
\sqrt{ \mathcal{V}(X^{n} \mid Y^{n}) } \int_{\Phi^{-1}(1 - \varepsilon)}^{\infty} \left( \mathbb{P}\left\{ \frac{ \iota(X^{n} \mid Y^{n}) - \mathcal{H}(X^{n} \mid Y^{n}) }{ \sqrt{ \mathcal{V}(X^{n} \mid Y^{n}) } }  > r \ \middle| \ Y^{n} \right\} - (1 - \Phi( r )) \right) \, \mathrm{d} r
\label{def:capital-G_yn}
\end{align}
\hrule
\end{figure*}
\item
(e) follows from \eqref{eq:cond-expectation_int} with the trivial $\sigma$-algebra $\sigma(W) = \{ \emptyset, \Omega \}$, and
\item
(f) follows by the definition of $f_{\mathrm{G}} : [0, 1] \to [0, 1/\sqrt{2 \pi}]$ stated in \eqref{def:Gaussian_f}.
\end{itemize}
Substituting \eqref{eq:second-identity_max} into \eqref{eq:first-identity_max}, we obtain
\begin{align}
&
\mathbb{E}[ \langle \iota(X^{n} \mid Y^{n}) \mid Y^{n} \rangle_{\varepsilon} \mid Y^{n} ]
\notag \\
& =
(1 - \varepsilon) \, \mathcal{H}(X^{n} \mid Y^{n})
\notag \\
& \quad {}
- \sqrt{ \mathcal{V}(X^{n} \mid Y^{n}) } \, \Big( f_{\mathrm{G}}(1 - \varepsilon) - \varepsilon \, \Phi^{-1}(1 - \varepsilon) \Big)
\notag \\
& \qquad
{} + B_{Y^{n}} - D_{Y^{n}} - \varepsilon \, \Big( \eta_{Y^{n}} - \mathcal{H}(X^{n} \mid Y^{n}) \Big)
\notag \\
& \overset{\mathclap{\text{(a)}}}{=}
(1 - \varepsilon) \, \mathcal{H}(X^{n} \mid Y^{n})
\notag \\
& \quad {}
- \sqrt{ \mathcal{V}(X^{n} \mid Y^{n}) } \, \Big( f_{\mathrm{G}}(1 - \varepsilon) - \varepsilon \, \Phi^{-1}(1 - \varepsilon) \Big)
\notag \\
& \qquad
{} + B_{Y^{n}} - D_{Y^{n}} - \varepsilon \, \Big( b_{Y^{n}} + \sqrt{ \mathcal{V}(X^{n} \mid Y^{n}) } \, \Phi^{-1}( 1 - \varepsilon ) \Big)
\notag \\
& =
(1 - \varepsilon) \, \mathcal{H}(X^{n} \mid Y^{n})
\notag \\
& \quad {}
- \sqrt{ \mathcal{V}(X^{n} \mid Y^{n}) } \, f_{\mathrm{G}}(1 - \varepsilon) + B_{Y^{n}} - D_{Y^{n}} - \varepsilon \, b_{Y^{n}} ,
\label{eq:third-identity_max}
\end{align}
where (a) follows by the definition of $b_{Y^{n}}$ stated in \eqref{def:b_yn}.
Taking expectations in both sides of \eqref{eq:third-identity_max}, we have
\begin{align}
\mathfrak{C}_{\mathrm{c}}^{\varepsilon}(X^{n} \mid Y^{n})
& =
n \, (1 - \varepsilon) \, H(X \mid Y)
\notag \\
& \quad {}
- \mathbb{E}\Big[ \sqrt{ \mathcal{V}(X^{n} \mid Y^{n}) } \Big] \, f_{\mathrm{G}}(1 - \varepsilon)
\notag \\
& \qquad {}
+ \mathbb{E}[ B_{Y^{n}} ] + \mathbb{E}[ D_{Y^{n}} ] - \varepsilon \, \mathbb{E}[ b_{Y^{n}} ] .
\label{eq:fourth-identity_max}
\end{align}

Finally, we shall prove that the last three terms in \eqref{eq:fourth-identity_max} can be scaled as $+\mathrm{O}( 1 )$ as $n \to \infty$.
By Hypotheses (a) and (b) in \thref{th:max}, there exist two positive constants $V_{\inf}$ and $T_{\sup}$ satisfying
\begin{align}
\mathcal{V}(X \mid Y)
& \ge
V_{\inf}
\qquad (\mathrm{a.s.}) ,
\label{eq:unif-lower-bound_V} \\
\mathcal{T}(X \mid Y)
& \le
T_{\sup}
\qquad (\mathrm{a.s.}) ,
\label{eq:unif-upper-bound_T}
\end{align}
respectively.
Using those constants, we state the following lemma.

\begin{lemma}
\label{lem:remainder_terms}
Suppose that Hypotheses \emph{(a)} and \emph{(b)} in \thref{th:max} hold.
Given $0 < \varepsilon < 1$, it holds that
\begin{align}
\Big| \mathbb{E}[ B_{Y^{n}} ] + \mathbb{E}[ D_{Y^{n}} ] - \varepsilon \, \mathbb{E}[ b_{Y^{n}} ] \Big|
\le
\frac{ A \, (1+\varepsilon) \, T_{\sup}^{4/3} }{ c \, V_{\inf}^{3/2} } + \frac{ 3 \, A \, T_{\sup} }{ V_{\inf} }
\end{align}
for every $n \ge n_{0}$, where $A > 0$ is an absolute constant, $c = c( \varepsilon ) > 0$ is a constant depending only on $\varepsilon$, and $n_{0} = n_{0}(\varepsilon, V_{\inf}, T_{\sup}) \ge 1$ is a constant depending on $\varepsilon$, $V_{\inf}$, and $T_{\max}$.
\end{lemma}

To prove \lemref{lem:remainder_terms}, we shall use the following non-uniform strengthened Berry--Esseen bound.

\begin{lemma}[{non-uniform Berry--Esseen bound \cite{bikelis_1996}; see also \cite[Theorem~10]{kostina_polyanskiy_verdu_2015}}]
\label{lem:non-unif_Berry-Esseen}
Let $n \ge 1$ be an integer, and $Z_{1}, Z_{2}, \dots, Z_{n}$ independent, but not necessarily identically distributed, real-valued r.v.'s.
Define the following two quantities:
\begin{align}
V_{n}
& \coloneqq
\sum_{i = 1}^{n} \mathbb{E}[ (Z_{i} - \mathbb{E}[Z_{i}])^{2} ] ,
\\
T_{n}
& \coloneqq
\sum_{i = 1}^{n} \mathbb{E}[ |Z_{i} - \mathbb{E}[Z_{i}]|^{3} ] .
\end{align}
Then, it holds that
\begin{align}
\Bigg| \mathbb{P}\bigg\{ \sum_{i = 1}^{n} (Z_{i} - \mathbb{E}[Z_{i}]) \le z \, \sqrt{ V_{n} } \bigg\} - \Phi( z ) \Bigg|
\le
\frac{ A \, T_{n} }{ (1 + |z|^{3}) \, V_{n}^{3/2} }
\end{align}
for every $z \in \mathbb{R}$, provided that $V_{n} > 0$ and $T_{n} < \infty$, where $A > 0$ is an absolute constant.
\end{lemma}

Note that \lemref{lem:non-unif_Berry-Esseen} can be readily reduced to the uniform Berry--Esseen bound:
\begin{align}
\sup_{z \in \mathbb{R}} \Bigg| \mathbb{P}\bigg\{ \sum_{i = 1}^{n} (Z_{i} - \mathbb{E}[Z_{i}]) \le z \, \sqrt{ V_{n} } \bigg\} - \Phi( z ) \Bigg|
\le
\frac{ A \, T_{n} }{ V_{n}^{3/2} } .
\label{eq:unif_Berry-Esseen}
\end{align}

\begin{IEEEproof}[Proof of \lemref{lem:remainder_terms}]
Since $\iota(X^{n} \mid Y^{n})$ is a real-valued r.v., we see that $\mathbb{P}\{ \iota(X^{n} \mid Y^{n}) \le r \mid Y^{n} \}$ forms a cumulative distribution function of $r \in \mathbb{R}$ a.s.\ (see, e.g., \cite[Theorem~10.2.2]{dudley_2002}).
Thus, noting that $\eta_{Y^{n}}$ given in \eqref{eq:epsilon_given-yn} is $\sigma(Y^{n})$-measurable, it follows from the Berry--Esseen bound stated in \eqref{eq:unif_Berry-Esseen} with an absolute constant $A > 0$ that
\begin{align}
&
\mathbb{P}\{ \iota(X^{n} \mid Y^{n}) \le \eta_{Y^{n}} \mid Y^{n} \}
\notag \\
& \le
\Phi\left( \frac{ \eta_{Y^{n}} - \mathcal{H}(X^{n} \mid Y^{n}) }{ \sqrt{ \mathcal{V}(X^{n} \mid Y^{n}) } } \right) + \frac{ A \, \mathcal{T}(X^{n} \mid Y^{n}) }{ \mathcal{V}(X^{n} \mid Y^{n})^{3/2} }
\quad (\mathrm{a.s.}) ,
\label{eq:upper_Berry-Esseen_max} \\
&
\mathbb{P}\{ \iota(X^{n} \mid Y^{n}) < \eta_{Y^{n}} \mid Y^{n} \}
\notag \\
& \ge
\Phi\left( \frac{ \eta_{Y^{n}} - \mathcal{H}(X^{n} \mid Y^{n}) }{ \sqrt{ \mathcal{V}(X^{n} \mid Y^{n}) } } \right) - \frac{ A \, \mathcal{T}(X^{n} \mid Y^{n}) }{ \mathcal{V}(X^{n} \mid Y^{n})^{3/2} }
\quad (\mathrm{a.s.}) .
\label{eq:lower_Berry-Esseen_max}
\end{align}
It is clear that there exists an $n_{0} = n_{0}(\varepsilon, V_{\inf}, T_{\sup}) \ge 1$ satisfying
\begin{align}
\frac{ A \, T_{\sup} }{ \sqrt{ n } \, V_{\inf}^{3/2} }
<
\frac{ \min\{ 1 - \varepsilon, \varepsilon \} }{ 2 }
\end{align}
for every $n \ge n_{0}$, where note that we have assumed that $0 < \varepsilon < 1$.
Since
\begin{align}
\frac{ A \, \mathcal{T}(X^{n} \mid Y^{n}) }{ \mathcal{V}(X^{n} \mid Y^{n})^{3/2} }
=
\frac{ A \sum_{i = 1}^{n} \mathcal{T}(X_{i} \mid Y_{i}) }{ (\sum_{i = 1}^{n} \mathcal{V}(X_{i} \mid Y_{i}) )^{3/2} }
\le
\frac{ A \, T_{\sup} }{ \sqrt{ n } \, V_{\inf}^{3/2} }
\end{align}
a.s.,
substituting \eqref{eq:upper_Berry-Esseen_max} and \eqref{eq:lower_Berry-Esseen_max} into \eqref{eq:epsilon_given-yn}, we get
\begin{align}
\Phi^{-1}\left( 1 - \varepsilon - \frac{ A \, T_{\sup} }{ \sqrt{ n } \, V_{\inf}^{3/2} } \right)
& \le
\frac{ \eta_{Y^{n}} - \mathcal{H}(X^{n} \mid Y^{n}) }{ \sqrt{ \mathcal{V}(X^{n} \mid Y^{n}) } }
\notag \\
& \le
\Phi^{-1}\left( 1 - \varepsilon + \frac{ A \, T_{\sup} }{ \sqrt{ n } \, V_{\inf}^{3/2} } \right)
\label{eq:eta_Berry-Esseen_max}
\end{align}
a.s.\ for every $n \ge n_{0}$.
In addition, it follows by Taylor's theorem (and the inverse function theorem) that
\begin{align}
\Phi^{-1}( t + u ) - \Phi^{-1}( t )
=
\frac{ u }{ f_{\mathrm{G}}( s ) }
\label{eq:Taylor_inverse-Gaussian-CDF}
\end{align}
for every $0 < t < 1$, every $u \in (-t, 1 - t)$, and some $s \in [t, u+t]$.
Applying \eqref{eq:Taylor_inverse-Gaussian-CDF} to \eqref{eq:eta_Berry-Esseen_max}, we have
\begin{align}
| b_{Y^{n}} |
& =
\Big| \eta_{Y^{n}} - \mathcal{H}(X^{n} \mid Y^{n}) - \sqrt{ \mathcal{V}(X^{n} \mid Y^{n}) } \, \Phi^{-1}( 1 - \varepsilon ) \Big|
\notag \\
& \le
\frac{ A \, T_{\sup} }{ c } \sqrt{ \frac{ \mathcal{V}(X^{n} \mid Y^{n}) }{ n \, V_{\inf}^{3} } }
\notag \\
& \le
\frac{ A \, T_{\sup}^{4/3} }{ c \, V_{\inf}^{3/2} }
\qquad (\mathrm{a.s.})
\label{eq:estimate_eta_each-yn}
\end{align}
for every $n \ge n_{0}$, where the constant $c > 0$ is given as
\begin{align}
c
=
c( \varepsilon )
\coloneqq
\begin{dcases}
f_{\mathrm{G}}\bigg( \frac{ \varepsilon }{ 2 } \bigg)
& \mathrm{if} \ 0 < \varepsilon \le \frac{ 1 }{ 2 } ,
\\
f_{\mathrm{G}}\bigg( \frac{ 1 + \varepsilon }{ 2 } \bigg)
& \mathrm{if} \ \frac{ 1 }{ 2 } < \varepsilon < 1 ,
\end{dcases}
\label{eq:const_c}
\end{align}
and the last inequality follows from the fact that%
\footnote{The first inequality in \eqref{eq:bound_on_V_by_T} can be verified by $(\mathbb{E}[ |Z|^{p} \mid \mathcal{G} ])^{1/p} \le (\mathbb{E}[ |Z|^{q} \mid \mathcal{G} ])^{1/q}$ for $1 \le p < q$ and every $\mathcal{G} \subset \mathcal{F}$.}
\begin{align}
\mathcal{V}(X^{n} \mid Y^{n})
& =
\sum_{i = 1}^{n} \mathcal{V}(X_{i} \mid Y_{i})
\notag \\
& \le
\sum_{i = 1}^{n} \mathcal{T}(X_{i} \mid Y_{i})^{2/3}
\notag \\
& \le
n \, T_{\sup}^{2/3}
\qquad (\mathrm{a.s.}) .
\label{eq:bound_on_V_by_T}
\end{align}
We readily see that $F_{n}$ is bounded away from zero for sufficiently large $n$.

Now, it follows by the definition of $B_{Y^{n}}$ stated in \eqref{def:capital-B_yn} that
\begin{align}
\mathbb{E}[ |B_{Y^{n}}| ]
\le
\mathbb{E}\left[ \left| \int_{\min\{ 0, b_{Y^{n}} \}}^{\max\{ 0, b_{Y^{n}} \}} \mathrm{d} r \right| \right]
=
\mathbb{E}[ |b_{Y^{n}}| ]
\label{eq:capital-B_small-b}
\end{align}
for every $n \ge 1$.
Moreover, it follows from \eqref{eq:estimate_eta_each-yn} that
\begin{align}
\mathbb{E}[ |b_{Y^{n}}| ]
& \le
\frac{ A \, T_{\sup}^{4/3} }{ c \, V_{\inf}^{3/2} }
\end{align}
for every $n \ge n_{0}$.
Furthermore, it follows by the definition of $D_{Y^{n}}$ stated in \eqref{def:capital-G_yn} that \eqref{eq:upper-bound_Gn} written in the top of the next page holds for every $n \ge 1$,
\begin{figure*}[!t]
\begin{align}
\mathbb{E}[ D_{Y^{n}} ]
& =
\mathbb{E} \left[ \sqrt{ \mathcal{V}(X^{n} \mid Y^{n}) } \int_{\Phi^{-1}(1 - \varepsilon)}^{\infty} \left( \mathbb{P}\left\{ \frac{ \iota(X^{n} \mid Y^{n}) - \mathcal{H}(X^{n} \mid Y^{n}) }{ \sqrt{ \mathcal{V}(X^{n} \mid Y^{n}) } }  > r \ \middle| \ Y^{n} \right\} - (1 - \Phi( r )) \right) \, \mathrm{d} r \right]
\notag \\
& \overset{\mathclap{\text{(a)}}}{\le}
A \, \mathbb{E} \left[ \frac{ \mathcal{T}(X^{n} \mid Y^{n}) }{ \mathcal{V}(X^{n} \mid Y^{n}) } \right] \int_{\Phi^{-1}(1 - \varepsilon)}^{\infty} \frac{ \mathrm{d} r }{ 1 + |r|^{3} }
\notag \\
& \overset{\mathclap{\text{(b)}}}{\le}
\frac{ A \, T_{\sup} }{ V_{\inf} } \int_{\Phi^{-1}(1 - \varepsilon)}^{\infty} \frac{ \mathrm{d} r }{ 1 + |r|^{3} }
\notag \\
& \le
\frac{ 2 \, A \, T_{\sup} }{ V_{\inf} } \int_{0}^{\infty} \frac{ \mathrm{d} r }{ 1 + r^{3} }
\notag \\
& \le
\frac{ 2 \, A \, T_{\sup} }{ V_{\inf} } \bigg( \int_{0}^{1} \mathrm{d} r + \int_{1}^{\infty} \frac{ \mathrm{d} r }{ r^{3} } \bigg)
\notag \\
& =
\frac{ 3 \, A \, T_{\sup} }{ V_{\inf} }
\label{eq:upper-bound_Gn}
\end{align}
\hrule
\end{figure*}
where
\begin{itemize}
\item
(a) follows by the non-uniform Berry--Esseen theorem stated in \lemref{lem:non-unif_Berry-Esseen} with an absolute constant $A > 0$, and
\item
(b) follows from \eqref{eq:unif-lower-bound_V} and \eqref{eq:unif-upper-bound_T}.
\end{itemize}
Analogously, we may see that
\begin{align}
\mathbb{E}[ D_{Y^{n}} ]
\ge
- \frac{ 3 \, A \, T_{\sup} }{ V_{\inf} }
\label{eq:lower-bound_Gn}
\end{align}
for every $n \ge 1$.
Combining \eqref{eq:capital-B_small-b}--\eqref{eq:lower-bound_Gn}, we obtain \lemref{lem:remainder_terms}, as desired.
\end{IEEEproof}

The proof of \lemref{lem:asympt_cond_eps-entropy} is completed by applying \lemref{lem:remainder_terms} to \eqref{eq:fourth-identity_max}.
\hfill\IEEEQEDhere

\section{Proof of \lemref{lem:max_V}}
\label{app:max_V}

To prove \lemref{lem:max_V}, we use the following lemma.

\begin{lemma}[{Bernstein's inequality; see, e.g., \cite[Equation~(2.10)]{boucheron_lugosi_massart_2013}}]
\label{lem:Bernstein}
Let $n$ be a positive integer and $Z_{1}, \dots, Z_{n}$ real-valued and independent r.v.'s.
Suppose that there exists a positive constant $c$ satisfying $|Z_{i}| \le c$ a.s.\ for each $1 \le i \le n$.
Then, it holds that
\begin{align}
\mathbb{P}\left\{ \sum_{i = 1}^{n} (Z_{i} - \mathbb{E}[Z_{i}]) \le - t \right\}
\le
\exp\left( - \frac{ t^{2} }{ 2 (\sum_{i = 1}^{n} \mathbb{E}[ Z_{i}^{2} ] + c \, t / 3) } \right)
\end{align}
for every positive real number $t$.
\end{lemma}

Assume without loss of generality that $\mathbb{P}\{ \mathcal{V}(X \mid Y) > 0 \} > 0$.
For each integer $n \ge 1$, define the event
\begin{align}
\mathcal{A}_{n}
\coloneqq
\left\{ \sum_{i = 1}^{n} \frac{ V_{\mathrm{c}}(X \mid Y) - \mathcal{V}(X_{i} \mid Y_{i}) }{ \sqrt{2 \, \mathbb{E}[ \mathcal{V}(X \mid Y)^{2} ] \, n \log n} } \le 1 \right\} .
\label{def:event-An}
\end{align}
Since we have assumed that there exists a positive constant $V_{\sup}$ satisfying $\mathcal{V}(X \mid Y) \le V_{\sup}$ a.s., it follows from \lemref{lem:Bernstein} that \eqref{eq:Bernstein} written in the top of the next page holds,
\begin{figure*}[!t]
\begin{align}
\mathbb{P}( \mathcal{A}_{n} )
& \ge
1 - \exp\left( - \frac{ 2 \, \mathbb{E}[ \mathcal{V}(X \mid Y)^{2} ] \, n \log n }{ 2 \, n \, \mathbb{E}[ \mathcal{V}(X \mid Y)^{2} ] + (2/3) \, V_{\sup} \sqrt{2 \, \mathbb{E}[ \mathcal{V}(X \mid Y)^{2} ] \, n \log n} } \right)
\notag \\
& \ge
1 - \exp\left( - \frac{ \log n }{ 1 + V_{\sup} \sqrt{ (2 \log n) / (9 \, n \, \mathbb{E}[ \mathcal{V}(X \mid Y)^{2} ]) } } \right)
\notag \\
& =
1 + \mathrm{O}\left( \frac{ 1 }{ \sqrt{n} } \right)
\qquad (\mathrm{as} \ n \to \infty)
\label{eq:Bernstein}
\end{align}
\hrule
\end{figure*}
where the last equality follows from the fact that
\begin{align}
V_{\sup} \sqrt{ \frac{ 2 \log n }{ 9 \, n \, \mathbb{E}[ \mathcal{V}(X \mid Y)^{2} ] } }
\le
1
\end{align}
for sufficiently large $n$.
On the other hand, for sufficiently large $n$ satisfying
\begin{align}
\sqrt{ \frac{ 2 \, \mathbb{E}[ \mathcal{V}(X \mid Y)^{2} ] \log n }{ n \, V_{\mathrm{c}}(X \mid Y)^{2} } }
\le
1 ,
\end{align}
we observe that
\begin{align}
&
\mathbb{E}\Big[ \sqrt{ \mathcal{V}(X^{n} \mid Y^{n}) } \Big]
\notag \\
& \quad \ge
\mathbb{E}\Big[ \sqrt{ \mathcal{V}(X^{n} \mid Y^{n}) } \, \bvec{1}_{\mathcal{A}_{n}} \Big]
\notag \\
& \quad \ge
\mathbb{E}\left[ \sqrt{ n \, V_{\mathrm{c}}(X \mid Y) \left( 1 - \sqrt{ \frac{ 2 \, \mathbb{E}[ \mathcal{V}(X \mid Y)^{2} ] \log n }{ n \, V_{\mathrm{c}}(X \mid Y)^{2} } } \right) } \, \bvec{1}_{\mathcal{A}_{n}} \right]
\notag \\
& \quad \ge
\sqrt{ n \, V_{\mathrm{c}}(X \mid Y) } \left( 1 - \sqrt{ \frac{ 2 \, \mathbb{E}[ \mathcal{V}(X \mid Y)^{2} ] \log n }{ n \, V_{\mathrm{c}}(X \mid Y)^{2} } } \right) \, \mathbb{P}( \mathcal{A}_{n} ) ,
\label{eq:LLN}
\end{align}
where the second inequality follows by the definition of $\mathcal{A}_{n}$ stated in \eqref{def:event-An}.
Combining \eqref{eq:Jensen_V}, \eqref{eq:Bernstein}, and \eqref{eq:LLN}, we obtain \eqref{eq:Vmax_asympt}, completing the proof.
\hfill\IEEEQEDhere

\section{Proof of \lemref{lem:optimal-code_avg}}
\label{app:optimal-code_avg}.

Consider an $(L, \varepsilon)_{\mathrm{avg}}$-code $(F, G)$ satisfying
\begin{align}
\mathbb{E}[ \ell( F(X, Y) ) ]
& \le
L ,
\\
\mathbb{P}\{ X \neq G(F(X, Y), Y) \}
& \le
\varepsilon .
\label{eq:avg-error_F0G0}
\end{align}
As similarly done in \eqref{eq:construction_F1} and \eqref{eq:construction_g1} of \appref{app:optimal-code_max}, construct another code $(F_{1}, g_{1})$ from the original code $(F, G)$ via the random maps $\Phi_{Y}$ and $\Psi_{Y}$ defined in \eqref{def:phi} and \eqref{def:psi}, respectively.
Obviously, the same derivations as \eqref{eq:L_F1_max} and \eqref{eq:image_max} yield
\begin{align}
\mathbb{E}[ \ell( F_{1}(X, Y) ) ]
& \le
L ,
\label{eq:L_F1_avg}
\\
1 - \varepsilon
& \le
\mathbb{E}\left[ \sum_{i =1}^{|\mathcal{B}(Y)|} \mathbb{P}\{ X = \Psi_{Y}(i) \mid Y \} \right]
\notag \\
& =
\sum_{i =1}^{\infty} \mathbb{P}\{ X = \Psi_{Y}(i) \} ,
\end{align}
respectively, where the last equality follows from the fact that $\Psi_{Y}(i) = 0$ whenever $i > |\mathcal{B}(Y)|$ a.s.\ (see \eqref{def:psi}).
Now, defining two parameters $\xi$ and $\zeta$ so that%
\footnote{Note that $\alpha = 0$ if $\mathbb{P}\{ X = \psi(1, Y) \} \ge 1-\varepsilon$; and $\alpha = \infty$ if $\sum_{i = 1}^{\infty} \mathbb{P}\{ X = \psi(i, Y) \} = 1-\varepsilon$ and $\mathbb{P}\{ X = \psi(i, Y) \} > 0$ for all $i \ge 1$.}
\begin{align}
\xi
& \coloneqq
\sup\left\{ k \ge 0 \ \middle| \ \sum_{i = 1}^{k} \mathbb{P}\{ X = \Psi_{Y}(i) \} \le 1-\varepsilon \right\} ,
\\
\zeta
& \coloneqq
1 - \varepsilon - \sum_{i = 1}^{\xi} \mathbb{P}\{ X = \Psi_{Y}(i) \} ,
\end{align}
respectively.
In addition, define two parameters $\kappa$ and $\gamma$ so that
\begin{align}
\kappa
& \coloneqq
\sup\left\{ k \ge 0 \ \middle| \ \sum_{x = 1}^{k} \mathbb{P}\{ X = \varsigma_{Y}(x) \} \le 1 - \varepsilon \right\} ,
\\
\gamma
& \coloneqq
1 - \varepsilon - \sum_{x = 1}^{\kappa} \mathbb{P}\{ X = \varsigma_{Y}(x) \} .
\end{align}
respectively, where $\varsigma_{Y}$ is given in \eqref{def:sigmaY}.
Furthermore, define
\begin{align}
q_{1}(k)
& \coloneqq
\begin{cases}
\mathbb{P}\{ X = \Psi_{Y}(k) \}
& \mathrm{if} \ 1 \le k \le \xi ,
\\
\zeta
& \mathrm{if} \ k = \xi + 1 ,
\\
0
& \mathrm{if} \ \xi + 2 \le k < \infty ,
\end{cases}
\\
q_{2}(x)
& \coloneqq
\begin{cases}
\mathbb{P}\{ X = \varsigma_{Y}(x) \}
& \mathrm{if} \ 1 \le x \le \kappa ,
\\
\gamma
& \mathrm{if} \ x = \kappa + 1 ,
\\
0
& \mathrm{if} \ \kappa + 2 \le x < \infty .
\end{cases}
\label{def:q2}
\end{align}
Similar to \eqref{eq:LB_L_F1g1_max_sum} and \eqref{eq:cond-eps_optimal_sum}, we observe that
\begin{align}
\mathbb{E}[ \ell( F_{1}(X, Y) ) ]
& \ge
\sum_{j = 1}^{\infty} \sum_{k = 2^{j}}^{\infty} q_{1}( k ) ,
\label{eq:LB_L_F1g1_avg_sum} \\
\mathbb{E}[ \langle \lfloor \log \varsigma_{Y}^{-1}(X) \rfloor \rangle_{\varepsilon} ]
& =
\sum_{j = 1}^{\infty} \sum_{x = 2^{j}}^{\infty} q_{2}( x ) .
\label{eq:uncond-eps_optimal_sum}
\end{align}
Since $\varsigma_{Y}$ defined in \eqref{def:sigmaY} ensures that
\begin{align}
\mathbb{P}\{ X = \varsigma_{Y}(1) \}
\ge
\mathbb{P}\{ X = \varsigma_{Y}(2) \}
\ge
\cdots ,
\end{align}
it can be verified that $q_{1}(\cdot)$ is majorized by $q_{2}(\cdot)$, i.e., it follows that
\begin{align}
\sum_{k = 1}^{l} q_{1}(k)
\le
\sum_{x = 1}^{l} q_{2}(x)
\qquad (\mathrm{a.s.})
\label{eq:majorization_avg}
\end{align}
for every $l \ge 1$, and
\begin{align}
\sum_{k = 1}^{\infty} q_{1}(k)
=
\sum_{x = 1}^{\infty} q_{2}(x)
=
1 - \varepsilon
\qquad (\mathrm{a.s.}) .
\label{eq:majorization_equality_avg}
\end{align}
Therefore, it follows from \eqref{eq:LB_L_F1g1_avg_sum}, \eqref{eq:uncond-eps_optimal_sum}, \eqref{eq:majorization_avg} and \eqref{eq:majorization_equality_avg} that
\begin{align}
\mathbb{E}[ \ell( F_{1}(X, Y) ) ]
\ge
\mathbb{E}[ \langle \lfloor \log \varsigma_{Y}^{-1}(X) \rfloor \rangle_{\varepsilon} ] .
\label{eq:LB_L_F1g1_cutoff-eps_avg}
\end{align}
Combining \eqref{eq:L_F1_avg} and \eqref{eq:LB_L_F1g1_cutoff-eps_avg}, we observe that the existence of an $(L, \varepsilon)_{\mathrm{avg}}$-code $(F, G)$ implies that
\begin{align}
L
\ge
\mathbb{E}[ \langle \lfloor \log \varsigma_{Y}^{-1}(X) \rfloor \rangle_{\varepsilon} ] ,
\label{eq:LB_L_avg}
\end{align}
which corresponds to the converse bound of \lemref{lem:optimal-code_avg}.

Finally, we shall show the existence of an $(L, \varepsilon)_{\mathrm{avg}}$-code meeting the equality in \eqref{eq:LB_L_avg}.
In fact, constructing a variable-length stochastic code $(F_{\mathrm{avg}}^{\ast}, g^{\ast})$ so that
\begin{align}
F_{\mathrm{avg}}^{\ast}(x, Y)
& \coloneqq
\begin{cases}
\bvec{b}_{\varsigma_{Y}^{-1}(x)}
& \mathrm{if} \ 1 \le \varsigma_{Y}^{-1}(x) \le \kappa ,
\\
B_{\mathrm{avg}}
& \mathrm{if} \ \varsigma_{Y}^{-1}(x) = \kappa + 1 ,
\\
\varnothing
& \mathrm{if} \ \kappa < \varsigma_{Y}^{-1}(x) < \infty ,
\end{cases}
\label{eq:optimal-F_avg}
\\
g^{\ast}(\bvec{b}, Y)
& \coloneqq
x
\quad \mathrm{if} \ \bvec{b} = \bvec{b}_{\varsigma_{Y}^{-1}(x)} \ \mathrm{for} \ \mathrm{some} \ x \in \mathcal{X} ,
\label{eq:optimal-g_avg}
\end{align}
where $B_{\mathrm{avg}}$ denotes a $\{ \varnothing, \bvec{b}_{\kappa+1} \}$-valued r.v.\ satisfying the condition that it is independent of $(X, Y)$ (i.e., that $B_{\mathrm{avg}} \Perp (X, Y)$), and
\begin{align}
B_{\mathrm{avg}}
=
\begin{cases}
\bvec{b}_{\kappa+1}
& \mathrm{with} \ \mathrm{probability} \ \gamma ,
\\
\varnothing
& \mathrm{with} \ \mathrm{probability} \ 1 - \gamma ,
\end{cases}
\end{align}
we readily see that
\begin{align}
\mathbb{E}[ \ell( F_{\mathrm{avg}}^{\ast}(X, Y) ) ]
& =
\mathbb{E}[ \langle \lfloor \log \varsigma_{Y}^{-1}(X) \rfloor \rangle_{\varepsilon} ] ,
\\
\mathbb{P}\{ X \neq g^{\ast}(F_{\mathrm{avg}}^{\ast}(X, Y), Y) \}
& =
\varepsilon .
\end{align}
This completes the proof of \lemref{lem:optimal-code_avg}.
\hfill\IEEEQEDhere

\section{Proof of \lemref{lem:coding_vs_guessing}}
\label{app:guessing}

\subsection{Proof of \eqref{eq:Nmax_vs_Lmax}}
\label{app:guessing_max}

By \lemref{lem:optimal-code_max} in \appref{app:eps-cutoff_max-err}, it suffices to show that
\begin{align}
- |\log c_{\mathrm{e}}|
& \le
N_{\max}^{\ast}(\varepsilon, X, Y) - \mathbb{E}[ \langle \lfloor \log \varsigma_{Y}^{-1}( X ) \rfloor \mid Y \rangle_{\varepsilon} ]
\notag \\
& \le
1 + |\log c_{\mathrm{e}}| ,
\label{eq:Cmax_vs_cutoff-floor}
\end{align}
where $N_{\max}^{\ast}(\varepsilon, X, Y) \coloneqq N_{\max}^{\ast}(1, \varepsilon, X, Y)$.
Throughout \appref{app:guessing}, we consider one-shot ($n=1$) guessing strategies as defined in \sectref{sect:guessing} (by taking $n$ therein to be $1$).
Specifically, we now consider an \emph{$(N, \varepsilon)_{\max}$-guessing strategy} $(\mathsf{g}, \pi(\cdot \mid \cdot))$ satisfying
\begin{align}
\mathbb{E}[ \log \mathsf{G}(X, Y) ]
& \le
N ,
\label{eq:N_Guess_max} \\
\mathbb{P}\{ \mathsf{G}(X, Y) \neq \mathsf{g}(X, Y) \mid Y \}
& \le
\varepsilon
\qquad (\mathrm{a.s.}) .
\label{eq:max-error_Guess}
\end{align}
It is clear that
\begin{align}
\big| \mathbb{E}[ (\log \mathsf{G}(X, Y)) \, \bvec{1}_{\{ \mathsf{G}(X, Y) \neq \mathsf{g}(X, Y) \}} ] \big|
\le
|\log c_{\mathrm{e}}| .
\label{eq:error_cost}
\end{align}
It follows from \eqref{eq:max-error_Guess} that
\begin{align}
1 - \varepsilon
& \le
1 - \mathbb{P}\{ \mathsf{G}(X, Y) \neq \mathsf{g}(X, Y) \mid Y \}
\notag \\
& =
\mathbb{P}\{ \mathsf{G}(X, Y) = \mathsf{g}(X, Y) \mid Y \}
\notag \\
& =
\sum_{k = 1}^{\infty} \mathbb{P}\{ \mathsf{G}(X, Y) = k \mid Y \}
\qquad (\mathrm{a.s.}) ,
\label{eq:sum_G=g}
\end{align}
where the last equality follows from the fact that $\mathsf{G}(X, Y) = k$ only if $\mathsf{g}(X, Y) = k$.
Based on \eqref{eq:sum_G=g}, define the $\sigma(Y)$-measurable real-valued r.v.'s $\nu_{Y}$ and $\upsilon_{Y}$ as follows:
\begin{align}
\nu_{Y}
& \coloneqq
\sup\left\{ k \ge 1 \ \middle| \ \sum_{x = 1}^{k} \mathbb{P}\{ \mathsf{G}(X, Y) = k \mid Y \} \le 1 - \varepsilon \right\} ,
\\
\upsilon_{Y}
& \coloneqq
1 - \varepsilon - \sum_{x = 1}^{\nu_{Y}} \mathbb{P}\{ \mathsf{G}(X, Y) = k \mid Y \} ,
\end{align}
respectively.
Furthermore, define
\begin{align}
p_{3}(k \mid Y)
& \coloneqq
\begin{cases}
\mathbb{P}\{ \mathsf{G}(X, Y) = k \mid Y \}
& \mathrm{if} \ 1 \le k \le \nu_{Y} ,
\\
\upsilon_{Y}
& \mathrm{if} \ k = \nu_{Y} + 1 ,
\\
0
& \mathrm{if} \ \nu_{Y} + 2 \le k < \infty .
\end{cases}
\end{align}
Similar to \eqref{eq:LB_L_F1g1_max_sum} in \appref{app:optimal-code_max}, we get
\begin{align}
\mathbb{E}[ \lfloor \log \mathsf{G}(X, Y) \rfloor \, \bvec{1}_{\{ \mathsf{G}(X, Y) = \mathsf{g}(X, Y) \}} \mid Y ]
& \ge
\sum_{j = 1}^{\infty} \sum_{i = 2^{j}}^{\infty} p_{3}(i \mid Y)
\label{eq:cond_log-guessing_sum}
\end{align}
a.s.
Moreover, in the same way as \eqref{eq:majorization_max} and \eqref{eq:majorization_equality_max} in \appref{app:optimal-code_max}, it can be verified that $p_{3}(\cdot \mid Y)$ is majorized by $p_{2}(\cdot \mid Y)$ a.s., where $p_{2}(\cdot \mid Y)$ is defined in \eqref{def:p2}.
Therefore, combining \eqref{eq:N_Guess_max}--\eqref{eq:error_cost} and \eqref{eq:cond_log-guessing_sum}, the existence of an $(N, \varepsilon)_{\max}$-guessing strategy implies that
\begin{align}
N + |\log c_{\mathrm{e}}|
\ge
\mathbb{E}[ \langle \lfloor \log \varsigma_{Y}^{-1}( X ) \rfloor \mid Y \rangle_{\varepsilon} ] ,
\end{align}
which corresponds to the left-hand inequality of \eqref{eq:Cmax_vs_cutoff-floor}.

Finally, considering the guessing strategy $(\mathsf{g}^{\ast}, \pi_{\max}^{\ast}(\cdot \mid \cdot))$ given by
\begin{align}
\mathsf{g}^{\ast}(x, Y)
& =
\varsigma_{Y}^{-1}( x )
\qquad (\mathrm{a.s.}) ,
\label{eq:optimal_guessing} \\
\pi_{\max}^{\ast}(x \mid Y)
& =
\begin{dcases}
0
& \mathrm{if} \ 1 \le x \le \kappa_{Y} ,
\\
1 - \frac{ \gamma_{Y} }{ P_{X|Y}(\varsigma_{Y}(x) \mid Y) }
& \mathrm{if} \ x = \kappa_{Y} + 1 ,
\\
1
& \mathrm{if} \ \kappa_{Y} + 2 \le x < \infty ,
\end{dcases}
\end{align}
and denoting by $\mathsf{G}_{\max}^{\ast} : \mathcal{X} \times \mathcal{Y} \to \mathcal{X}$ the giving-up guessing function induced by the strategy $(\mathsf{g}^{\ast}, \pi_{\max}^{\ast}(\cdot \mid \cdot))$, we obtain after some algebra that
\begin{align}
\mathbb{P}\{ \mathsf{G}_{\max}^{\ast}(X, Y) \neq \mathsf{g}^{\ast}(X, Y) \mid Y \}
& =
\varepsilon
\qquad (\mathrm{a.s.})
\end{align}
and
\begin{align}
&
\mathbb{E}[ (\log \mathsf{G}_{\max}^{\ast}(X, Y)) \, \bvec{1}_{\{ \mathsf{G}(X, Y) = \mathsf{g}(X, Y) \}} ]
\notag \\
& \qquad \le
\mathbb{E}[ \lfloor \log \mathsf{G}_{\max}^{\ast}(X, Y) \rfloor \, \bvec{1}_{\{ \mathsf{G}(X, Y) = \mathsf{g}(X, Y) \}} ] + 1
\notag \\
& \qquad =
\mathbb{E}[ \langle \lfloor \log \varsigma_{Y}^{-1}(X) \rfloor \mid Y \rangle_{\varepsilon} ] + 1 ,
\end{align}
which, together with \eqref{eq:error_cost} and \eqref{eq:cond_log-guessing_sum}, imply the right-hand inequality of \eqref{eq:Cmax_vs_cutoff-floor}.
This completes the proof of \eqref{eq:Nmax_vs_Lmax}.
\hfill\IEEEQEDhere

\subsection{Proof of \eqref{eq:Navg_vs_Lavg}}
\label{app:guessing_avg}

By \lemref{lem:optimal-code_avg}, it suffices to show that
\begin{align}
- |\log c_{\mathrm{e}}|
& \le
N_{\mathrm{avg}}^{\ast}(\varepsilon, X, Y) - \mathbb{E}[ \langle \lfloor \log \varsigma_{Y}^{-1}( X ) \rfloor \mid Y \rangle_{\varepsilon} ]
\notag \\
& \le
1 + |\log c_{\mathrm{e}}| ,
\label{eq:Cavg_vs_cutoff-floor}
\end{align}
where $N_{\mathrm{avg}}^{\ast}(\varepsilon, X, Y) \coloneqq N_{\mathrm{avg}}^{\ast}(1, \varepsilon, X, Y)$.
Consider an \emph{$(N, \varepsilon)_{\mathrm{avg}}$-guessing strategy} $(\mathsf{g}, \pi(\cdot \mid \cdot))$ satisfying
\begin{align}
\mathbb{E}[ \log \mathsf{G}(X, Y) ]
& \le
N ,
\label{eq:N_Guess_avg} \\
\mathbb{P}\{ \mathsf{G}(X, Y) \neq \mathsf{g}(X, Y) \}
& \le
\varepsilon ,
\label{eq:avg-error_Guess}
\end{align}
where $\mathsf{G} : \mathcal{X} \times \mathcal{Y} \to \mathcal{X}$ is the giving-up guessing function induced by the strategy $(\mathsf{g}, \pi(\cdot \mid \cdot))$.
Similar to \eqref{eq:sum_G=g}, one has
\begin{align}
1 - \varepsilon
& \le
\sum_{k = 1}^{\infty} \mathbb{P}\{ \mathsf{G}(X, Y) = k \} .
\label{eq:sum_G=g_avg}
\end{align}
Based on \eqref{eq:sum_G=g_avg}, define two parameters $\nu$ and $\upsilon$ by
\begin{align}
\nu
& \coloneqq
\sup\left\{ k \ge 1 \ \middle| \ \sum_{x = 1}^{k} \mathbb{P}\{ \mathsf{G}(X, Y) = k \} \le 1 - \varepsilon \right\} ,
\\
\upsilon
& \coloneqq
1 - \varepsilon - \sum_{x = 1}^{\nu} \mathbb{P}\{ \mathsf{G}(X, Y) = k \} ,
\end{align}
respectively.
Moreover, define
\begin{align}
q_{3}(k)
& \coloneqq
\begin{cases}
\mathbb{P}\{ \mathsf{G}(X, Y) = k \}
& \mathrm{if} \ 1 \le k \le \nu ,
\\
\upsilon
& \mathrm{if} \ k = \nu + 1 ,
\\
0
& \mathrm{if} \ \nu + 2 \le k < \infty ,
\end{cases}
\end{align}
Similar to \eqref{eq:cond_log-guessing_sum}, we observe that
\begin{align}
\mathbb{E}[ \lfloor \log \mathsf{G}(X, Y) \rfloor \, \bvec{1}_{\{ \mathsf{G}(X, Y) = \mathsf{g}(X, Y) \}} ]
& \ge
\sum_{j = 1}^{\infty} \sum_{k = 2^{j}}^{\infty} q_{3}( k ) ,
\label{eq:uncond_log-guessing strategy_sum}
\end{align}
Moreover, in the same way as \eqref{eq:majorization_avg} and \eqref{eq:majorization_equality_avg} in \appref{app:optimal-code_avg}, it can be verified that $q_{3}(\cdot)$ is majorized by $q_{2}(\cdot)$, where $q_{2}(\cdot)$ is defined in \eqref{def:q2}.
Therefore, combining \eqref{eq:error_cost}, \eqref{eq:N_Guess_avg}, \eqref{eq:avg-error_Guess}, and \eqref{eq:uncond_log-guessing strategy_sum}, the existence of an $(N, \varepsilon)_{\mathrm{avg}}$-guessing strategy implies that
\begin{align}
N + |\log c_{\mathrm{e}}|
\ge
\mathbb{E}[ \langle \lfloor \log \varsigma_{Y}^{-1}( X ) \rfloor \rangle_{\varepsilon} ] ,
\end{align}
which corresponds to the left-hand inequality of \eqref{eq:Cavg_vs_cutoff-floor}.

Finally, considering the guessing strategy $(\mathsf{g}^{\ast}, \pi_{\mathrm{avg}}^{\ast}(\cdot \mid \cdot))$ so that $\mathsf{g}^{\ast}$ is given as \eqref{eq:optimal_guessing} and
\begin{align}
\pi_{\mathrm{avg}}^{\ast}(x \mid Y)
& =
\begin{dcases}
0
& \mathrm{if} \ 1 \le x \le \kappa ,
\\
1 - \frac{ \gamma }{ \mathbb{P}\{ X = \varsigma_{Y}(x) \} }
& \mathrm{if} \ x = \kappa + 1 ,
\\
1
& \mathrm{if} \ \kappa + 2 \le x < \infty ,
\end{dcases}
\end{align}
and denoting by $\mathsf{G}_{\mathrm{avg}}^{\ast} : \mathcal{X} \times \mathcal{Y} \to \mathcal{X}$ the giving-up guessing function induced by the strategy $(\mathsf{g}^{\ast}, \pi_{\mathrm{avg}}^{\ast}(\cdot \mid \cdot))$, we obtain after some algebra that
\begin{align}
\mathbb{P}\{ \mathsf{G}_{\mathrm{avg}}^{\ast}(X, Y) \neq \mathsf{g}^{\ast}(X, Y) \}
& =
\varepsilon
\end{align}
and
\begin{align}
&
\mathbb{E}[ (\log \mathsf{G}_{\mathrm{avg}}^{\ast}(X, Y)) \, \bvec{1}_{\{ \mathsf{G}(X, Y) = \mathsf{g}(X, Y) \}} ]
\notag \\
& \qquad \le
\mathbb{E}[ (\lfloor \log \mathsf{G}_{\mathrm{avg}}^{\ast}(X, Y) \rfloor) \, \bvec{1}_{\{ \mathsf{G}(X, Y) = \mathsf{g}(X, Y) \}} ] + 1
\notag \\
& \qquad =
\mathbb{E}[ \langle \lfloor \log \varsigma_{Y}^{-1}(X) \rfloor \rangle_{\varepsilon} ] + 1 ,
\end{align}
which, together with \eqref{eq:error_cost} and \eqref{eq:uncond_log-guessing strategy_sum}, imply the right-hand inequality of \eqref{eq:Cavg_vs_cutoff-floor}.
This completes the proof of \eqref{eq:Navg_vs_Lavg}.
\hfill\IEEEQEDhere

\section{Relaxations of Hypotheses in \thref{th:max}}
\label{app:relaxation}

\subsection{Proof of \propref{prop:finite-X}}
\label{app:finite-X}

Suppose that $\mathbb{P}\{ X \in \mathcal{A} \} = 1$ for some finite $\mathcal{A} \subset \mathcal{X}$.
Then, we readily see that
\begin{align}
\sum_{x \in \mathcal{A}} P_{X|Y}(x \mid Y)
=
1
\qquad (\mathrm{a.s.}) .
\label{eq:finite-sum_is_unity}
\end{align}
Hence, the well-known upper bound on the Shannon entropy shows that
\begin{align}
&
\sum_{x \in \mathcal{X}} P_{X|Y}(x \mid Y) \log \frac{ 1 }{ P_{X|Y}(x \mid Y) }
\notag \\
& \qquad =
\sum_{x \in \mathcal{A}} P_{X|Y}(x \mid Y) \log \frac{ 1 }{ P_{X|Y}(x \mid Y) }
\notag \\
& \qquad \le
\log |\mathcal{A}|
\qquad (\mathrm{a.s.}) .
\end{align}
Therefore, it can be verified by the dominated convergence theorem for the conditional expectation that
\begin{align}
\mathcal{H}(X \mid Y)
\le
\log |\mathcal{A}|
\qquad (\mathrm{a.s.}) .
\label{eq:DCT_H}
\end{align}

On the other hand, we get
\begin{align}
&
\sum_{x \in \mathcal{A}} P_{X|Y}(x \mid Y) \, \left| \log \frac{ 1 }{ P_{X|Y}(x \mid Y) } - \mathcal{H}(X \mid Y) \right|^{3}
\notag \\
& \quad \le
\sum_{x \in \mathcal{A}} P_{X|Y}(x \mid Y) \log^{3} \frac{ 1 }{ P_{X|Y}(x \mid Y) } + \mathcal{H}(X \mid Y)^{3} .
\label{eq:cond-T_Y}
\end{align}
Since the mapping $u \mapsto u \ln^{3}(1/u)$ on $[0, 1]$ is maximized at $u = \mathrm{e}^{-3}$, it follows from \eqref{eq:finite-sum_is_unity} and \eqref{eq:cond-T_Y} that
\begin{align}
\sum_{x \in \mathcal{A}} P_{X|Y}(x \mid Y) \log^{3} \frac{ 1 }{ P_{X|Y}(x \mid Y) }
\le
\frac{ 27 \, |\mathcal{A}| }{ (\mathrm{e} \ln 2)^{3} }
\label{eq:maximize-bound}
\end{align}
a.s., where $\ln$ stands for the natural logarithm.
Combining \eqref{eq:DCT_H}--\eqref{eq:maximize-bound}, we can obtain from the dominated convergence theorem that
\begin{align}
\mathcal{T}(X \mid Y)
\le
\frac{ 27 \, |\mathcal{A}| }{ (\mathrm{e} \ln 2)^{3} } + \log^{3} |\mathcal{A}|
\qquad (\mathrm{a.s.}) ,
\end{align}
which implies that $\mathcal{T}(X \mid Y)$ is bounded away from infinity a.s.
This completes the proof of \propref{prop:finite-X}.
\hfill\IEEEQEDhere

\subsection{Proof of \propref{prop:finite-Y}}
\label{app:finite-Y}

Assume without loss of generality that $P_{Y}( y )$ is positive for each $y \in \mathcal{Y}$.
Hypothesis~(a) in \thref{th:max} is used only to ensure \lemref{lem:asympt_cond_eps-entropy}; thus, it suffices to prove \lemref{lem:asympt_cond_eps-entropy} without Hypothesis~(a) in \thref{th:max}.

Let $\eta( y ) \ge 0$ be a real number satisfying
\begin{align}
&
\mathbb{P}\{ \iota(X \mid Y) > \eta(y) \mid Y = y \}
\notag \\
& \qquad {}
+ \beta(y) \, \mathbb{P}\{ \iota(X \mid Y) = \eta(y) \mid Y = y \} = \varepsilon
\end{align}
for some $0 \le \beta(y) < 1$.
Define
\begin{align}
\eta_{\max}
& =
\max_{y \in \mathcal{Y}} \eta( y ) .
\end{align}
Since we now do not assume that Hypothesis~(a) in \thref{th:max} holds, there may exist a $y \in \mathcal{Y}$ satisfying $\mathcal{V}(X \mid y) = 0$.
If $\mathcal{V}(X \mid y) = 0$ for every $y \in \mathcal{Y}$, then we readily see that
\begin{align}
\mathbb{E}[ \langle \iota(X^{n} \mid Y^{n}) \mid Y^{n} \rangle_{\varepsilon} \mid Y^{n} = \bvec{y} ]
=
(1 - \varepsilon) \sum_{i = 1}^{n} \mathcal{H}(X \mid y_{i})
\end{align}
for every $n \ge 1$ and $\bvec{y} \in \mathcal{Y}^{n}$.
Therefore, \lemref{lem:asympt_cond_eps-entropy} also holds even if $V(X \mid y) = 0$ for every $y \in \mathcal{Y}$.

In the following, we assume that there exists a $y \in \mathcal{Y}$ satisfying $V(X \mid y) > 0$.
Define
\begin{align}
H_{\max}
& \coloneqq
\max_{y \in \mathcal{Y}} \mathcal{H}(X \mid y) ,
\\
V_{\max}
& \coloneqq
\max_{y \in \mathcal{Y}} \mathcal{V}(X \mid y) ,
\\
V_{\min}
& \coloneqq
\min_{\substack{ y \in \mathcal{Y} : \\ \mathcal{V}(X \mid y) > 0 }} \mathcal{V}(X \mid y) ,
\\
T_{\max}
& \coloneqq
\max_{y \in \mathcal{Y}} \mathcal{T}(X \mid y) .
\end{align}
Note that the three numbers $H_{\max}$, $V_{\max}$, and $T_{\max}$ are finite since Hypothesis~(b) in \thref{th:max} is satisfied.
To prove \lemref{lem:asympt_cond_eps-entropy} without Hypothesis~(a) in \thref{th:max}, it suffices to prove an analog of \lemref{lem:remainder_terms} stated in \appref{app:asympt_cond_eps-entropy} without Hypothesis~(a) in \thref{th:max}.
More precisely, we shall prove the following lemma.

\begin{lemma}
\label{lem:remainder_terms_finite-Y}
Recall that $b_{Y^{n}}$, $B_{Y^{n}}$, and $D_{Y^{n}}$ are defined in \eqref{def:b_yn}, \eqref{def:capital-B_yn}, and \eqref{def:capital-G_yn}, respectively (see \appref{app:asympt_cond_eps-entropy}).
Suppose that the following hold:
(i) $\mathcal{Y}$ is a finite alphabet,
(ii) there exists a $y \in \mathcal{Y}$ satisfying $V( P_{X|Y=y} ) > 0$, and
(iii) Hypothesis \emph{(b)} in \thref{th:max} holds.
Given $0 < \varepsilon < 1$, it holds that
\begin{align}
&
\Big| \mathbb{E}[ B_{Y^{n}} ] + \mathbb{E}[ D_{Y^{n}} ] - \varepsilon \, \mathbb{E}[ b_{Y^{n}} ] \Big|
\notag \\
& \quad \le
(1+\varepsilon) \left( \frac{ A \, T_{\max}^{4/3} }{ c \, V_{\min}^{3/2} } + K \, \Big( \eta_{\max} + H_{\max} + V_{\max} \, \Phi^{-1}(1 - \varepsilon) \Big) \right)
\notag \\
& \qquad {}
+ \frac{ 3 \, A \, T_{\max} }{ V_{\min} }
\end{align}
for every $n \ge K$, where $A > 0$ is an absolute constant, $c = c( \varepsilon ) > 0$ is a constant depending only on $\varepsilon$, and $K = K(\varepsilon, V_{\min}, T_{\max}) > 0$ is a constant depending on $\varepsilon$, $V_{\min}$, and $T_{\max}$. 
\end{lemma}

\begin{IEEEproof}[Proof of \lemref{lem:remainder_terms_finite-Y}]
Fix an infinite sequence $\bvec{y} = (y_{1}, y_{2}, \dots) \in \mathcal{Y}^{\mathbb{N}}$ arbitrarily.
For each $n \ge 1$, denote by $\bvec{y}^{(n)} = (y_{1}, \dots, y_{n})$ the $n$-length prefix of $\bvec{y}$. 
For each $n \ge 1$, consider two parameters $\eta_{n} \ge 0$ and $0 \le \beta_{n} < 1$ given so that
\begin{align}
&
\mathbb{P}\{ \iota(X^{n} \mid Y^{n}) > \eta_{n} \mid Y^{n} = \bvec{y}^{(n)} \}
\notag \\
& \qquad {}
+ \beta_{n} \, \mathbb{P}\{ \iota(X^{n} \mid Y^{n}) = \eta_{n} \mid Y^{n} = \bvec{y}^{(n)} \}
=
\varepsilon .
\end{align}
Let $\mathcal{K} \subset \mathbb{N}$ be the subset in which for all $k \in \mathcal{K}$, there exists a finite $\mathcal{A}_{k} \subset \mathcal{X}$ satisfying $P_{X|Y}(x \mid y_{k}) = 1/|\mathcal{A}_{k}|$ for each $x \in \mathcal{A}_{k}$.
Define $k(n) \coloneqq |\{ 1 \le k \le n \mid k \notin \mathcal{K} \}|$ for each $n \ge 1$.
Moreover, let $n_{1} \ge 1$ be chosen so that
\begin{align}
n_{1}
\coloneqq
\sup\left\{ n \ge 1 \ \middle| \ \frac{ A \, T_{\max} }{ \sqrt{k(n) + 1} \, V_{\min} } \ge \frac{ \min\{ 1 - \varepsilon, \varepsilon \} }{ 2 } \right\} .
\end{align}
Since $\mathcal{V}(X \mid y_{k}) = \mathcal{T}(X \mid y_{k}) = 0$ for every $k \in \mathcal{K}$, and since $\mathcal{V}(X \mid y_{i}) \, \mathcal{T}(X \mid y_{i}) > 0$ for every $i \in \mathbb{N} \setminus \mathcal{K}$, we observe that
\begin{align}
\frac{ A \, \mathcal{T}(X^{n} \mid \bvec{y}^{(n)}) }{ \mathcal{V}(X^{n} \mid \bvec{y}^{(n)})^{3/2} }
& =
\frac{ A \sum_{i = 1 : i \notin \mathcal{K}}^{n} \mathcal{T}(X \mid y_{i}) }{ (\sum_{i = 1 : i \notin \mathcal{K}}^{n} \mathcal{V}(X \mid| y_{i}) )^{3/2} }
\notag \\
& \le
\frac{ A \, T_{\max} }{ \sqrt{k(n)} \, V_{\min}^{3/2} } ;
\end{align}
therefore, it can be shown by the same way as \eqref{eq:estimate_eta_each-yn} that
\begin{align}
\Big| \eta_{n} - \mathcal{H}(X^{n} \mid \bvec{y}^{(n)}) - \sqrt{ \mathcal{V}(X^{n} \mid \bvec{y}^{(n)}) } \, \Phi^{-1}( 1 - \varepsilon ) \Big|
\le
\frac{ A \, T_{\max}^{4/3} }{ c \, V_{\min}^{3/2} }
\label{eq:bound_by_case1}
\end{align}
for every $n \ge n_{1}$, provided that $n_{1} < \infty$, where $A > 0$ is an absolute constant that appears in \lemref{lem:non-unif_Berry-Esseen}, and $c = c( \varepsilon ) > 0$ a constant is given in \eqref{eq:const_c}.

Now, consider the case where $n_{1} = \infty$.
Note that $n_{1} = \infty$ if and only if
\begin{align}
\lim_{n \to \infty} k( n )
\le
\left( \frac{ 2 \, A \, T_{\max} }{ V_{\min} \, \min\{ 1 - \varepsilon, \varepsilon \} } \right)^{2} - 1
\eqqcolon
K(\varepsilon, V_{\min}, T_{\max}) .
\label{eq:bound_kn}
\end{align}
It is clear from the definition of $\mathcal{K}$ that
\begin{align}
\mathcal{H}(X^{n} \mid \bvec{y}^{(n)})
=
\sum_{i = 1 : i \notin \mathcal{K}}^{n} \mathcal{H}(X \mid y_{i}) + \sum_{j = 1 : j \in \mathcal{K}}^{n} \log |\mathcal{A}_{j}|
\label{eq:sum_cond-H}
\end{align}
for each $n \ge 1$.
If $1 \in \mathcal{K}$, then
\begin{align}
\mathbb{P}\{ \iota(X_{1} \mid Y_{1}) > t \mid Y_{1} = y_{1} \}
=
\begin{cases}
0
& \mathrm{if} \ t \le \log |\mathcal{A}_{1}| ,
\\
1
& \mathrm{if} \ t > \log |\mathcal{A}_{1}| ,
\end{cases}
\end{align}
implying that
\begin{align}
\eta_{1}
=
\log |\mathcal{A}_{1}| .
\label{eq:const_eta-1}
\end{align}
Moreover, for each $k \in \mathcal{K}$ satisfying $k \ge 2$, we see that
\begin{align}
&
\mathbb{P}\{ \iota(X^{k} \mid Y^{k}) > \eta_{k} \mid Y^{k} = \bvec{y}^{(n)} \}
\notag \\
& \quad =
\mathbb{P}\{ \iota(X^{k-1} \mid Y^{k-1}) > \eta_{k-1} + \log |\mathcal{A}| \mid Y^{k} = \bvec{y}^{(n)} \}
\label{eq:const_eta-k}
\end{align}
for every $t \in \mathbb{R}$, implying that
\begin{align}
\eta_{k}
=
\eta_{k-1} + \log |\mathcal{A}_{k}| .
\end{align}
Therefore, since $\mathcal{V}(X \mid y_{k}) = 0$ for each $k \in \mathcal{K}$, it follows from \eqref{eq:bound_kn}, \eqref{eq:sum_cond-H}, \eqref{eq:const_eta-1}, and \eqref{eq:const_eta-k} that
\begin{align}
&
\Big| \eta_{n} - \mathcal{H}(X^{n} \mid \bvec{y}^{(n)}) - \sqrt{ \mathcal{V}(X^{n} \mid \bvec{y}^{(n)}) } \, \Phi^{-1}( 1 - \varepsilon ) \Big|
\notag \\
& \ \le
K(\varepsilon, V_{\min}, T_{\max}) \, \Big( \eta_{\max} + H_{\max} + V_{\max} \, \Phi^{-1}(1 - \varepsilon) \Big)
\label{eq:bound_by_case2}
\end{align}
for every $n \ge 1$, provided that $n_{1} = \infty$.

Combining \eqref{eq:bound_by_case1} and \eqref{eq:bound_by_case2}, we obtain
\begin{align}
&
\Big| \eta_{n} - \mathcal{H}(X^{n} \mid \bvec{y}^{(n)}) - \sqrt{ \mathcal{V}(X^{n} \mid \bvec{y}^{(n)}) } \, \Phi^{-1}( 1 - \varepsilon ) \Big|
\notag \\
& \le
\frac{ A \, T_{\max}^{4/3} }{ c \, V_{\min}^{3/2} } + K(\varepsilon, V_{\min}, T_{\max}) \, \Big( \eta_{\max} + H_{\max} + V_{\max} \, \Phi^{-1}(1 - \varepsilon) \Big)
\end{align}
for every $n \ge K(\varepsilon, V_{\min}, T_{\max})$.
Therefore, since the infinite sequence $\bvec{y} = (y_{1}, y_{2}, \dots)$ is arbitrary, we have
\begin{align}
&
\mathbb{E}[ |b_{Y^{n}}| ]
\le
\frac{ A \, T_{\max}^{4/3} }{ c \, V_{\min}^{3/2} }
\notag \\
& \quad {}
+ K(\varepsilon, V_{\min}, T_{\max}) \, \Big( \eta_{\max} + H_{\max} + V_{\max} \, \Phi^{-1}(1 - \varepsilon) \Big)
\label{eq:absolute-E_b_yn}
\end{align}
for every $n \ge K(\varepsilon, V_{\min}, T_{\max})$.
The proof of \lemref{lem:remainder_terms_finite-Y} is finally completed by combining \eqref{eq:capital-B_small-b}, \eqref{eq:upper-bound_Gn}, \eqref{eq:lower-bound_Gn}, and \eqref{eq:absolute-E_b_yn}.
\end{IEEEproof}

\lemref{lem:asympt_cond_eps-entropy} without Hypothesis~(a) in \thref{th:max} is now ensured by applying \lemref{lem:remainder_terms_finite-Y} to \eqref{eq:fourth-identity_max}.
This completes the proof of \propref{prop:finite-Y}.
\hfill\IEEEQEDhere

\section{A Counterexample of \\ Converse Part in \cite[Section~II-A]{kostina_polyanskiy_verdu_2015}}
\label{app:counterexample}

In this appendix, we consider a variable-length compression problem for a discrete memoryless source in the \emph{absence} of side-information.
Here, we use similar notations to those in \cite{kostina_polyanskiy_verdu_2015}.
Fix a real number $0 < \epsilon < 1$.
Suppose that the source $S$ defined as a positive-integer-valued r.v.\ satisfies that $P_{S}( 1 ) \ge P_{S}( 2 ) \ge \cdots$.
As written in \remref{eq:another_stochastic-code}, consider a stochastic encoder $P_{W|S}$ defined as a conditional probability distribution of $S$ given a $\{ 0, 1 \}^{\ast}$-valued r.v.\ $W$.
In addition, consider a deterministic decoder $\mathsf{g} : \{ 0, 1 \}^{\ast} \to \{ 1, 2, \dots \}$.

In \cite[Section~II-A]{kostina_polyanskiy_verdu_2015}, it was stated that an optimal variable-length stochastic code $(P_{W|S}^{\ast}, \mathsf{g}^{\ast})$ like \eqref{eq:optimal-F_max} and \eqref{eq:optimal-g_max}, or like \eqref{eq:optimal-F_avg} and \eqref{eq:optimal-g_avg}, has an average codeword length and error probability that are both no larger than those of the given variable-length stochastic code $(P_{W|S}, \mathsf{g})$.
This is an unconditional version of Lemmas~\ref{lem:optimal-code_max} and~\ref{lem:optimal-code_avg}.
In the particular case of $S$ satisfying $P_{S}( 1 ) + P_{S}( 2 ) + \cdots + P_{S}( M ) = 1 - \epsilon$ for some $M \ge 1$, its proof is written in Lines~3--19 of the left column of Page~4318 of \cite[Section~II-A]{kostina_polyanskiy_verdu_2015} as follows:

``\emph{Formally, for a given encoder $P_{W|S}$, the optimal decoder is always deterministic and we denote it by $\mathsf{g}$.
Consider \emph{[sic]} $w_{0} \in \{ 0, 1 \}^{\ast} \setminus \varnothing$ and source realization $m$ with $P_{W|S=m}( w_{0} ) > 0$.
If $\mathsf{g}( w_{0} ) \neq m$, the average length can be decreased, without affecting the probability of error, by setting $P_{W|S=m}( w_{0} ) = 0$ and adjusting $P_{W|S=m}( \varnothing )$ accordingly.
This argument implies that the optimal encoder has at most one source realization $m$ mapping to each $w_{0} \neq \varnothing$.
Next, let $m_{0} = \mathsf{g}( \varnothing )$ and by a similar argument conclude that $P_{W|S=m_{0}}( \varnothing ) = 1$.
But then, interchanging $m_{0}$ and $1$ leads to the same or better probability of error and shorter average length, which implies that the optimal encoder maps $1$ to $\varnothing$.
Continuing in the same manner for $m_{0} = \mathsf{g}(0), \mathsf{g}(1), \dots , \mathsf{g}( \mathsf{f}_{S}^{\ast}(M))$, we conclude that the optimal code maps $\mathsf{f}(m) = \mathsf{f}_{S}^{\ast}(m)$, $m = 1, .\dots , M$.
Finally, assigning the remaining source outcomes whose total mass is $\epsilon$ to $\varnothing$ shortens the average length without affecting the error probability, so $\mathsf{f}(m) = \varnothing$, $m > M$ is optimal.}''

In the above proof, note that $\mathsf{f} : \{ 1, 2, \dots \} \to \{ 0, 1 \}^{\ast}$ denotes a deterministic encoder induced by the stochastic encoder $P_{W|S}$ constructed by the above procedures, and $\mathsf{f}_{S}^{\ast} : \{ 1, 2, \dots \} \to \{ 0, 1 \}^{\ast}$ stands for the optimal one-to-one code satisfying $\mathsf{f}_{S}^{\ast}( m ) = \bvec{b}_{m}$ for every $m \ge 1$ (cf.\ \cite{alon_orlitsky_1994}), where $\{ \bvec{b}_{i} \}_{i = 1}^{\infty}$ denotes the lexicographical ordering of the strings in $\{ 0, 1 \}^{\ast}$ (cf.\ \appref{app:optimal-code_max}).

One can, however, show a counterexample of the triplet $(X, P_{B|X}, \mathsf{g})$ in which the above proof strategy fails to prove the desired assertion.
Let $\epsilon = 1/6$ be fixed, and $P_{S}$ the source distribution given as
\begin{align}
P_{S}( m )
=
\begin{dcases}
\frac{ 1 }{ 2 }
& \mathrm{if} \ m = 1 ,
\\
\frac{ 1 }{ 3 }
& \mathrm{if} \ m = 2 ,
\\
\frac{ 1 }{ 6 }
& \mathrm{if} \ m = 3 ,
\\
0
& \mathrm{otherwise} .
\end{dcases}
\end{align}
Since $P_{S}( 1 ) + P_{S}( 2 ) = 1 - \epsilon$, it is clear that $M = 2$ in this case.
Consider the stochastic encoder $P_{W|S}$ defined as
\begin{align}
P_{W|S = m}( w )
=
\begin{dcases}
\frac{ 5 }{ 6 }
& \mathrm{if} \ m = 1 \ \mathrm{and} \ w = 0 ,
\\
\frac{ 1 }{ 6 }
& \mathrm{if} \ m = 1 \ \mathrm{and} \ w = 1
\\
1
& \mathrm{if} \ m = 2 \ \mathrm{and} \ w = \varnothing ,
\\
\frac{ 1 }{ 2 }
& \mathrm{if} \ m = 3 \ \mathrm{and} \ w \in \{ 0, 1 \} ,
\\
0
& \mathrm{otherwise} ,
\end{dcases}
\label{eq:counterexample_enc}
\end{align}
and the decoder $\mathsf{g}$ given as
\begin{align}
\mathsf{g}( w )
=
\begin{dcases}
1
& \mathrm{if} \ w = 0 ,
\\
2
& \mathrm{if} \ w = \varnothing ,
\\
3
& \mathrm{if} \ w = 1 ,
\\
\mathrm{arbitrary}
& \mathrm{otherwise} .
\end{dcases}
\label{eq:counterexample_dec}
\end{align}
Then, the error probability and the average codeword length can be calculated as $\mathbb{P}\{ S \neq \hat{S} \} = 1/6 = \epsilon$ and $\mathbb{E}[ \ell( W ) ] = 2/3$, respectively, where $\hat{S} = \mathsf{g}( W )$.
By applying the above proof strategy to the given variable-length stochastic code $(P_{W|S}, \mathsf{g})$, we construct the optimal encoder and decoder:
\begin{align}
P_{W|S = m}^{\ast}( w )
& =
\begin{dcases}
1
& \mathrm{if} \ m = 1 \ \mathrm{and} \ w = \varnothing ,
\\
1
& \mathrm{if} \ m = 2 \ \mathrm{and} \ w = 0 ,
\\
1
& \mathrm{if} \ m = 3 \ \mathrm{and} \ w = \varnothing ,
\\
0
& \mathrm{otherwise} ,
\end{dcases}
\\
\mathsf{g}^{\ast}( w )
& =
\begin{dcases}
1
& \mathrm{if} \ w = \varnothing ,
\\
2
& \mathrm{if} \ w = 0 ,
\\
\mathrm{arbitrary}
& \mathrm{otherwise} ,
\end{dcases}
\end{align}
respectively, yielding the optimal performance:
$\mathbb{P}\{ S \neq \hat{S} \} = 1/6 = \epsilon$ and $\mathbb{E}[ \ell(W) ] = 1/3$.

By the procedure ``\emph{Consider \emph{[sic]} $w_{0} \in \{ 0, 1 \}^{\ast} \setminus \varnothing$ and source realization $m$ with $P_{W|S=m}( w_{0} ) > 0$.
If $\mathsf{g}( w_{0} ) \neq m$, the average length can be decreased, without affecting the probability of error, by setting $P_{W|S=m}( w_{0} ) = 0$ and adjusting $P_{W|S=m}( \varnothing )$ accordingly,}'' we modify the stochastic encoder $P_{W|S}$ to another $P_{W|S}^{(1)}$ so that $P_{W|S = m}^{(1)}( w_{0} ) = 0$ if $w_{0} \neq \varnothing$ and $\mathsf{g}( w_{0} ) \neq m$; and $P_{W|S = m}^{(1)}( w_{0} ) = P_{W|S = m}( w_{0} )$ if $w_{0} \neq \varnothing$ and $\mathsf{g}( w_{0} ) = m$; and
\begin{align}
P_{W|S = m}^{(1)}( \varnothing )
& =
P_{W|S = m}( \varnothing )
\notag \\
& \qquad {}
+ \sum_{\substack{ w \in \{ 0, 1 \}^{\ast} \setminus \{ \varnothing \} : \\ \mathsf{g}( w ) \neq m }} P_{W|S = m}( w )
\end{align}
for every $m \ge 1$.
Namely, we see from \eqref{eq:counterexample_enc} that
\begin{align}
P_{W|S = m}^{(1)}( w )
=
\begin{dcases}
\frac{ 5 }{ 6 }
& \mathrm{if} \ m = 1 \ \mathrm{and} \ w = 0 ,
\\
\frac{ 1 }{ 6 }
& \mathrm{if} \ m = 1 \ \mathrm{and} \ w = \varnothing ,
\\
1
& \mathrm{if} \ m = 2 \ \mathrm{and} \ w = \varnothing ,
\\
\frac{ 1 }{ 2 }
& \mathrm{if} \ m = 3 \ \mathrm{and} \ w \in \{ \varnothing, 1 \} ,
\\
0
& \mathrm{otherwise} .
\end{dcases}
\end{align}
As mentioned as ``\emph{the average length can be decreased, without affecting the probability of error,}'' the error probability and the average codeword length are then given as $\mathbb{P}\{ S \neq \hat{S} \} = 1/6 = \epsilon$ and $\mathbb{E}[ \ell( W ) ] = 1/2$, respectively.

Next, consider the procedure ``\emph{Next, let $m_{0} = \mathsf{g}( \varnothing )$ and by a similar argument conclude that $P_{W|S=m_{0}}( \varnothing ) = 1$.
But then, interchanging $m_{0}$ and $1$ leads to the same or better probability of error and shorter average length, which implies that the optimal encoder maps $1$ to $\varnothing$.}''
In this example, we see from \eqref{eq:counterexample_dec} that $m_{0} = 2$, and interchanging the source symbols $m_{0} = 2$ and $1$ results in the following encoder and decoder:
\begin{align}
P_{W|S = m}^{(2)}( w )
& =
\begin{dcases}
1
& \mathrm{if} \ m = 1 \ \mathrm{and} \ w = \varnothing ,
\\
\frac{ 5 }{ 6 }
& \mathrm{if} \ m = 2 \ \mathrm{and} \ w = 0 ,
\\
\frac{ 1 }{ 6 }
& \mathrm{if} \ m = 2 \ \mathrm{and} \ w = \varnothing ,
\\
\frac{ 1 }{ 2 }
& \mathrm{if} \ m = 3 \ \mathrm{and} \ w \in \{ \varnothing, 1 \} ,
\\
0
& \mathrm{otherwise} ,
\end{dcases}
\label{eq:counterexample_f2} \\
\mathsf{g}^{(2)}( w )
& =
\begin{dcases}
1
& \mathrm{if} \ w = \varnothing ,
\\
2
& \mathrm{if} \ w = 0 ,
\\
3
& \mathrm{if} \ w = 1 ,
\\
\mathrm{arbitrary}
& \mathrm{otherwise} ,
\end{dcases}
\end{align}
respectively,
and we readily see that
$\mathbb{P}\{ S \neq \hat{S} \} = 5/36 < \epsilon$ and $\mathbb{E}[ \ell( W ) ] = 13/36$.

Moreover, consider the iterative procedure ``\emph{Continuing in the same manner for $m_{0} = \mathsf{g}(0), \mathsf{g}(1), \dots , \mathsf{g}( \mathsf{f}_{S}^{\ast}(M))$, we conclude that the optimal code maps $\mathsf{f}(m) = \mathsf{f}_{S}^{\ast}(m)$, $m = 1, .\dots , M$.}''
However, since $M = 2$ and $\mathsf{g}^{(2)} \circ \mathsf{f}_{S}^{\ast}( 2 ) = 2$, this iterative procedure does not change the variable-length stochastic code $(P_{W|S}^{(2)}, \mathsf{g}^{(2)})$ at all.
Namely, since the stochastic encoder $P_{W|S}^{(2)}$ given in \eqref{eq:counterexample_f2} stochastically maps the source symbols $m = 1$ and $m = 3$, we cannot conclude that ``\emph{the optimal code maps $\mathsf{f}(m) = \mathsf{f}_{S}^{\ast}(m)$, $m = 1, \dots , M$}'' in this example.

Finally, consider the procedure ``\emph{Finally, assigning the remaining source outcomes whose total mass is $\epsilon$ to $\varnothing$ shortens the average length without affecting the error probability, so $\mathsf{f}(m) = \varnothing$, $m > M$ is optimal.}''
Then, we have the following encoder:
\begin{align}
P_{W|S = m}^{(3)}( w )
& =
\begin{dcases}
1
& \mathrm{if} \ m = 1 \ \mathrm{and} \ w = \varnothing ,
\\
\frac{ 5 }{ 6 }
& \mathrm{if} \ m = 2 \ \mathrm{and} \ w = 0 ,
\\
\frac{ 1 }{ 6 }
& \mathrm{if} \ m = 2 \ \mathrm{and} \ w = \varnothing ,
\\
1
& \mathrm{if} \ m = 3 \ \mathrm{and} \ w = \varnothing ,
\\
0
& \mathrm{otherwise} ,
\end{dcases}
\end{align}
and we observe from the variable-length stochastic code $(P_{W|S}^{(3)}, \mathsf{g}^{(2)})$ that $\mathbb{P}\{ S \neq \hat{S} \} = 2/9 > \epsilon$ and $\mathbb{E}[ \ell( W ) ] = 5/18$.
This calculation, however, violates the error probability constraint $\mathbb{P}\{ S \neq \hat{S} \} \le \epsilon$, which justifies our claim that there is an error in the proof in \cite[Section~II-A]{kostina_polyanskiy_verdu_2015}.

\section*{Acknowledgement}

The authors would like to thank Dr.\ Lei Yu for helpful discussions.
The authors are also grateful to the two reviewers for their valuable and detailed comments.

\bibliographystyle{IEEEtran}
\bibliography{IEEEabrv,mybib}

\newpage

\begin{IEEEbiographynophoto}{Yuta Sakai}(Member, IEEE)
was born in Japan in 1992.
He is currently a Research Fellow in the Department of Electrical and Computer Engineering at the National University of Singapore (NUS).
He received the B.E.\ and M.E.\ degrees in the Department of Information Science from the University of Fukui in 2014 and 2016, respectively, and the Ph.D.\ degree in the Advanced Interdisciplinary Science and Technology from the University of Fukui in 2018.
His research interests include information theory and coding theory.
\end{IEEEbiographynophoto}

\begin{IEEEbiographynophoto}{Vincent Y.\ F.\ Tan}(Senior~Member, IEEE)
was born in Singapore in 1981. He is currently a Dean's Chair Associate Professor in the Department of Electrical and Computer Engineering and the Department of Mathematics at the National University of Singapore (NUS).
He received the B.A.\ and M.Eng.\ degrees in Electrical and Information Sciences from Cambridge University in 2005 and the Ph.D.\ degree in Electrical Engineering and Computer Science (EECS) from the Massachusetts Institute of Technology (MIT)  in 2011.  His research interests include information theory, machine learning, and statistical signal processing.

Dr.\ Tan received the MIT EECS Jin-Au Kong outstanding doctoral thesis prize in 2011, the NUS Young Investigator Award in 2014,  the Singapore National Research Foundation (NRF) Fellowship (Class of 2018) and the NUS Young Researcher Award in 2019. He was also an IEEE Information Theory Society Distinguished Lecturer during the years 2018 to 2019. He is currently serving as an Associate Editor of the IEEE Transactions on Signal Processing and an Associate Editor of Machine Learning for the IEEE Transactions on Information Theory.
\end{IEEEbiographynophoto}

\end{document}